\documentclass[aip,graphicx]{revtex4-1}

\usepackage[english]{babel}

\usepackage{changes} 

\usepackage{amsmath}
\usepackage{embedfile}
\usepackage{graphicx}
\usepackage[colorlinks=true, allcolors=blue]{hyperref}
\usepackage{wrapfig}
\usepackage{multirow}
\usepackage{eurosym}
\usepackage{url}
\usepackage{array}
\usepackage{longtable}
\usepackage{float}

\begin{document}
\title{\large{STE-QUEST} \\ \normalsize{Space Time Explorer and QUantum Equivalence principle Space Test: The 2022 medium-class mission concept}}


\author{Naceur Gaaloul}
\affiliation{Institute of Quantum Optics, Leibniz University Hannover, Welfengarten 1, 30167 Hannover, Germany}

\author{Holger Ahlers}
\affiliation{German Aerospace Center (DLR), Institute for Satellite Geodesy and Inertial
Sensing, Callinstr. 30b, 30167 Hannover, Germany}

\author{Leonardo Badurina}
\affiliation{Walter Burke Institute for Theoretical Physics, California Institute of Technology, 1200 East California Boulevard, Pasadena, CA 91125, USA}

\author{Angelo Bassi}
\affiliation{Department of Physics, University of Trieste, Strada Costiera 11, 34151 Trieste, Italy}
\affiliation{Trieste Section, Istituto Nazionale di Fisica Nucleare (INFN), Via Valerio 2, 34127 Trieste, Italy}

\author{Baptiste Battelier}
\affiliation{Laboratoire Photonique, Numérique et Nanosciences (LP2N), Institut d'Optique Graduate School, CNRS, Université de Bordeaux, Rue François Mitterrand, 33400 Talence, France}

\author{Quentin Beaufils}
\affiliation{ LTE, Observatoire de Paris, Universit\'e PSL, Sorbonne Universit\'e, Universit\'e de Lille, LNE, CNRS 61 Avenue de l’Observatoire, 75014 Paris, France}

\author{Kai Bongs}
\affiliation{Institute of Quantum Technologies, German Aerospace Center (DLR), Wilhelm-Runge-Straße 10, 89081 Ulm, Germany}

\author{Philippe Bouyer}
\affiliation{Quantum Delta NL, University of Amsterdam, Technical University Eindhoven, Nether-
lands}

\author{Claus Braxmaier}
\affiliation{Institute of Quantum Technologies, German Aerospace Center (DLR), Wilhelm-Runge-Straße 10, 89081 Ulm, Germany}
\affiliation{Institut für Mikroelektronik, Universität Ulm, Albert-Einstein-Allee 45, 89069 Ulm, Germany}

\author{Oliver Buchmueller}
\affiliation{Physics Department, Imperial College London, Prince Consort Road, London SW7 2AZ, United Kingdom}
\affiliation{Department of Physics, University of Oxford, Parks Road, Oxford OX1 3PU, United Kingdom}

\author{Matteo Carlesso}
\affiliation{Department of Physics, University of Trieste, Strada Costiera 11, 34151 Trieste, Italy}
\affiliation{Trieste Section, Istituto Nazionale di Fisica Nucleare (INFN), Via Valerio 2, 34127 Trieste, Italy}

\author{Eric Charron}
\affiliation{Université Paris-Saclay, CNRS, Institut des Sciences Moléculaires d'Orsay, F-91405 Orsay, France}

\author{Maria Luisa Chiofalo}
\affiliation{Department of Physics "Enrico Fermi", University of Pisa, Largo Bruno Pontecorvo 3, 56126 Pisa, Italy}
\affiliation{INFN-Pisa, Largo Bruno Pontecorvo 3, 56127 Pisa, Italy}

\author{Robin Corgier}
\affiliation{ LTE, Observatoire de Paris, Universit\'e PSL, Sorbonne Universit\'e, Universit\'e de Lille, LNE, CNRS 61 Avenue de l’Observatoire, 75014 Paris, France}

\author{Sandro Donadi}
\affiliation{Centre for Quantum Materials and Technologies, School of Mathematics and Physics, Queen’s University Belfast, BT7 1NN, United Kingdom}
\affiliation{Trieste Section, Istituto Nazionale di Fisica Nucleare (INFN), Via Valerio 2, 34127 Trieste, Italy}

\author{Fabien Droz}
\affiliation{Centre Suisse d'Electronique et de Microtechnique (CSEM), Rue Jaquet-Droz 1, 2002 Neuchâtel, Switzerland}

\author{John Ellis}
\affiliation{Physics Department, King's College London, Strand, London WC2R 2LS, United Kingdom}

\author{Frédéric Estève}
\affiliation{Centre National d'Études Spatiales (CNES), 18 Avenue Édouard Belin, 31400 Toulouse, France}

\author{Enno Giese}
\affiliation{Technische Universit{\"a}t Darmstadt, Fachbereich Physik, Institut f{\"u}r Angewandte Physik, Schlo{\ss}gartenstr. 7, 64289 Darmstadt, Germany}

\author{Jens Grosse}
\affiliation{Center of Applied Space Technology and Microgravity (ZARM), University of Bremen, Am Fallturm 2, 28359 Bremen, Germany}

\author{Aurélien Hees}
\affiliation{ LTE, Observatoire de Paris, Universit\'e PSL, Sorbonne Universit\'e, Universit\'e de Lille, LNE, CNRS 61 Avenue de l’Observatoire, 75014 Paris, France}

\author{Thomas A. Hensel}
\affiliation{Max Planck Institute for Multidisciplinary Sciences, Am Faßberg 11, 37077 Göttingen, Germany}

\author{Waldemar Herr}
\affiliation{German Aerospace Center (DLR), Institute for Satellite Geodesy and Inertial
Sensing, Callinstr. 30b, 30167 Hannover, Germany}

\author{Philippe Jetzer}
\affiliation{Department of Physics, University of Zurich, Winterthurerstrasse 190, 8057 Zurich, Switzerland}

\author{Gina Kleinsteinberg}
\affiliation{Institute of Quantum Optics, Leibniz University Hannover, Welfengarten 1, 30167 Hannover, Germany}

\author{Carsten Klempt}
\affiliation{German Aerospace Center (DLR), Institute for Satellite Geodesy and Inertial
Sensing, Callinstr. 30b, 30167 Hannover, Germany}

\author{Steve Lecomte}
\affiliation{Centre Suisse d'Electronique et de Microtechnique (CSEM), Rue Jaquet-Droz 1, 2002 Neuchâtel, Switzerland}

\author{Louise Lopes}
\affiliation{CNES – Toulouse Space Center, 18 Avenue Édouard Belin, 31400 Toulouse, France}

\author{Sina Loriani}
\affiliation{Institute of Meteorology and Climate Research, University of Potsdam, Potsdam, Germany}

\author{Victor Martín}
\affiliation{Institute of Space Sciences (ICE, CSIC), Institute of Space Studies of Catalonia (IEEC), Barcelona, Spain}

\author{Gilles Métris}
\affiliation{Université Côte d'Azur, Observatoire de la Côte d'Azur, CNRS, IRD, Géoazur, France}

\author{Gabriel Müller}
\affiliation{Institute of Quantum Optics, Leibniz University Hannover, Germany}

\author{Miquel Nofrarias}
\affiliation{Institute of Space Sciences (ICE, CSIC), Institute of Space Studies of Catalonia (IEEC), Barcelona, Spain}

\author{Franck Pereira Dos Santos}
\affiliation{ LTE, Observatoire de Paris, Universit\'e PSL, Sorbonne Universit\'e, Universit\'e de Lille, LNE, CNRS 61 Avenue de l’Observatoire, 75014 Paris, France}

\author{Ernst M. Rasel}
\affiliation{Institute of Quantum Optics, Leibniz University Hannover, Germany}

\author{Mike Salter}
\affiliation{RAL Space, UKRI-STFC Rutherford Appleton Laboratory, Didcot, OX11 0QX, United Kingdom}

\author{Dennis Schlippert}
\affiliation{Institute of Quantum Optics, Leibniz University Hannover, Germany}

\author{Christian Schubert}
\affiliation{German Aerospace Center (DLR), Institute for Satellite Geodesy and Inertial
Sensing, Callinstr. 30b, 30167 Hannover, Germany}

\author{Thilo Schuldt}
\affiliation{Institute of Quantum Technologies, German Aerospace Center (DLR), Wilhelm-Runge-Straße 10, 89081 Ulm, Germany}

\author{Carlos F. Sopuerta}
\affiliation{Institute of Space Sciences (ICE, CSIC), Institute of Space Studies of Catalonia (IEEC), Barcelona, Spain}

\author{Christian Struckmann}
\affiliation{Institute of Quantum Optics, Leibniz University Hannover, Germany}

\author{Guglielmo M. Tino}
\affiliation{Dipartimento di Fisica e Astronomia and LENS, Università di Firenze, INO-CNR, INFN Sezione di Firenze, via Sansone 1, I-50019 Sesto Fiorentino (FI), Italy}

\author{Tristan Valenzuela}
\affiliation{RAL Space, UKRI-STFC Rutherford Appleton Laboratory, Didcot, OX11 0QX, United Kingdom}

\author{Wolf von Klitzing}
\affiliation{Institute of Electronic Structure and Laser (IESL), Foundation for Research and Technology – Hellas (FORTH), Heraklion, Greece}

\author{Lisa Wörner}
\affiliation{German Aerospace Center (DLR), Institute for Satellite Geodesy and Inertial
Sensing, Callinstr. 30b, 30167 Hannover, Germany}

\author{Nan Yu}
\affiliation{Jet Propulsion Laboratory, California Institute of Technology, Pasadena, CA, USA}

\author{Martin Zelan}
\affiliation{Measurement Science and Technology, RISE Research Institutes of Sweden, Borås, Sweden}

\author{Peter Wolf}
\affiliation{ LTE, Observatoire de Paris, Universit\'e PSL, Sorbonne Universit\'e, Universit\'e de Lille, LNE, CNRS 61 Avenue de l’Observatoire, 75014 Paris, France}

\date{\today}

\maketitle

\newpage

\textbf{Abstract:} Space-borne quantum technologies, particularly those based on atom interferometry, are heralding a new era of strategic and robust space exploration. The unique conditions of space, characterized by low noise and low gravity environments, open up diverse possibilities for applications ranging from precise time and frequency transfer to Earth Observation and the search of new Physics.
In this paper, we summarise the M-class mission proposal in response to the 2022 call in ESA's science program: Space-Time Explorer and Quantum Equivalence Principle Space Test (STE-QUEST). It consists in a satellite mission featuring a dual-species atom interferometer operating over extended durations. This mission aims to tackle three of the most fundamental questions in Physics: \textit{(i)} testing the universality of free fall with an accuracy better than one part in $10^{-17}$, \textit{(ii)} exploring various forms of Ultra-Light Dark Matter, and \textit{(iii)} scrutinizing the foundations of Quantum Mechanics.

\begin{figure}[ht]
\centering
\vspace{1.5cm}
\includegraphics[width=0.6\textwidth]{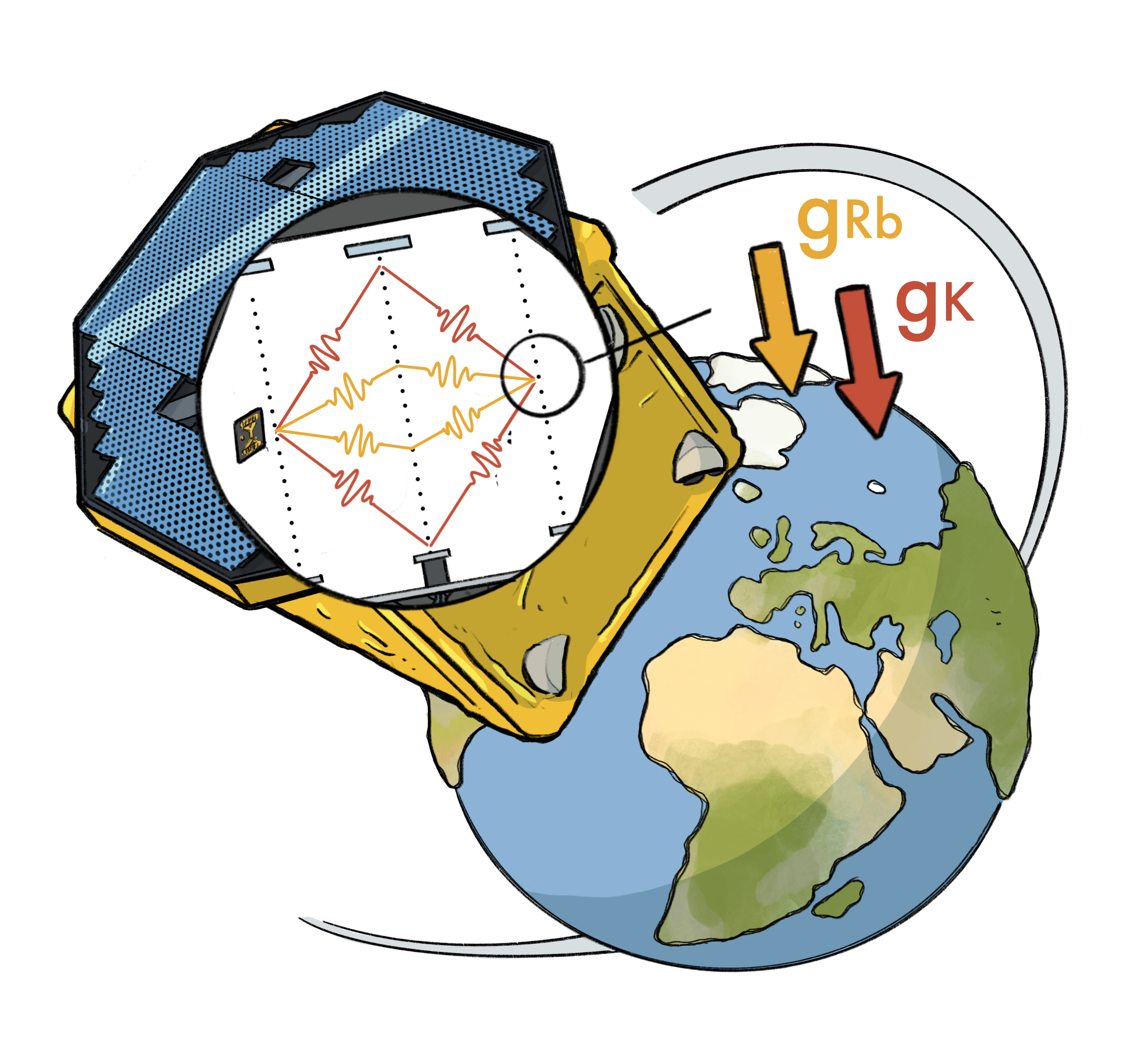}
\label{fig:title}
\end{figure}

 \newpage

\section{Introduction}

Einstein’s theory of general relativity (GR) is a cornerstone of our current description of the physical world. It is used to understand the flow of time in the presence of gravity, the motion of bodies from satellites to galaxy clusters, the propagation of electromagnetic waves in the presence of massive bodies, the evolution of stars, and the dynamics of the Universe as a whole. Although very successful so far, general relativity as well as numerous other alternative or more general theories of gravitation are classical theories. As such, they are fundamentally incomplete, because they do not include quantum effects. A theory solving this problem would represent a crucial step towards the unification of all fundamental forces of Nature. Several concepts have been proposed and are currently under investigation (e.g., string theory, quantum gravity, extra spatial dimensions) to bridge this gap and most of them lead to violations of the basic principles of GR. Therefore, a full understanding of gravity will require observations or experiments able to determine the relationship of gravity with the quantum world. This topic is a prominent field of activity with repercussions covering the complete range of physical phenomena, from particle and nuclear physics to galaxies and the Universe as a whole including dark matter and dark energy.

STE-QUEST (summarized in Tab. \ref{tab:exec_summary}) will address the most fundamental cornerstone of GR, the Einstein Equivalence Principle (EEP) by testing two of its three sub-principles: the Universality of Free Fall (UFF) and the Local Lorentz Invariance (LLI) using the most sensitive quantum sensors, i.e., atom interferometers, thereby also exploring the relationship between gravitation and the quantum world. 

The on board dual-species atom interferometer will use $^{41}$K and $^{87}$Rb atoms in quantum degenerate gases (Bose-Einstein Condensates) and in quantum states that have no classical analogues, i.e., coherent superposition states with macroscopic separations ($\le$ 130 cm) which are up to 3 orders of magnitude larger than the sizes of the individual wave-packets. A differential interferometric test will detect or constrain a violation of the UFF down to the $10^{-17}$ level.

The exceptional sensitivity of the STE-QUEST interferometer makes it possible to explore another challenge of modern physics, namely the detection of Dark Matter (DM), specifically ultralight dark matter (ULDM) candidates with masses below an eV that have recently gained much interest due to the lack of detection of more massive candidates at the Large Hadron Collider at CERN. 
STE-QUEST will expand the search for ULDM by extending the parameter space probed by several orders of magnitude for a large class of models (both scalar and vector particles) and offers the possibility of a groundbreaking discovery by providing the first direct detection of DM. 

The extremely low level of expansion energies ($\approx$ 10 pK) accessible with the atomic ensembles, and the long free fall times ($\leq 50$ s) used in STE-QUEST unlock the potential of
an additional scientific objective, namely to test the foundations of Quantum Mechanics by probing the limits of validity of the quantum superposition principle for larger systems. STE-QUEST will extend tests of the quantum superposition principle by probing Continuous Spontaneous Localization (CSL) collapse models with a 4-order-of-magnitude improvement over the state of the art, reaching the Ghirardi, Rimini and Weber (GRW) limit. Other modifications of quantum mechanics, such as the Di\'osi-Penrose gravitational collapse model, will also be tested.

The genesis of the STE-QUEST science case dates back to the consultation process conducted in 2009 by the ESA-appointed “Fundamental Physics Roadmap Advisory Team” (FPR-AT). FPR-AT was convened to draw up recommendations on the scientific and technological roadmap necessary to lead Europe toward the realization of future fundamental physics missions in the framework of the Cosmic Vision 2015-2025 plan. In the resulting roadmap document, FPR-AT recommended the concept of a medium-class mission testing the Einstein Equivalence Principle (EEP), specifically addressing UFF tests by tracking the propagation of matter waves in a differential atom interferometer and thereby addressing the quantum counterpart of classical tests based on macroscopic masses. As a result of the FPR-AT recommendation, ESA initiated a Concurrent Design Facility (CDF) study to investigate the feasibility of a clock mission testing the Einstein Equivalence Principle through the gravitational red-shift effect. The study, denominated STE, laid the foundations for the STE-QUEST mission concept, which was complemented with a dual atom interferometer performing a UFF test on quantum matter waves. Submitted in reply to the 2010 M3 call, STE-QUEST was recommended by the ESA advisory structure and finally selected by the agency for a 3 year assessment study. The assessment study gave rise to the assessment study report, with a more detailed description of the science objectives published in reference~\cite{Altschul2015}. In early 2015 STE-QUEST was re-submitted to the M4 call. The M4 version was designed around a core payload consisting of an Atom Interferometer (ATI) and a MicroWave Link (MWL). 

In the current proposal submitted as a response to the 2022 call in ESA's science program for M-class missions, the STE-QUEST core payload was simplified to the ATI alone, in order to focus on the UFF test and related measurements. The spacecraft and low-Earth, circular orbit were optimized accordingly in terms of mass accommodation, power requirements and perturbations (thermal, magnetic, gravity gradients, etc.). This has led to significant improvements in performance, further helped by recent developments on gravity-gradient control~\cite{Roura2017, Loriani2020}, one of the main limiting systematic effects in the M4 proposal. 

Finally, the current STE-QUEST proposal benefits from significant recent technological/mission heritage. MICROSCOPE and LISA-Pathfinder have demonstrated adequate drag-free and attitude control technology, and more generally provided a wealth of data and experience on related perturbations and systematic effects, and their mitigation. Cold atom accelerometers have been flown in microgravity settings (Zero-g Airbus, Drop tower, sounding rockets), within projects like ICE~\cite{Condon2019,barrett2016,Geiger2011}, QUANTUS~\cite{Deppner2021,Rudolph2015,vanZoest2010}, PRIMUS~\cite{Vogt2020,Kulas2017} and MAIUS~\cite{Lachmann2021,becker2018}, and are being actively developed and qualified for space, e.g., in the framework of the EU CARIOQA-PMP project~\cite{carioqa-pmp-approved2022}. More generally, cold atom and quantum technologies in space have been the center of much recent interest in diverse communities ranging from particle physics through Earth observation to cosmology. STE-QUEST is well embedded in that context as an integral part of the recently formulated community roadmap for cold atoms in space~\cite{CommunityWorkshop}.


During the selection process in 2022 STE-QUEST, for the first time, passed the technological and programmatic (TRL and cost) screening, which shows that the technology has significantly matured over the years and is now considered mature for space. However, the scientific selection panel, weighing the science of planetary, astronomical, or fundamental physics missions against each other, did not consider the “more likely, negative” outcome sufficient to justify an M-class budget. In response, hundreds of scientists signed an open letter to ESA's Director of Science to emphasize that even a null result (no violation detected) at STE-QUEST's unprecedented sensitivity ($\eta\approx10^{-17}$, two orders of magnitude beyond MICROSCOPE's $\eta\approx1\times10^{-15}$) would significantly constrain quantum gravity theories and shape our understanding of gravity, quantum mechanics, and their unification for decades.
\newpage
        \begin{longtable}{m{4 cm}|m{12 cm}}
          	\hline \hline
          	\multicolumn{2}{c}{\bf SCIENTIFIC OBJECTIVES}	\\
			\hline
          	\multicolumn{2}{l}{\bf Tests of the Einstein Equivalence Principle}	\\
          	\hline
          	UFF - Free fall of Quantum Matter Waves & Test the Universality of Free Fall (UFF) with a sensitivity of $\eta \leq 10^{-17}$ using ultra-cold $^{87}$Rb and $^{41}$K atoms in quantum superposition.\\
          	\hline
          	LLI - Local Lorentz Invariance & Search for Lorentz violation in the Standard Model Extension, with 3 orders of magnitude improvement on present sensitivities. \\
          	\hline
          	\multicolumn{2}{l}{\bf Ultralight Dark Matter (DM) searches}	\\
          	\hline
          	Scalar DM, linear coupling & Extend the sensitivity to DM couplings by up to 1.5 orders of magnitude for masses $\leq 10^{-11}$~eV. \\
          	\hline
          	Scalar DM, quadratic coupling & Extend the sensitivity to DM couplings by 1.5  - 3 orders of magnitude for masses $\geq 10^{-20}$~eV. \\
          	\hline
          	Vector DM & Extend the sensitivity to DM couplings by 1.5  - 3 orders of magnitude for masses $\leq 10^{-11}$~eV. \\
          	\hline
          	Other DM models & Potential to explore other/new  DM models (relaxion, spin-2, \dots).   \\
          	\hline
          	\multicolumn{2}{l}{\bf Tests of Quantum Mechanics}	\\
          	\hline
          	Continuous Spontaneous Localization model & Improve sensitivity by up to 4 orders of magnitude. Reach the theoretically motivated GRW value.\\
          	\hline
          	Di\'osi - Penrose model & Improve best current sensitivity by more than an order of magnitude.\\
          	\hline
          	Other models & Large superpositions ($\leq$ 1.3~m) and long free fall times ($\leq$  50~s) are well suited to explore other/new modifications of quantum mechanics (see e.g.~\cite{Bassi2013rev}).\\
            \hline \hline
            \multicolumn{2}{c}{\bf PAYLOAD}	\\
			\hline
          	Dual Atom Interferometer & $^{87}$Rb vs $^{41}$K differential acceleration ($\Delta a$) measurement with $\sqrt{S_a(f)} \leq 4.8\times 10^{-13}\, \rm m/s^2/\sqrt{Hz}$. Systematics at signal frequency/phase $\leq 6.6\times 10^{-17}\, \rm m/s^2$.\\
          	\hline
          	GNSS receiver & Dual-band receiver with modest performance requirements ($\approx$ 200~m).	\\
            \hline \hline
            \multicolumn{2}{c}{\bf MISSION PROFILE}	\\
            \hline
            Orbit & SSO circular orbit, 1400~km altitude. \\
            \hline
            Launcher & Direct orbit injection with VEGA-C from Kourou. Launch window available all year.\\
            \hline
            Mission Duration & 3 yrs with 80\% science availability, including 6 months commissioning.\\
            \hline
            End of life & Solid fuel propulsion for controlled re-entry manoeuvre.\\ 
			\hline \hline
            \multicolumn{2}{c}{\bf SPACECRAFT}	\\
			\hline \hline
			S/C design & Cylindrical with body mounted solar panels. STE-QUEST M3/M4 and LISA-Pathfinder (LPF) heritage.\\
			\hline
			DFACS & Drag-free and attitude control using cold-gas microthrusters/ inertial measurement unit/ star trackers. Req.: $\sqrt{S_a(f)} \leq 4.0\times 10^{-10}\, \rm m/s^2/\sqrt{Hz}$ and  $\sqrt{S_{\dot{\Omega}}(f)} \leq 3.2\times 10^{-7}\, \rm rad/s^2/\sqrt{Hz}$ (see Tab. \ref{tab:SC_acc_rot}). MICROSCOPE and LPF heritage. \\
			\hline
			Mass & 1187~kg wet mass, all margins included. \\
			Power & 1235 W average consumption, all margins included. \\
			Communications & S/X band up/downlinks. Req.: $\leq 110$~kbps science data in downlink. \\
			\hline \hline
        \caption{\it Summary of STE-QUEST M7 mission proposal.}
        \label{tab:exec_summary}
    \end{longtable}

\newpage
\section{STE-QUEST science}


\begin{wrapfigure}{lb}{0.5\textwidth}
\includegraphics[width=0.48\textwidth]{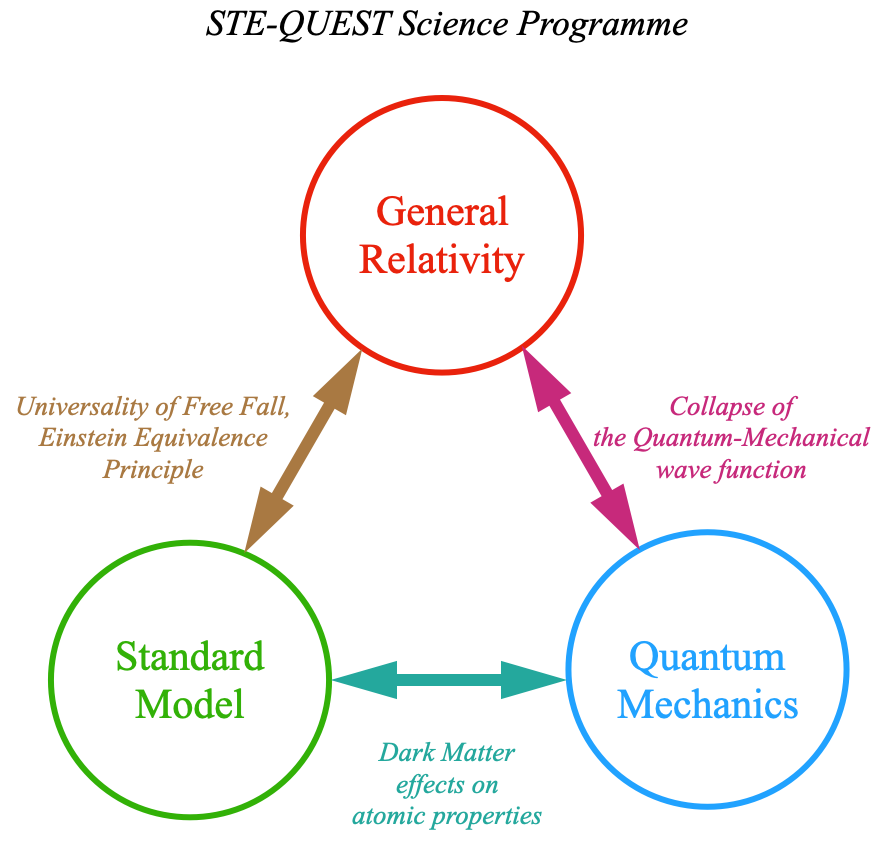}
\caption{\it The STE-QUEST science programme targets the interfaces between the current building-blocks
of fundamental physics, namely General Relativity, the Standard Model of particles, and Quantum Mechanics.}
\label{fig:Science_Programme}
\end{wrapfigure}

Our current description of fundamental physics is based upon three basic building-blocks, namely
General Relativity, the Standard Model of particle physics, and Quantum Mechanics. Each of these
is very successful within its domain of applicability, but we have no unified description of all
physical phenomena from the atomic scale to astrophysics and cosmology. For example, we do not know
how well General Relativity works at the atomic level, or whether there might be additional forces
at that scale. Although the Standard Model describes very well the visible matter in the Universe,
astronomers tell us there is much more invisible Dark Matter, whose quantum effects may affect the
properties of atoms. At the atomic scale, where Quantum Mechanics works so well, there are fundamental questions such as the measurement problem and the transition from quantum to classical behavior. And the greatest unsolved theoretical problem may be the reconciliation of
Quantum Mechanics with General Relativity, which may require modifying one or the other or both of these fundamental theories.

As illustrated in Fig.~\ref{fig:Science_Programme}, the scientific programme of STE-QUEST targets
directly the interfaces between the three building-blocks of fundamental physics, addressing these
puzzling questions by using advanced quantum sensors based on cold atoms. The deployment of cold atoms in space will enable STE-QUEST to
make unprecedented advances in science at the interface between general relativity, atomic physics 
and quantum mechanics. It will do so by exploiting the advances made in recent years in the
development of cold atom technologies in terrestrial experiments and applications, which will
provide synergies between the deployment of quantum technologies in space for exploring 
fundamental physics, as proposed here, and for applications such as Earth observation, geodesy, time-keeping and 
navigation. As such, it is an integral part of the recent community road map for cold atoms in space
that was authored by over 250 scientists worldwide~\cite{Alonso2022}.

The present proposal builds upon a White Paper submitted to the ESA Voyage-2050 call~\cite{Battelier2021},
whose main science objective was a test of the Universality of Free Fall (UFF) and the Einstein Equivalence
Principle (EEP)\footnote{Generally UFF (also known as the Weak Equivalence Principle, WEP) is considered as one of the three fundamental sub-principles of EEP~\cite{Will2018}, the other two being Local Lorentz Invariance and Local Position Invariance. In this proposal we are mostly concerned with UFF so ``EEP'' will in general refer to that, unless explicitly stated otherwise.} using ultracold atoms in quantum superposition states. STE-QUEST offers a sensitivity about 
three orders of magnitude beyond the best existing result obtained by the MICROSCOPE space mission in 
2017 ~\cite{Touboul2017} (only two orders of magnitude after publication of the last MICROSCOPE results of Ref. ~\cite{Touboul2022}). Other science goals are accessible with the same instrument. One is the search for the effects of coherent
waves of Dark Matter fields on atomic properties, and another is to test the foundations of 
quantum mechanics by probing mechanisms for the collapse of the Quantum-Mechanical wave function that have
been motivated by models of quantum gravity and the measurement problem. Each of these STE-QUEST 
objectives has the potential to revolutionize our understanding of physics and the Universe, or 
advance significantly our knowledge about the validity of our best current theories and models at the 
most fundamental level.

This STE-QUEST proposal inherits experience from the previous Cosmic Vision M3 proposal, 
and also that for the M4 call, with three important advantages. One is the broader physics programme
outlined above, and another is the widespread support it has attracted in the global cold atom community.
The third advantage is that significant progress has been made on the payload Technical Readiness Level (TRL),
thanks to experiments on the ground and in microgravity (drop-tower~\cite{drop, Deppner2021}, 0-g flights~\cite{Nyman2006}, sounding rockets~\cite{becker2018, Lachmann_2021} and the International Space Station~\cite{aveline2020}), as well as on the control of the main 
systematic effects~\cite{Loriani2020a}. Also, the current proposal concentrates on the core science objectives 
with enhanced performance by de-scoping payloads related to secondary objectives 
and optimizing the orbit for the primary objectives (SSO circular orbit @ 1400 km) leading to further cost savings 
and minimizing extraneous risk by using a Vega launcher instead of Soyuz. Finally, STE-QUEST will take advantage of the immense technological heritage from recent missions that use drag-free and precise attitude control (MICROSCOPE and LISA-Pathfinder) as well as the technology development for the upcoming LISA mission.

STE-QUEST will put ESA at the forefront of fundamental physics in space, opening the way for unprecedented discoveries 
at the frontiers of General Relativity, Dark Matter and Quantum Mechanics, firmly establishing Europe as the leader 
of the quantum revolution in space.


\subsection{Test of the equivalence principle at the $10^{-17}$ level}

As outlined above, our current description of the physical Universe at the most fundamental level, is based on three theories: 
the classical theory of General Relativity, whose exploration has reached a new level with the direct observation of gravitational
waves and imaging of photon rings surrounding massive black holes, Quantum Mechanics, whose principles underpin our understanding
of microscopic phenomena via quantum field theory in particular, and the Standard Model, which describes the subatomic
structure of the visible matter in the Universe and predicted successfully the existence of the (Englert-Brout-)Higgs boson that
was discovered 10 years ago.

Despite their individual successes, these theories have not yet provided a unified description of physical phenomena,
but have apparent contradictions and leave gaps in our understanding of the Universe. For example, many attempts to unify
gravity with the other fundamental forces described by the Standard Model, such as string-inspired models,
suggest violations of the UFF and the EEP due, for example, 
to so-called fifth forces that have not yet been detected. Secondly, the Standard Model has many shortcomings such as its
failure to explain astrophysical and cosmological observations that require the existence of a quantity of invisible dark matter
that is greater than that of the visible matter, or possibly some modification of our theory of gravity. Quantum interactions
of dark matter with Standard Matter particles could have signatures that could appear to violate the UFF and the EEP~\cite{damour:2010zr,Bertone:2016nfn,Hui:2016ltb,hees:2018aa,Rogers:2020ltq}.
Thirdly, it has been argued on the
basis of theoretical studies of black holes that there is a contradiction with the basic principle of quantum mechanics.
There is no generally-agreed resolution of this contradiction, but theories addressing this problem typically modify
either General Relativity and/or Quantum Mechanics in an essential way. For example, it has been suggested that
quantum-gravitational effects may cause the collapse of the wave function, with potential implications for the
measurement problem of Quantum Mechanics~\cite{EHNS,diosi1987universal,diosi1989models,penrose1996gravity,belenchia2021test,gasbarri2021testing} [cf.~Sec.~\ref{sec.QM}].

Central to all these issues is Einstein’s theory of General Relativity (GR), which is the cornerstone of our current description 
of the physical world at macroscopic scales. It describes successfully the motions of bodies from satellites to galaxy clusters, 
the propagation of electromagnetic waves in the vicinity of massive bodies, the flow of time in the presence of gravity, 
the evolution of stars, gravitational waves, gravitational forces within a few Schwarzschild radii of massive black holes
and the dynamics of the Universe as a whole. 
However, GR and many more general theories of gravitation are classical theories that are fundamentally incomplete, 
because they do not include quantum effects, whereas any theory seeking to unify all fundamental forces of Nature and include the Standard Model of particle physics must reconcile GR and Quantum Mechanics. Several proposals for such a reconciliation
are currently under investigation, including string theory, loop quantum gravity, and extra spatial dimensions,
most of which predict violations of the basic principles of GR.
The science program of STE-QUEST will provide the most powerful
probes of the most sensitive aspect of GR, namely the EEP, by testing the UFF and LLI.

The EEP is not a fundamental symmetry of physics like
the principle of local gauge invariance in particle physics. Rather,
the EEP is a fundamental feature of all theories of gravity that 
describe it as a geometrical phenomenon, i.e., as the curvature of space-time.
In such a theory space-time has a position-dependent dynamical metric $g_{\mu \nu}$ that defines the separations
between events:
\begin{equation}
    {\rm d}s^2 \; = \; g_{\mu \nu}(x) \, {\rm d}x^\mu {\rm d}x^\nu, 
\end{equation}
in a space-time manifold parametrized by coordinates $x^\mu$. In such theories, freely-falling test bodies move along
geodesics of extremal length:
\begin{equation}
    \delta \int {\rm d}s \; = \; 0,
    \label{eq:UFF}
\end{equation}
that are independent of the bodies' compositions, i.e., free fall is universal. 
Additionally, clocks measure proper time along their trajectories,
\begin{equation}
    {\rm d}\tau^2 \; = \; - \frac{1}{c^2} \, {\rm d}s^2,
\end{equation}
independent of the types of clocks used.
Moreover, the other laws of physics satisfy the principle of special
relativity in local freely-falling reference frames, i.e., they are Lorentz-invariant.
The universal coupling to all sources of mass and energy that is implicit in the EEP is necessary for all metric theories of gravitation, 
including many other theories in addition to GR. As such, the EEP is one of the fundamental principles
of modern physics. 

Since the conceptual basis of the EEP is very different from that of the gauge symmetries that have proven so
successful in the Standard Model, probes of the EEP are conceptually independent of current tests of the Standard Model. 
Indeed, many theories that go beyond the Standard Model and GR entail some violation of the EEP~\cite{Damour_2012}. A broad
class of such theories invoke the existence of one or more ultralight bosonic fields whose couplings to Standard Model particles
are not constrained to be universal, and may be accessible to STE-QUEST. The discovery of the Higgs boson, an apparently elementary scalar particle with non-universal
couplings to other particles, may be considered a prototype for such light scalar fields. Examples of such fields
present in fundamental theories include
the moduli and dilaton fields appearing in generic compactification of string theory. Examples also appear in dynamical models of
Dark Energy, such as quintessence fields. Coherent waves of such light scalar fields may also provide Dark Matter and cause
apparent variations in fundamental constants and the EEP, as discussed in the following Section. As also elaborated there, light vector
fields could also exist, and would in general also have non-universal couplings to Standard Model particles and hence
generate apparent violations of the EEP. These may also appear in other extensions and modifications of GR such as models
of extended gravity.

The best-known aspect of the EEP is the universality of free fall (UFF, sometimes also referred to as the weak equivalence principle, WEP),
see (\ref{eq:UFF}). A convenient figure of merit for all UFF/EEP tests is the E\"otv\"os ratio $\eta_{AB}$ for two test masses
$A$ and $B$ in the gravitational field of a specified source mass:
\begin{equation}\label{etaAB}
\eta_{AB} = 2\,\frac{a_A - a_B}{a_A + a_B}\,,
\end{equation}
where $a_i$ ($i=A,B$) is the gravitational acceleration of object $i$ with respect to the source mass. 
We note that the data from any given experiment can be interpreted by reference to different source masses, 
with correspondingly different results for $\eta_{AB}$. Also, though $\eta_{AB}$ is a useful tool for comparing different experiments,
it cannot account for the diversity of possible underlying theories, e.g., different types of couplings depending on the source and test objects, 
or couplings to space-time-varying background fields other than local gravity. Thus, not only is the best performance in terms of the 
E\"otv\"os ratio required, but also a large diversity of test objects and source masses. 

The history of experimental tests of the UFF dates back at least as far as the 16th century and Galileo Galilei. Since then, 
tremendous efforts have been carried out to push laboratory tests to uncertainties as low as parts in $10^{-13}$~\cite{Wagner:2012ui,Touboul2019,Asenbaum2020}. 
However, ground tests are ultimately limited by the Earth’s gravitational environment, and future progress in probing the UFF
will come from space experiments~\cite{belenchia2022quantum}, such as the MICROSCOPE experiment~\cite{Touboul2017},
which pioneered tests of the UFF in space between 2016 and 2018. Table \ref{tab:StateArt} presents the state of the art in UFF/EEP tests, 
sorted into different classes depending on the techniques and the types of test-masses employed. In particular, we distinguish 
classical tests using macroscopic test masses from hybrid tests and atom interferometry (ATI) tests that use matter waves in a quantum superposition, 
possibly condensed into quantum-degenerate states of a Bose-Einstein Condensate (BEC) with coherence lengths $\geq \mu$m. 
\begin{wrapfigure}[22]{h}{0.42\textwidth}
\includegraphics[width=0.37\textwidth]{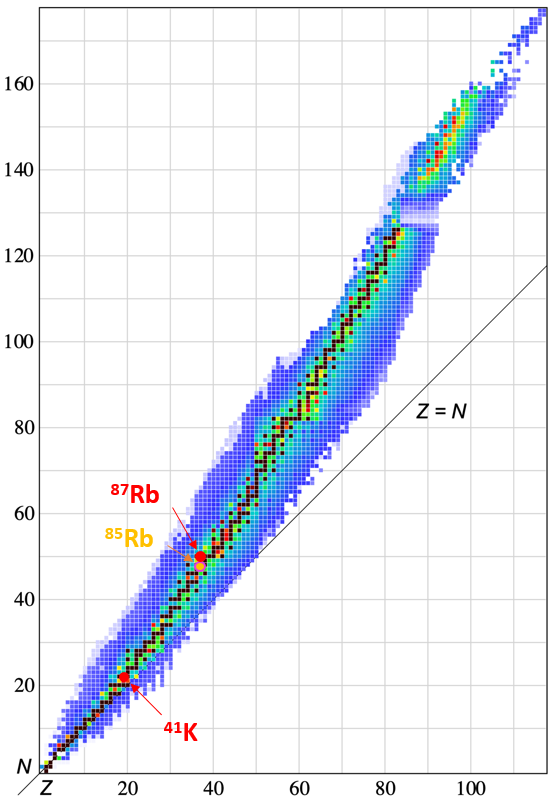}
\caption{\it The valley of nuclei in the (N,Z) plane, with the $^{85}\rm Rb$-$^{87}\rm Rb$ combination from the M3 proposal in orange-red and the $^{41}\rm K$-$^{87}\rm Rb$ one from this STE-QUEST proposal in red.}
\label{fig:species}
\end{wrapfigure}
The game-changing results of the MICROSCOPE mission demonstrate the potential of exploiting a quiet and well-controlled space environment, 
with relatively long free-fall times. Similarly, the recent leap in the sensitivities of quantum tests of the UFF
by four orders of magnitude is largely due to the much longer free-fall times attained in a 10~m drop tower~\cite{Asenbaum2020}.
These advances give indications of the improvements that can be expected from a space mission such as STE-QUEST. 
\begin{table}[h]
\hspace{-5mm}
\begin{center}
\hspace{-5mm}
\begin{tabular}{ccccc}
\hline 
{\bf Class} & {\bf Elements} & {\bf $\eta$} & {\bf Year} & {\bf Comments} \\
\hline\hline
\multirow{4}{*}{Classical} & Be - Ti & $2\times10^{-13}$ & 2008 & Torsion balance \\
& Pt - Ti & $1\times10^{-14}$ & 2017 & MICROSCOPE first results \\
& Pt - Ti & $2.7\times10^{-15}$ & 2022 & MICROSCOPE full data \\
\hline
\multirow{3}{*}{Hybrid} & $^{133}$Cs - CC & $7\times10^{-9}$ & 2001 & Atom Interferometry\\ 
& $^{87}$Rb - CC & $7\times10^{-9}$ & 2010 & and macroscopic corner cube (CC) \\
\hline
\multirow{5}{*}{Quantum} & $^{39}$K - $^{87}$Rb & $3\times10^{-7}$ & 2020 & different elements \\
& $^{87}$Sr - $^{88}$Sr & $2\times10^{-7}$ & 2014 & same element, fermion vs. boson  \\
& $^{85}$Rb - $^{87}$Rb & $3\times10^{-8}$ & 2015 & same element, different isotopes  \\
& $^{85}$Rb - $^{87}$Rb & $3.8\times 10^{-12}$ & 2020 & 10 m drop tower \\
& \textcolor{red}{{$^{41}$K - $^{87}$Rb}} & \textcolor{red}{{$(10^{-17})$}} &  & \textcolor{red}{STE-QUEST} \\
\hline
Antimatter & $\overline{\rm H}$ - H & ($10^{-2}$) & 2023+ & under construction at CERN \\
\hline
\end{tabular}
\caption{\it State of the art in UFF/EEP tests. Numbers in brackets are results expected in the future, including STE-QUEST
(shown in red). Table adapted from ~\cite{Alonso2022}, where the original references can be found.} \label{tab:StateArt}
\end{center}
\end{table}

The current STE-QUEST proposal aims at measuring the differential acceleration between $^{41}$K and $^{87}$Rb. These two atomic species are well separated in the (N,Z) plane, see Fig.~\ref{fig:species}, which is desirable since it enhances the EP-violating signal predicted in some theoretical models, see e.g.~\cite{damour:2010zr,Fayet2018}.

In particular, this choice of species explores a way larger part of the (N,Z) plane compared to the ones considered in the initial STE-QUEST M3 proposal, which was one of the drivers for this choice. The other main driver was the technological readiness for both species. Indeed they have very similar atomic structure, which means that e.g. the laser and trapping technology can be identical (frequency doubled telecom lasers) as the wavelengths are very close (see Tab. \ref{tab:params}). Also both species are actively used and explored in ground based and 0-g cold atom interferometry experiments. 
STE-QUEST is designed to improve on the best present results by about three orders of magnitude, reaching a sensitivity in the low $10^{-17}$ range, as discussed in more detail in the following sections. Such sensitivity is considered to be impossible for ground experiments because of the limited free-fall times and the local environmental, gravitational, and inertial perturbations.


\paragraph{Search for a breaking of Lorentz symmetry}
STE-QUEST will also allow to probe Lorentz symmetry, which is an essential ingredient of both of our current best theories of physics: General Relativity and Quantum Mechanics. Lorentz symmetry  stipulates that the results of experiments do not depend on the orientation of the laboratory or on its velocity. It has been suggested that Lorentz symmetry may not be a fundamental symmetry of Nature and may be broken at some level. While some early motivations came from string theories~\cite{kostelecky:1989jk}, breaking of Lorentz symmetry also appears in loop quantum gravity, non commutative geometry, multiverses, brane-world scenarios and others (for a review, see e.g.~\cite{tasson:2014qv,mattingly:2005uq}).  In particular, a dedicated effective field theoretic framework has been developed in order to consider systematically all hypothetical violations of Lorentz invariance. This framework is known as the Standard-Model Extension (SME)~\cite{colladay:1997vn,colladay:1998ys} and covers all fields of physics. It contains the Standard Model of particle physics, GR and all possible Lorentz-violating terms that can be constructed at the level of the Lagrangian, introducing a  number of new coefficients that can be constrained experimentally. Of prime interest for this project, the SME framework includes a matter-gravity sector which contains 12 parameters that are directly related to a breaking of the UFF, the so-called $\bar a_\mathrm{eff}^\mu$ coefficients for the electron, proton and neutron~\cite{kostelecky:2009jk,kostelecky:2011kx}. The current best constraints on the SME matter-gravity couplings come from a dedicated analysis of the MICROSCOPE data~\cite{bars:2019aa}. These couplings will induce an orientation-dependent violation of the UFF. In the context of STE-QUEST, this leads to a phenomenology, which, additionally to the ``standard'' UFF-violating signature, contains an annual modulation due to the trajectory of the Earth in the Solar System.  The expected sensitivity of STE-QUEST to these SME gravity-matter couplings improves over the current best constraints obtained by MICROSCOPE by 3 orders of magnitude, thereby opening another possibility for a glimpse of physics beyond GR and the Standard Model.

\subsection{Search for Dark Matter}\label{sec:DM}

Many astrophysical and cosmological measurements at different scales ranging from galaxies to the cosmological background radiation point to the presence of dark matter (DM) with a density several times larger than that of the ordinary matter that is described by the Standard Model of particle physics~\cite{Bertone:2016nfn}. This dark matter is invisible, but has gravitational interactions and may have other interactions with ordinary matter. Understanding the true microscopic nature of dark matter constitutes one of the greatest challenges in modern physics.

\begin{wrapfigure}{L}{0.5\textwidth}
\includegraphics[width=0.5\textwidth]{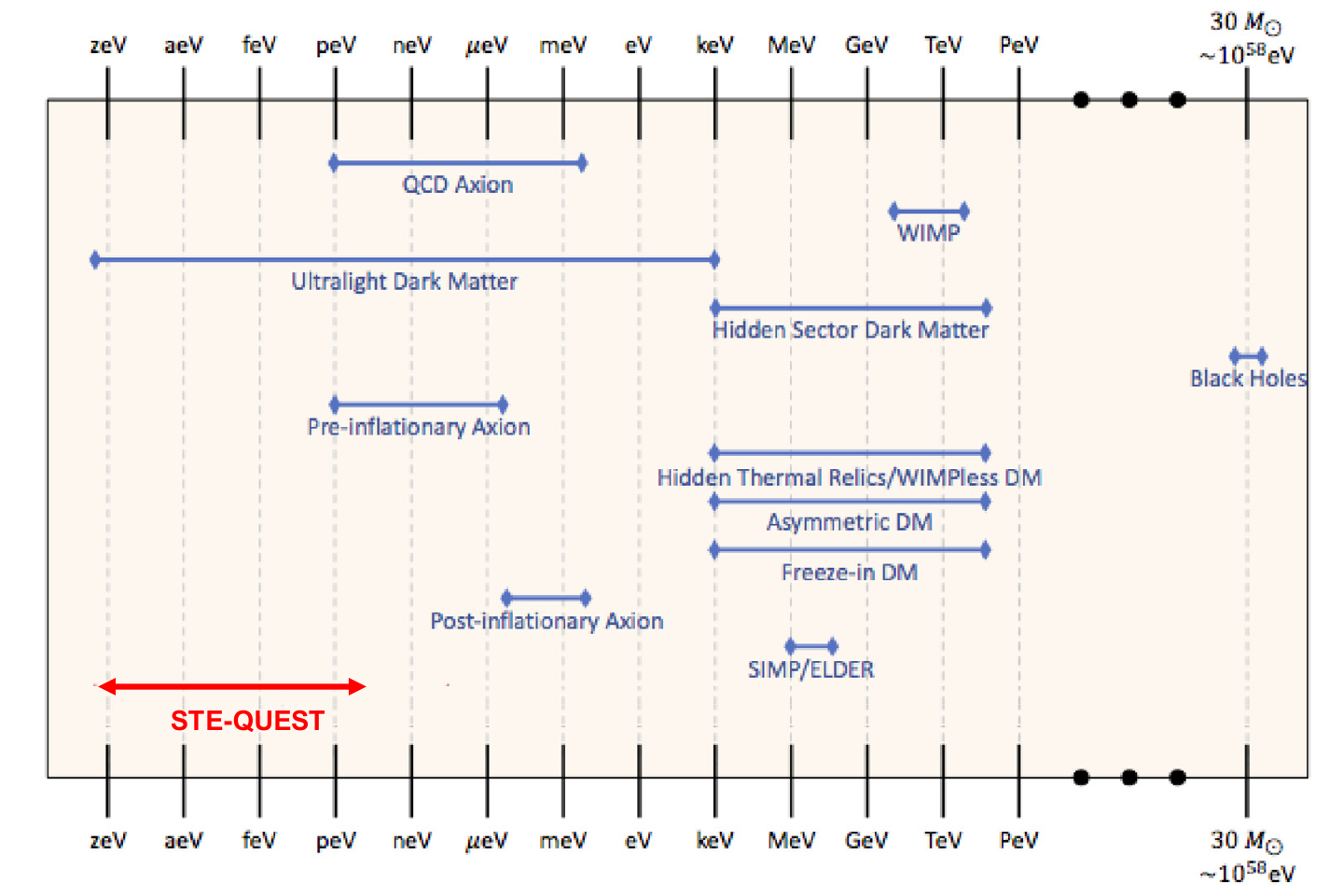}
\caption{\it Summary of some DM candidates and their masses, see~\cite{belenchia2022quantum,battaglieri:2017tk} for more details. The mass range explored by STE-QUEST is indicated in red. }
\label{fig:DM}
\end{wrapfigure}

In the past decades, there have been many searches for dark matter over a huge mass range considering various theoretical scenarios, see Fig.~\ref{fig:DM}. While the weakly interacting massive particle (WIMP) paradigm, i.e.,~massive particles that could be observed at colliders, was extremely popular previously, it now suffers from a lack of direct detection at the Large Hadron Collider at CERN. For this reason, alternative dark matter models such as ultralight dark matter (ULDM) have recently gained increased scientific interest.

ULDM refers to dark matter candidates whose mass is typically below an eV. For such a mass range, the occupation number (i.e., the number of particles per unit of phase-space volume) corresponding to the dark matter distribution in our Galaxy is larger than one. The Pauli exclusion principle therefore implies that such dark matter candidates are necessarily bosonic particles~\cite{baldeschi:1983vz}. ULDM candidates therefore encompasses scalar fields (spin 0), pseudo-scalar fields (e.g., the axion), vector fields (spin 1), tensor fields (spin 2), etc. 

On cosmological scales, a massive spin 0 or 1 field of mass $m$ will  oscillate at its Compton frequency~\cite{arvanitaki:2015qy,stadnik:2015yu,horns:2013wf}, i.e.
\begin{equation}\label{eq:DM_osc}
    \varphi = \varphi_0 \cos mt \, \qquad \mathrm{or} \qquad \vec X = \vec X_0 \cos mt \, ,
\end{equation}
where $\varphi$ is a scalar (spin 0) field and $\vec X$ is a vector (spin 1) field. 

\noindent The averaged stress-energy tensor related to these two dark matter candidates shows that they behave as pressureless fluids whose energy density is directly related to the amplitude of the oscillations~\cite{arvanitaki:2015qy,stadnik:2015yu,horns:2013wf}
\begin{equation}
    \rho = \frac{m^2 \varphi_0^2}{2}\, \qquad \mathrm{and}\qquad \rho = \frac{m^2 \left| \vec X_0\right|^2}{2}\, .
\end{equation}

If a new (scalar or vector) field is added to the Standard Model, it is expected to couple to ordinary matter, except if a fundamental symmetry prevents it. This coupling will in general impact the behavior of standard matter and produce observational signatures that are characteristic of a breaking of the Einstein equivalence principle. The precise signature of this violation of the equivalence principle depends on the specific theoretical model considered (the spin of the new particle, the type of coupling with standard matter, etc.). As a consequence, the phenomenology arising from ULDM is extremely rich. 

 STE-QUEST is sensitive to a possible interaction of ordinary matter with ULDM, which may either induce a new fifth force or produce coherent waves that could induce apparent variations in fundamental constants and atomic energy levels. It will expand the search for DM by extending the parameter space probed by several orders of magnitude for a large class of models. 
 Thus STE-QUEST offers the possibility of a groundbreaking discovery by providing the first direct detection of DM. Even in the absence of a positive detection, 
 STE-QUEST will improve dramatically constraints on various DM models. In the following, we present the prospect for STE-QUEST to search for two very well established and motivated DM candidates: 
 a scalar particle and a hidden photon (vector particle).

\subsubsection{Scalar Dark Matter}
A new spin 0 particle is the simplest model of ULDM~\cite{arvanitaki:2015qy,stadnik:2015yu}. Such candidates for scalar ULDM fields include the moduli and dilaton fields appearing in string theory~\cite{damour:1994fk}, and the relaxion that appears in attempts to understand the hierarchy of fundamental mass scales in physics~\cite{graham:2015tl}.

A useful way to parametrize the interaction between a scalar field and the Standard Model is provided by the following Lagrangian~\cite{damour:2010zr}
\begin{equation}
    \mathcal L = \varphi^n  \Bigg[\frac{d^{(n)}_e}{4\mu_0}F^2-\frac{d_g^{(n)}\beta_3}{2g_3}\left(F^A\right)^2 -\sum_{i=e,u,d}\Big(d_{m_i}^{(n)}+\gamma_{m_j}d_g^{(n)}\Big)m_i\bar\psi_i\psi_i\Bigg]\, ,
\end{equation}
where $d_X^{(n)}$ are the coupling strength of the new scalar interaction with various components of the Standard Model: $F_{\mu\nu}$ is the standard electromagnetic Faraday tensor, $\mu_0$ is the magnetic permeability, $F^A_{\mu\nu}$ is the gluon strength tensor, $g_3$ is the quantum chromodynamics (QCD) gauge coupling, $\beta_3$ is  the $\beta$ function for the running of $g_3$, $m_j$ are the masses of matter fermions (the electron and light quarks), $\gamma_{m_j}$ is the QCD anomalous dimension giving the running with energy of the masses of the strongly-interacting fermions and $\psi_j$ are the fermion spinors. Two types of coupling have been considered in the literature: a linear coupling ($n=1$)~\cite{damour:2010zr} and a quadratic coupling ($n=2$) motivated by a $Z_2$ symmetry~\cite{stadnik:2015yu}. These two couplings lead to different observational signatures, as discussed in Ref. ~\cite{hees:2018aa}.

\begin{wrapfigure}{r}{0.5\textwidth}
\vspace{-0.3cm}
\includegraphics[width=0.5\textwidth]{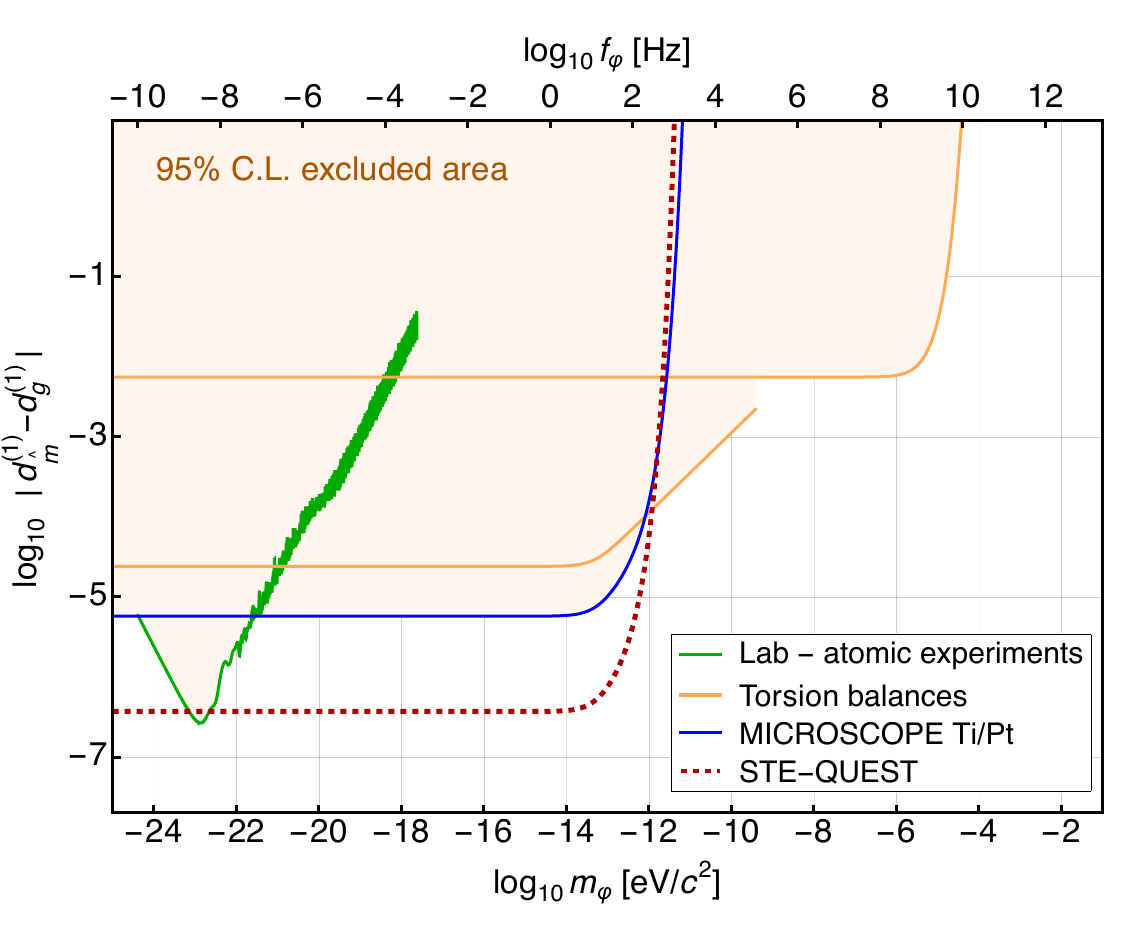}
\caption{\it The sensitivity of STE-QUEST (dashed line) to a linear coupling of scalar ULDM to quarks, 
compared to those of current experiments (shaded region) including atomic clocks~\cite{Hees2016}, the MICROSCOPE experiment~\cite{Berge:2017ovy} and torsion balances~\cite{Wagner:2012ui}.}
\vspace{-0.5cm}
\label{fig:ULDM}
\end{wrapfigure}

The linearly coupled scalar field has two distinct signatures: (i) the oscillatory behavior of the DM candidate, see Eq.~(\ref{eq:DM_osc}), and (ii) a composition-dependent Yukawa-type modification of the effective potential interaction~\cite{Berge:2017ovy}. While atomic clocks are sensitive to the first of these signatures, STE-QUEST will be sensitive to the second one, which will produce a static violation of the UFF in the gravitational field of the Earth.  The prospective STE-QUEST sensitivity to a linear coupling of a scalar ULDM field $\varphi$ to quark fields is shown in Fig.~\ref{fig:ULDM} 
as a function of the mass, $m_\varphi$, of the ULDM field. Also shown as shaded regions are the current constraints on the ULDM-quark coupling provided by atomic clocks~\cite{Hees2016}, the MICROSCOPE experiment~\cite{Berge:2017ovy} and torsion balances~\cite{Wagner:2012ui}. 
We see that STE-QUEST will provide better sensitivity for $m_\varphi$ between about $10^{-22}$~eV  - below which ULDM would be in tension with observational
constraints on the `fuzziness' of small-scale astrophysical structures~\cite{Rogers:2020ltq} - 
and $m_\varphi \sim 10^{-12}$~eV - above which STE-QUEST loses sensitivity and torsion balance experiments become competitive. We note that over 10
orders of magnitude in $m_\varphi$ the STE-QUEST sensitivity will exceed that of the current world-leading MICROSCOPE experiment~\cite{Touboul2017}
by some 1.5 orders of magnitude in the linear ULDM-proton coupling $d_{\hat{m}}-d_g$. Similar improvements are expected for the other coupling parameters.

\begin{wrapfigure}[16]{R}{0.5\textwidth}
\includegraphics[width=0.5\textwidth]{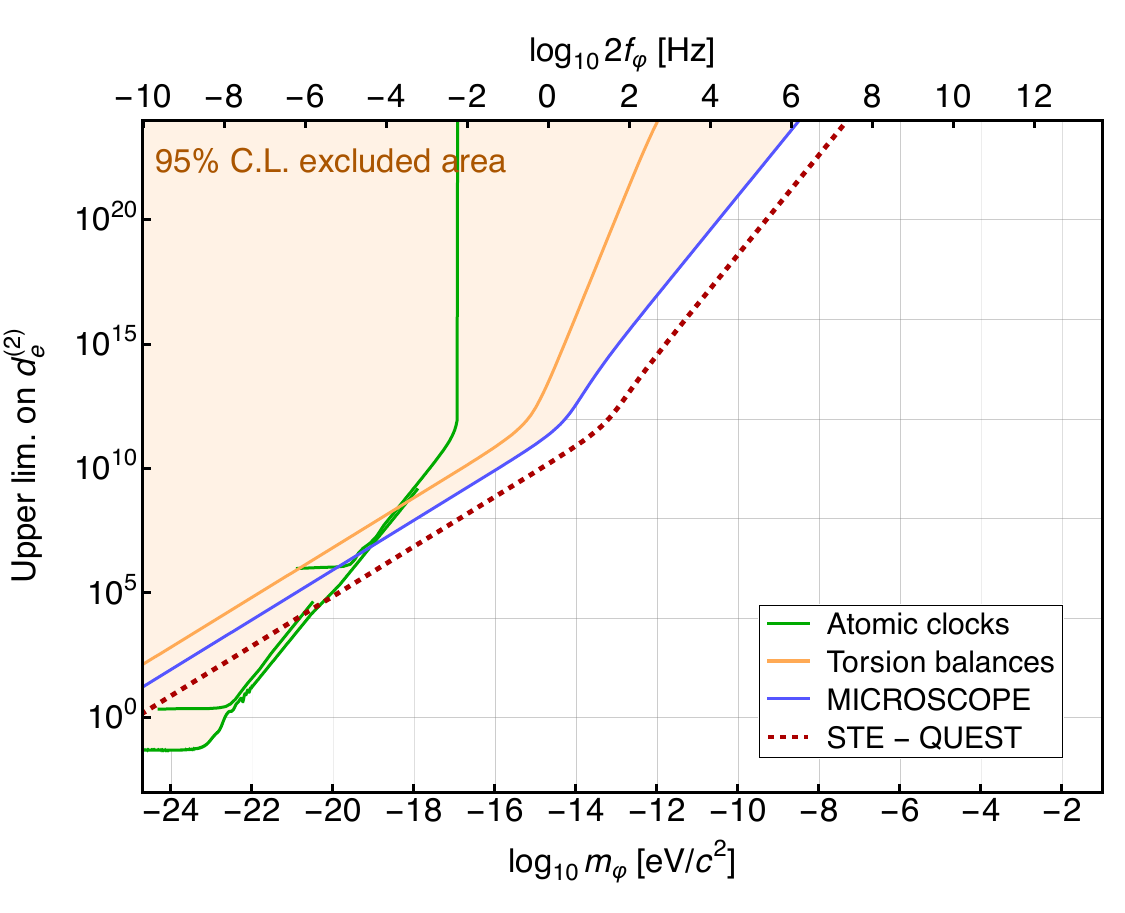}
\caption{\it The sensitivity of STE-QUEST (dashed line) to a quadratic coupling of scalar ULDM to electromagnetism, compared to those of current experiments (shaded region) including atomic clocks~\cite{hees:2018aa}, the MICROSCOPE experiment~\cite{Touboul2017} and torsion balances~\cite{Wagner:2012ui}.}
\label{fig:ULDMquad}
\end{wrapfigure}
The case of a quadratic coupling between matter and the scalar particle leads to richer phenomenology. 
In such a scenario, the leading observational effect is a static UFF violation whose amplitude depends on the distance to the central body and on the amplitude $\varphi_0$ of the oscillations
of the scalar field. The non-linearity of the theory implies that the  oscillation amplitude from the DM candidate will be affected by the presence of a body like the Earth. In particular, if the coupling parameters are positive, the amplitude of the oscillations can be strongly reduced close to the Earth, a phenomenon known as a ``screening mechanism''~\cite{hees:2018aa}. On the other hand, a negative coupling can lead to an amplification of the scalar field, a phenomenon known as ``scalarization''. This means that, for a positive coupling parameter, the violation of the equivalence principle is strongly suppressed close to the Earth, making it difficult to detect with on-ground experiments. The prospective STE-QUEST sensitivity to a quadratic coupling of a scalar ULDM field $\varphi$ to the electromagnetic field is shown in Fig.~\ref{fig:ULDMquad} as a function of the mass, $m_\varphi$, of the ULDM field. Except in the very low mass regime $m_\varphi<10^{-20}$ eV, which is best constrained by atomic clock experiments, STE-QUEST will improve significantly on the current best searches for DM. More precisely, for a mass $m_\varphi$ between  $10^{-19}$ and $10^{-15}$ eV, STE-QUEST is expected to provide a 1.5 order of magnitude improvement compared to the current MICROSCOPE result. Moreover, as a consequence of the screening mechanism, for masses larger than  $10^{-14}$ eV, the STE-QUEST outcome is expected to improve the MICROSCOPE result by 3 orders of magnitude. Similar improvements are expected for the other coupling parameters.


\subsubsection{Vector Dark Matter: dark photon}\label{sec:Vec-DM}
Another DM candidate that has recently gained a lot of scientific interest consists of an additional U(1) gauge boson, sometimes known as a dark photon or U-boson. Such an additional spin 1 particle appears naturally in theories involving grand unification, su\-per\-sym\-met\-ry, inflation, string theories, see, e.g.,~\cite{fayet:1977ts,fayet:1990tu,Jaeckel:2012mjv,Fayet2018} and references therein. Regardless of their possible origin, the couplings of a spin-1 particle are generally expected to obey a gauge symmetry principle, which implies that the coupling to regular matter is expected to be proportional to a linear combination of the $B$ (ba\-ry\-on\-ic), $L$ (leptonic) and $Y$ (hy\-per\-charge) currents~\cite{fayet:1990tu}. We note that well motivated su\-per\-sym\-met\-ric models predict a coupling of the vector particle to the $B-L$ current, parametrized by a coupling strength $\varepsilon_\mathrm{B-L}$.
 \begin{wrapfigure}{R}{0.5\textwidth}
\includegraphics[width=0.48\textwidth]{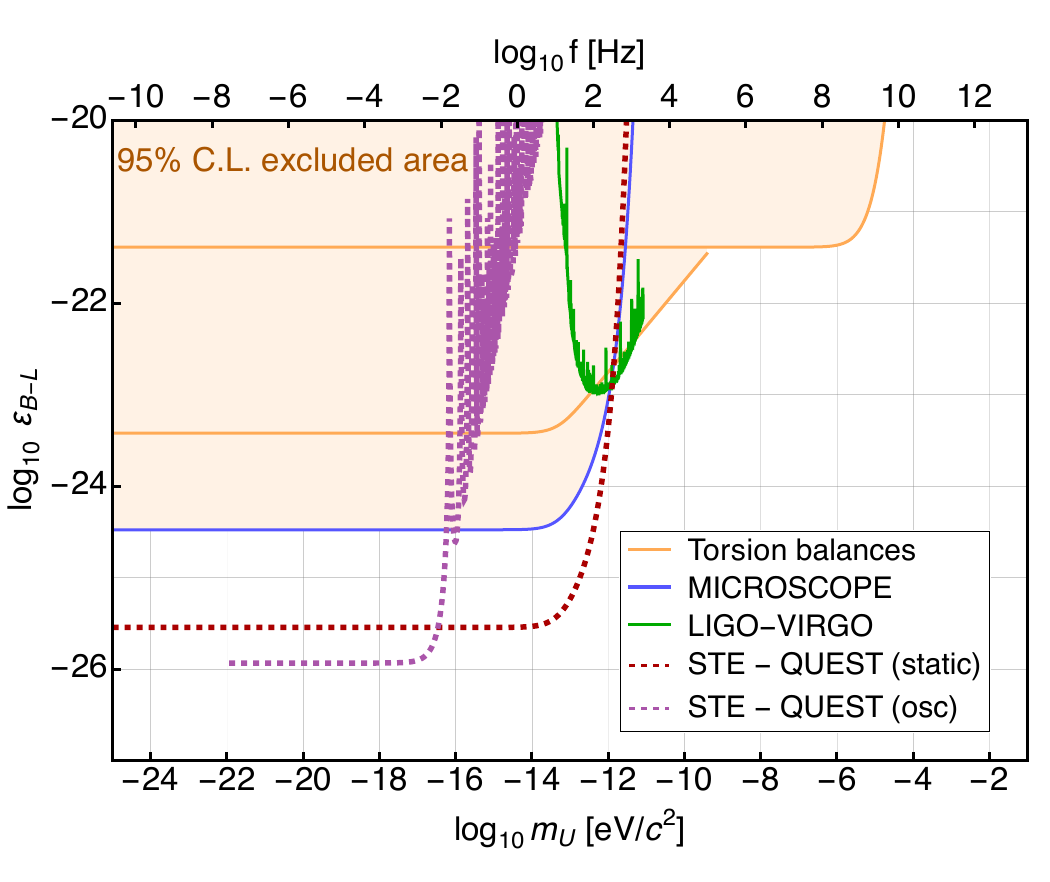}
\caption{\it The sensitivity of STE-QUEST (dashed lines) to the coupling strength of a U(1) dark matter candidate to the $B-L$ current. The dashed red line corresponds to the sensitivity to the modification of the effective potential interaction, while the purple dashed curve presents the sensitivity to the oscillatory signature of the DM candidate. For comparison, current   experiments are also shown (shaded region), including the MICROSCOPE experiment~\cite{Berge:2017ovy}, torsion balances~\cite{Wagner:2012ui} and gravitational wave detector searches~\cite{abbott:2022va}.}
\label{fig:Vec-DM}
\end{wrapfigure}
Two distinct signatures can arise from a massive vector field~\cite{graham:2016ab}: (i) a composition-dependent Yukawa-type modification of the 2-body interaction and (ii) the oscillatory behavior seen in Eq.~(\ref{eq:DM_osc}). Fig.~\ref{fig:Vec-DM} presents the expected sensitivity of STE-QUEST to a new B-L-coupled spin 1 gauge boson. The red dashed curve presents the sensitivity to the modification of the effective potential interaction. It improves on the recent MICROSCOPE result by 1.5 orders of magnitude and also improves on the recent  results from the LIGO-VIRGO-KAGRA collaboration~\cite{abbott:2022va}. STE-QUEST loses sensitivity for $m_U \geq 10^{-13}$ eV, for which the range of interaction of the new U(1) force becomes shorter than the altitude of the STE-QUEST satellite. STE-QUEST will also be sensitive to the oscillatory behavior of a dark photon field, as shown by the purple dashed curve in Fig.~\ref{fig:Vec-DM}. The expected sensitivity of STE-QUEST to such a signature improves on the current best limit from MICROSCOPE by $\sim$ 2 orders of magnitude in the mass range $m_U\sim 10^{-22}-10^{-17}$ eV, where the sensitivity of the instrument is optimal. 
Similar improvements are expected for couplings to the $B$ or $L$ currents. 

We underline that STE-QUEST will explore most of the range of UFF violation
predicted by models containing a light U(1) gauge boson associated with grand unification, compactification, inflation, and supersymmetry-breaking, which predict a violation of the UFF at the level of $\eta \approx 10^{-18}-10^{-12}$ ~\cite{Fayet2018}.

In conclusion, the microscopic nature of DM remains one of the most exciting open question of modern physics. Following the lack of direct detection of WIMPs using particle accelerators, a large class of alternative DM candidates have been revived. Among them,  models of bosonic ultralight particles predicted in various theoretical scenarios such as string theories, supersymetric models, etc.  have gained increased scientific interest. STE-QUEST will offer a unique opportunity to push the search for such Dark Matter candidates into unexplored regions of parameter space. In the most well-motivated models discussed above, STE-QUEST will extend the searched parameter space by 1.5 to 3 orders of magnitude over a mass range that extends over 10 orders of magnitude. In addition, STE-QUEST will also be able to probe some well-motivated specific models that predict a violation of the UFF~\cite{Fayet2018} at a level reachable with this project. For these reasons, STE-QUEST offers the possibility of a ground-breaking discovery in the field of DM and, even in the absence of a positive result, it will constrain severely many theoretical models. 

\subsection{Test of quantum mechanics} 
\label{sec.QM}

An additional scientific objective of STE-QUEST is to test the foundations of Quantum Mechanics, specifically the limits of validity of the quantum superposition principle for larger systems. The reason why  quantum properties of microscopic systems (in particular, the possibility of being in the superposition of two states at once) do not carry over to macroscopic objects has been subject of intense debates over the last century~\cite{penrose1996gravity,Adler:2004wc,Leggett871,weinberg2017trouble}. Its possible resolution  could be a progressive breakdown of the superposition principle when moving from the microscopic to the macroscopic regime. The most important consequence would be to change fundamentally our understanding of Quantum Mechanics --- now commonly considered as a fundamental theory of Nature --- as an effective theory appearing only as the limiting case of a more general one~\cite{Adler2009rev}. Several models have been proposed to account for such a breakdown of the quantum superposition principle. They go under the common name of (wavefunction) collapse models~\cite{Bassi2003rev,Adler2009rev,Bassi2013rev}, and modify the standard Schr\"odinger dynamics by adding collapse terms whose action leads to the localization of the wavefunction in a chosen basis. 

Another suggested motivation for collapse models, beyond having a universal theory whose validity stretches from the microscopic world to the macroscopic world,
comes from a cosmological perspective. Collapse models have been proposed to justify the emergence of cosmic structures in the Universe, whose signatures are imprinted in the Cosmic Microwave Background (CMB) in the form of temperature anisotropies~\cite{perez2006quantum, PhysRevD.85.123001,PhysRevD.88.085020}. Moreover, collapse models were also proposed as possible candidates to implement an effective cosmological constant, thus explaining the acceleration of the expansion of the Universe~\cite{PhysRevLett.118.021102}. The application of collapse models to cosmology is however not straightforward, as it requires a relativistic generalization of the non-relativistic models discussed below. How to build these relativistic generalizations of collapse models is still not clear: several proposals have been suggested~\cite{PhysRevLett.124.080402,PhysRevLett.127.091302,Bengochea:2020vr,jones2021impossibility,jones2021mass}, but each has limitations and the debate in the theoretical community is still open.

\paragraph{Continuous Spontaneous Localization (CSL) model --}  The most studied collapse model is CSL~\cite{pearle1989combining,ghirardi1990markov}, a phenomenological model that treats the system under scrutiny as fundamentally quantum but subject to the weak and continuous action of some measurement-like dynamics. The full dynamical equation for the wavefunction $|\psi_t\rangle$ is:
\begin{equation}\label{eq.csl}
\begin{split}
\operatorname{d}\! |{\psi_t}\rangle\!=\! \Bigg[ &- \frac{i}{\hbar} \hat{H} \operatorname{d}\! t+\! \int\! \operatorname{d}^3\! {\bf x}\,\left(\hat{M}({\bf x})-\langle \hat{M}({\bf x}) \rangle_t \right)
\operatorname{d}\! W_t({\bf x}) \Bigg.
\\
&\Bigg.  -  \tfrac{1}{2}\!\int\! \operatorname{d}^3\!{\bf x}\operatorname{d}^3\!{\bf y}\,{\mathcal D}({\bf x}-{\bf y})\!\prod_{{\bf q}={\bf x},{\bf y}}\!
\left(\hat{M}({\bf q})-\langle{ \hat{M}({\bf q}) }\rangle_t \right)
\,\operatorname{d}\! t  \Bigg]\! |{\psi_t}\rangle{.}
\end{split}
\end{equation}
The first term describes the standard Schr\"odinger dynamics of the system, governed by its quantum Hamiltonian $\hat H$, whereas the second and third terms in Eq.~\eqref{eq.csl} describe the 
wavefunction collapse, which is driven by a family of white noise terms $\operatorname{d}\!W_t({\bf x})/\operatorname{d}\!t$ (one for each point of space ${\bf x}$) with spatial correlation $\mathcal D({\bf x}-{\bf y})=\frac{\lambda}{m_0^2}\exp(-|{\bf x}-{\bf y}|^2/4r_\text{\tiny C}^2)$, where $m_0$ is a reference mass taken to be that of a nucleon. Both terms depend on the difference between the mass density operator $\hat M({\bf x})$ and its expectation value $\langle \hat{M}({\bf x}) \rangle_t=\langle\psi_t| \hat{M}({\bf x})|\psi_t \rangle$. The presence of these expectation values, which makes the equation non-linear in the state $|\psi_t\rangle$, is fundamental for generating the collapse. The motivation for $\hat M({\bf x})$ as the collapse operator is twofold: on the one hand, it provides a localization of the state in the position basis that is used to measure the properties of physical systems; on the other hand, it provides automatically an amplification mechanism such that microscopic systems are essentially left untouched by the collapse, while macroscopic ones are strongly affected, with a scaling given by a monotonically growing function of the mass. 
It can  be also shown that, when the collapse dynamics is dominant, the probabilities of collapsing at a point ${\bf x}$ is given, with excellent approximation, by the Born rule. 

\begin{wrapfigure}{tb}{0.5\textwidth}
\includegraphics[width=0.5\textwidth]{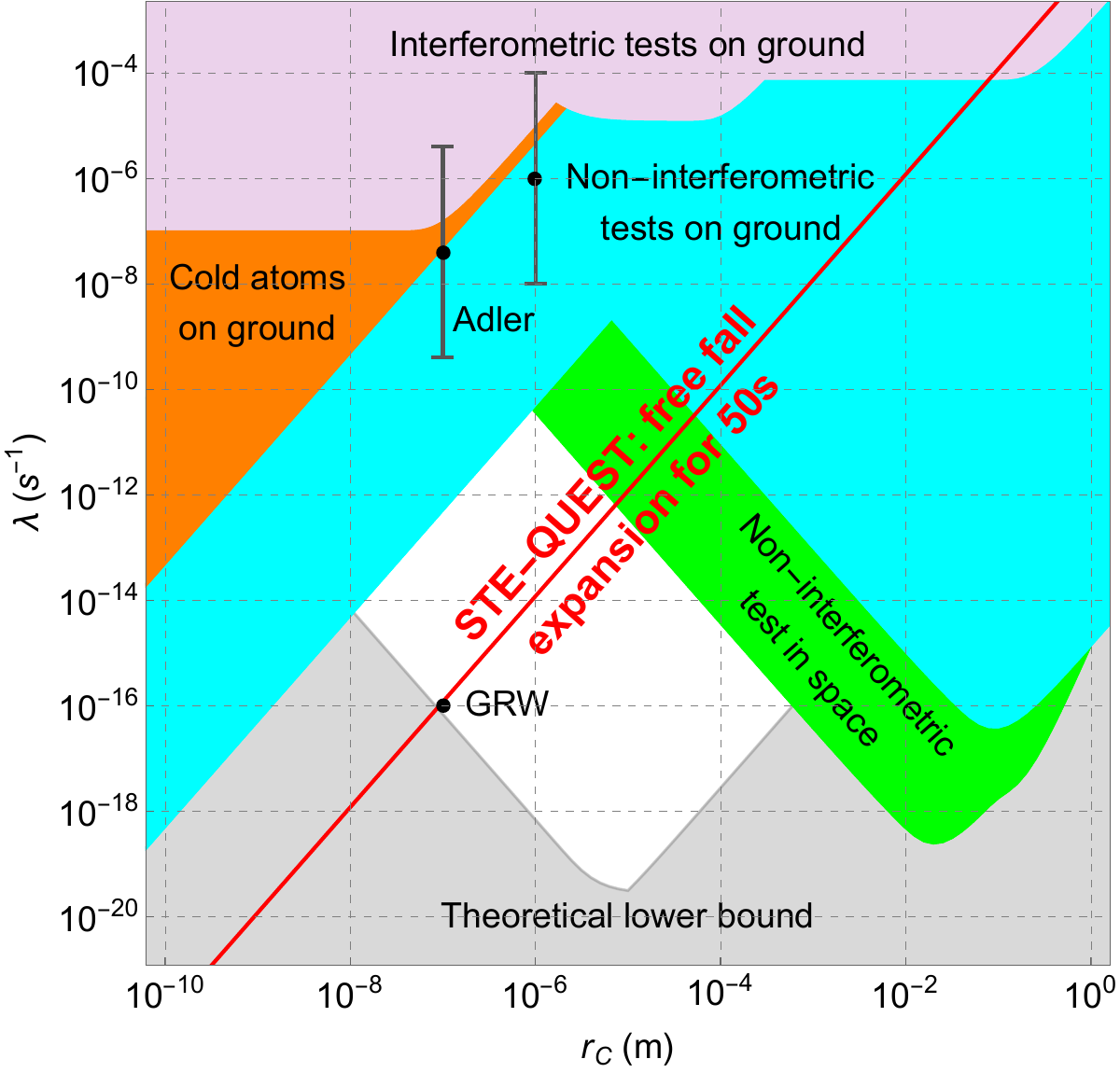}
\caption{\it Comparison between  state-of-the-art bounds on the CSL model for wave-function collapse~\cite{arndt2014testing,carlesso2022present} and what can be achieved by STE-QUEST. The red line is the sensitivity foreseen for STE-QUEST with free expansion for 50\,s. The pink region is excluded by interferometric experiments on the ground~\cite{Hornberger:2004aa,Belli:2016aa,torovs2017colored,fein2019quantum}, the blue area is excluded  by non-interferometric experiments on the ground~\cite{carlesso2016experimental,vinante2017improved,adler2019testing,vinante2020narrowing,donadi2021novel}, the orange region is excluded  by previous cold atom experiments on the ground~\cite{Bilardello:2016aa}, the green region is excluded by a non-interferometric experiment in space~\cite{carlesso2016experimental,PhysRevD.95.084054, altamura_improved_2025}, and the grey region is excluded theoretically assuming a collapse at the macroscopic scale being the basic requirement of the model~\cite{torovs2017colored}. The GRW and Adler values are reported as a black dot and an interval, respectively.}
\label{fig:CSL}
\end{wrapfigure}

The CSL model is characterized by two free parameters: the collapse rate $\lambda$, which characterizes the strength of the collapse, and the correlation length of the collapse noise $r_\text{\tiny C}$, which is the length-scale  defining the spatial resolution of the collapse and thus characterizing the transition between the micro and macro domains. Although extensive research over the past 20 years has set ever stronger upper bounds on these parameters~\cite{arndt2014testing,carlesso2022present}, there is still a wide unexplored region in the parameter space, as illustrated in Fig.~\ref{fig:CSL}. The parameter values labelled there as GRW after Ghirardi, Rimini and Weber ($\lambda =10^{-16}$\,s$^{-1}$ and $r_\text{\tiny C}=10^{-7}$\,m) were proposed theoretically so as to guarantee the effective collapse of macroscopic systems.
Conversely, the values of $\lambda =4\times 10^{-8\pm2}$\,s$^{-1}$ and $r_\text{\tiny C}=10^{-7}$\,m were proposed by Adler~\cite{Adler:2007ab} so that the collapse would take place at the mesoscopic scale instead. 
\noindent
While Adler's values have already been excluded experimentally, those of GRW are  yet to be tested and are commonly regarded as targets to reach for fully probing the model. Details on the excluded values of $\lambda$ and $r_\text{\tiny C}$ are reported in the caption of Fig.~\ref{fig:CSL}.
Since the structure of the CSL dynamics resembles that of a weak continuous Gaussian measurement (at zero efficiency, since the outcome of the measurement is not recorded) --- which is a quite general framework --- one typically regards it as a figure of merit for a wide class of collapse models.   

\paragraph{Di\'osi-Penrose (DP) model --} Also worthy of mention is the Di\'osi-Penrose (DP) model~\cite{diosi1987universal,penrose1996gravity}, which is also considered among the most important collapse models.
 The DP model predicts the breakdown of the superposition principle when gravitational effects are strong enough. Penrose provided several arguments why there is a fundamental tension between the principle of general covariance in General Relativity and the superposition principle of Quantum Mechanics~\cite{penrose1996gravity, penrose2014gravitization}, suggesting that systems in spatial superposition should collapse spontaneously to localized states and that this effect should get stronger the larger the mass of the system. A model that describes this effect through an equation of the form of Eq.~\eqref{eq.csl} was introduced by Di\'osi in Ref. ~\cite{diosi1989models} and is known as the DP model. The corresponding spatial correlation in Eq.~\eqref{eq.csl} is of the form $\mathcal D({\bf x}-{\bf y})=\tfrac{G}{\hbar}\tfrac{1}{|{\bf x}-{\bf y}|}$, where $G$ is the gravitational constant, so that the model is free from any fitting parameter. However, due to the standard divergences of the Newtonian potential in $D({\bf x}-{\bf y})$  at small distances, the collapse rate for a point-like particle diverges, irrespective of its mass. This implies an instantaneous collapse even for microscopic particles, in contrast to the requirements of the model.  
To avoid this divergence, a point-like mass distribution may be replaced by an extended mass distribution with a size given by a fixed minimum length $R_0$, which then becomes the only free parameter of the DP model. Several experiments set lower bounds on $R_0$~\cite{carlesso2022present, vinante2021gravity},  and the strongest bound is given currently by a search for spontaneous radiation emission from germanium~\cite{donadi2021underground}.

\paragraph{Position variance expansion --} The direct way to test collapse models is to quantify the loss of quantum coherence in interferometric experiments with particles as massive as possible, so as to magnify the collapse effects on the superposition~\cite{arndt2014testing}. Currently, the most massive particle that has been placed in a superposition has had a mass around $2.5\times 10^4$\,amu~\cite{fein2019quantum}. The corresponding bound is, however, around 9 orders of magnitude away from testing the GRW values. With the aim of testing such values, one would need to prepare superpositions with masses around $10^9$\,amu on a time-scale of 10\,s ~\cite{gasbarri2021testing}, which is far beyond the current capabilities of the state-of-the-art and near-future technology. 
In parallel to the interferometric approach, alternative strategies have been developed, which provide stronger bounds, without necessarily requiring the creation of a superposition state. They are based on indirect effects of the modifications collapse models introduce into quantum dynamics~\cite{carlesso2022present}, such as extra heating and diffusion or spontaneous radiation emission. Among them, the measurement of the variance in position $\sigma_t^2$ of a non-interacting BEC in free fall is of interest in STE-QUEST. It may be expressed as
\begin{equation}\label{eq.csl.position}
\sigma_t^2=\sigma_{\text{\tiny QM},t}^2+\frac{\hbar^2}{6m_0^2 r_\text{\tiny C}^2}\lambda t^3.
\end{equation}
The variance is enhanced by the action of collapse models on the BEC with respect to that predicted by quantum mechanics $\sigma_{\text{\tiny QM},t}^2\propto t^2$, exhibiting a different scaling that is proportional to the cube of the free evolution time.
This test can be implemented directly in STE-QUEST without requiring additional instrumentation beyond what is already envisioned for the interferometric experiment.    
A study of BEC expansion has already set a competitive bound on CSL~\cite{Bilardello:2016aa}, which excludes the  orange region in Fig.~\ref{fig:CSL}. This experiment was performed on the ground~\cite{Kovachy:2015ab}, where the major limitation was provided by gravity, which constrains the total duration of the experiments to a few seconds. In such an experiment, a BEC is created in a vertically-oriented quadrupole trap, allowed to evolve freely and cooled down through the use of a delta-kick technique to make $\sigma_{\text{\tiny QM},t}$ as small as possible. Finally, it is again allowed to evolve freely, and eventually its position variance is measured.
As suggested in Ref. ~\cite{kaltenbaek2016macroscopic,gasbarri2021testing,belenchia2021test,belenchia2022quantum}, operating such an experiment in space allows one to extend considerably the free-fall evolution time, and opens up the  possibility of making a competitive test capable of improving significantly the bounds on the CSL and DP models. By
measuring the BEC expansion over long free-fall times of the order of 50\,s and assuming a position variance accuracy of $\mu$m, the expected sensitivities to the CSL parameters are around 4 orders of magnitude stronger than those reached by state-of-the-art ground-based experiments, as seen in Fig.~\ref{fig:CSL}. Likewise, the bounds on the DP model can be improved by more than an order of magnitude.

\paragraph{Robustness of the bound --}
Another important point to address is the robustness of the bound when one considers deviations from the white spectrum for the collapse noise. Indeed, if the latter has a physical source, it becomes natural to assume that it will be characterized by a cutoff frequency $\Omega$, above which the collapse action is strongly suppressed~\cite{adler2007collapse}. To be quantitative, if one assumes that the collapse noise has a cosmological origin, a reasonable estimate is $\Omega\sim10^{11}-10^{12}\,$Hz ~\cite{ciufolini2010general}. While interferometric experiments are fairly robust to such modifications~\cite{arndt2014testing,torovs2017colored}, non-interferometric experiments are strongly dependent on the relation between $\Omega$ and the characteristic frequency/time-scale of the experiment~\cite{carlesso2018colored}. A prominent example is the spontaneous radiation emission in the X-ray band from germanium, whose corresponding bound is strongly suppressed when considering also large values for $\Omega$ up to the X-ray characteristic frequencies  $\sim 10^{18}\,$Hz. Experiments such as those involving the free expansion of BECs are strongly robust to such modifications, as their time-scale is quite long. In particular, the bound provided by STE-QUEST will not significantly change with respect to that in Fig.~\ref{fig:CSL} for $\Omega \geq 10^{-2}\,$Hz, which is many orders of magnitude smaller than the value obtained from the cosmological estimate.

\subsection{Other measurements of interest}
With the advance of the mission in its definition and with the increase of the scientific community around STE-QUEST we expect that new measurements and experiments of scientific interest will arise. Those could add to the scientific achievements of STE-QUEST and may be carried out provided they require no, or only very minor, changes to the instruments. They also need to fit into the overall measurement scheme and mission duration. One example could be a measurement of the gravitational Aharonov-Bohm effect following the recent first measurement~\cite{Overstreet2022}, but with potentially much better accuracy. To do so one would make use of the self gravity of the S/C (c.f. Sec. \ref{sec:self_grav}) and the long free evolution times and corresponding large superpositions of the wave packets. Rather than modulating the self-gravity as in Ref. ~\cite{Overstreet2022} one would instead modulate the free-evolution times and take differential measurements between the two species to measure the effect. At this stage, such a measurement is only a potential additional science objective to be explored, but it shows that new ideas may well arise in the coming years as STE-QUEST is developed in more detail.


\subsection{Possible de-scoping options}

STE-QUEST being essentially a single payload mission, de-scoping options cannot be implemented by removing individual instruments as may be the case e.g. in planetary missions or for focal plane instruments of astronomy missions. De-scoping then concerns simplifying the payload and/or reducing the mission profile/duration. 

This is likely to come at the cost of less ambitious goals of one or several science objectives. Without going into details, we note that all three major science objectives can afford a loss in performance of up to an order of magnitude, arguably even more, and still be of interest for a broad scientific community. Additionally, even with reduced objectives, STE-QUEST will still fully play its role as a vital element in the development of cold atom and quantum technologies in space, for future missions in fundamental physics or Earth observation~\cite{Alonso2022}.

As an example, if the imaging and/or wave-packet control requirements for the quantum mechanics test (c.f. Sec. \ref{sec.non.interf} below) turn out to be too challenging or costly, the ambitions of that test could be reduced, or in the worst case, that particular science objective could be abandoned.

Similarly, in some DM models, like the vector DM discussed in Sec. \ref{sec:Vec-DM}, bounds are set by searching for oscillations of the differential acceleration of Rb-K in a wide range of frequencies rather than only around the orbital frequency, $f_{orb}$. Indeed, the curve labeled ``STE-QUEST (Osc)'' in Fig. \ref{fig:Vec-DM} indicates the sensitivity when searching for oscillations over a range of frequencies rather than only $f_{orb}$, at which the Yukawa-term would manifest itself (labeled ``STE-QUEST (static)'' in Fig. \ref{fig:Vec-DM}). That in turn may lead to additional complexity in terms of control of noise and systematics. In that case the broadband search could be abandoned or carried out with less sensitivity, but as can be seen on Fig. \ref{fig:Vec-DM}, the loss in science remains tolerable.

In summary, de-scoping options concern mainly a reduction of the sensitivity and the breadth of the science objectives, which could lead to some reduction in complexity and corresponding cost of the payload and S/C.  

\section{Scientific requirements}

\subsection{Reference mission parameters}\label{sec:ref_mission}
The scientific requirements are derived from the primary mission objectives of the previous section using the reference mission parameters summarised in Tab. \ref{tab:ref_mission}. 

\begin{wraptable}{R}{0.5\textwidth}
    \centering
    \begin{tabular}{c|c}
        \multicolumn{2}{c}{Parameters}	\\
        \hline \\ [-1.5ex]
        Orbit & 1400~km SSO 6h\\
        Attitude & Inertial + modulation \\
        $g_0$ & 6.6~m/s$^2$ \\
        $\partial g_0/(2\partial r)$ & $8.5\times 10^{-7}$~s$^{-2}$\\
        $f_{orb}$ & $1.46\times 10^{-4}$~Hz \\
        Eclipses & none \\
        Mission duration $T_M$ & 3 years \\
        Science time $T_{sc}$ & 24 months \\
        \hline
    \end{tabular}
    \caption{\it\it Reference mission parameters.}
    \label{tab:ref_mission}
\end{wraptable}

\noindent In inertial attitude with the sensitive axis of the instrument in the orbital plane, the expected UFF-violating signal is modulated at orbital frequency, $f_{orb}$. For further de-correlation from systematic effects, we will modify the orientation of the sensitive axis by irregular (every 50 orbits on average) rotations of $\approx 10^\circ$ in the orbital plane, leading to an additional phase modulation of the expected signal \footnote{The optimal modulation sequence is yet to be determined, the sequence described here is a first estimate.}. The total mission duration of 3 years includes 6 months commissioning and a science duty-cycle of 80\% for the remaining 2.5~years.
 

\subsection{Experimental sequence and operational parameters}\label{sec:Instr_op_param}
\subsubsection{Atom source engineering}

\begin{figure}[h]
\includegraphics[width=0.6\textwidth]{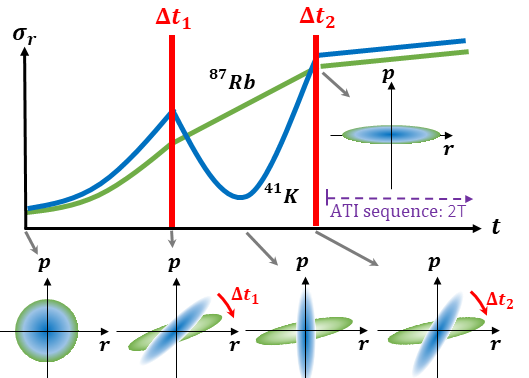}
\caption{\it Principle of the dual delta-kick collimation for a binary mixture following~\cite{Corgier2020}. After a pre-expansion of $t_{\rm exp,1}$, a first DKC pulse is switched on during $\Delta t_1$ to ensure a slow down of the $^{87}$Rb expansion and to focus $^{41}$K. This step is followed by a second free-expansion of $t_{\rm exp,2}$, long enough to let the $^{41}$K expand. A second DKC pulse of duration $\Delta t_2$ collimates simultaneously the two species and the ATI sequence can start. As an example, starting from a final trap of angular frequency $\omega_{Rb}=2\pi \times 10$\,Hz, $\omega_{K}=\sqrt{m_{Rb}/m_K}\omega_{Rb}$, an expansion speed below the 10\,pK can be reached for $t_{\rm exp,1}=300$\,ms, $\Delta t_1= 578\,\mu$s, $t_{\rm exp,2}=2036.4$\,ms and $\Delta t_2 \approx 82\,\mu$s. Simpler schemes with less pulses could be implemented in specific cases~\cite{Meister2022,Molecular-toolbox-decadal2022}.}
\label{fig:Dual_DKC}
\end{figure}

Two Bose-Einstein condensates of $^{41}$K and $^{87}$Rb are first simultaneously produced with $2.5\times 10^6$ atoms in each. An atom-chip-based magneto-optical trap (MOT) fed by a 2D$^{+}$-MOT captures and cools down the atoms of both species~\cite{Rudolph2015, Piest2021}. Later, atoms are transferred into a magnetic trap generated by the chip and pre-evaporated. This ensures high transfer efficiency from the magnetic trap to a crossed optical dipole trap (ODT), which is the next step of the atomic sample preparation. The ODT is used for further evaporation and for reaching the BEC phase. To ensure the miscibility of the two condensates, a magnetic field (about 70 G) is used to tune the inter-species scattering length~\cite{Thalhammer2008, Ferrari2002}. An optimized delta-kick collimation (DKC)~\cite{Chu1986, Ammann1997, muentinga2013, Kovachy2015} stage is applied in combination with the ODT after release and leads to the targeted expansion energy of 10\,pK. Shortly after, the magnetic field can be switched-off without any noticeable further effect on the expansion dynamics. This preparation of the binary quantum mixture is studied in detail in reference~\cite{Corgier2020} including the collisional mean field dynamics and is illustrated in Fig.~\ref{fig:Dual_DKC}. The required DKC performance is in line with droptower experiments achieving 3D expansion energies as low as 38\,pK ~\cite{Deppner2021}.

\subsubsection{Atom interferometry sequence}
 
The first operational mode of the main instrument is the dual-species atom interferometer (ATI). It is the standard ``interferometer mode'' where the atoms cooled to degeneracy are separated and recombined in a Mach-Zehnder scheme. 

\begin{wraptable}{R}{0.5\textwidth}
    \centering
    \begin{tabular}{c|c}
        
        \hline \\ [-1.5ex]
        Atom number $N$ & $2.5 \times 10^6$\\
        $k_\text{eff}$ for Rb & $8\pi/(780$~nm) \\
        $k_\text{eff}$ for K & $8\pi/(767$~nm)\\
        Free evolution time $T$ & 25~s\\
        Max. separation Rb & 0.59~m \\
        Max. separation K & 1.27~m \\
        Cycle time $T_c$ & 60~s \\
        Contrast $C$ & 1\\
        Expansion energy & 10~pK \\
        Expansion velocity $\sigma_{v,Rb}$ & 31~$\mu$m/s \\
        Expansion velocity $\sigma_{v,K}$ & 45~$\mu$m/s \\
        Init. pos. spread $\sigma_{r}$ & 500~$\mu$m \\ 
        Init. diff. position $\Delta r$ & 1~$\mu$m \\
        Init. diff. velocity $\Delta v$ & 0.1~$\mu$m/s \\
        Indiv. Velocity (in S/C frame) $v$ & 1~$\mu$m/s \\
        \hline
    \end{tabular}
    \caption{\it Operational parameters of the ATI.}
    \label{tab:params}
\end{wraptable}

\noindent In this mode, atoms are separated by a standard $\pi/2-\pi-\pi/2$ sequence, with a moderate beam splitting of order $2$ in order to limit the spatial extent of the interferometers (IFOs). The lasers probing $^{41}$K and $^{87}$Rb are appropriately detuned from the respective two-photon transitions to minimize spontaneous emission. The interferometer is realized by three laser pulses, which symmetrically split, reflect, and recombine the BECs trajectories. We anticipate the use of double diffraction schemes~\cite{Leveque2009,Ahlers2016} to take advantage of the high intrinsic symmetry that they ensure in the spatial splitting of the atomic paths. At this stage, we keep the option to operate Raman or Bragg double-diffraction open since each technique has advantages and drawbacks that need to be carefully studied~\cite{Hartmann2020}. This choice does however not impact the Laser System (LS) technological 
solutions discussed in the payload section which could be adapted to the one or other technique in a straightforward way.
The total interferometer sequence takes $2T = 50$~s, adding to 10\,s detection and preparation time, for a total cycle time of $T_c=60$\,s. 
ATI parameters that allow reaching the scientific objectives (initial kinematics and expansion velocity) are constrained by the systematics study of section~\ref{sec:UFFrequirements} and are given in Tab.~\ref{tab:params}. The atomic densities i.e. interactions at the first interferometry pulse have for example to be sufficiently low to avoid mean field statistical and systematic effects. This imposes to start with an initial size of the sample of half a mm. Due to the requirement on the relative positioning and differential velocities of the two atomic ensembles, the quantum mixture needs to be prepared when the system is in the miscible regime. The sequence chosen in our study optimizes the overlap of the two atomic gases and allows to control their expansion dynamics at the desired level. The overlap between the two atomic clouds will be measured by spatial imaging and continuously verified during the mission.

The contrast can be assumed to be near unity, since major sources of contrast loss, such as gravity gradients, can be mitigated as outlined in Ref. ~\cite{Roura2017,Loriani2020}. Unless stated otherwise, parameters of Tab. \ref{tab:params} are the same for the two species.

Each atom-light interaction process imprints on the atomic wave function information on the distance between the atom and a common retro-reflecting mirror that acts as the phase reference for both species. This information, depending on the motion of the matter waves with respect to the common mirror, can be read out in terms of atomic population at the output ports of the simultaneous atom interferometers.
As a consequence, any acceleration of the freely falling atoms with respect to that mirror along the sensitive axis (parallel to the lasers) leads to a phase shift equal to $\Phi = k_{eff} a T^2$, and given that the two-photon effective wave vector $k_{eff}$ is slightly different for the two species will lead to a residual parasitic signal in the differential measurement. To avoid that, the ``acceleration free'' combination of the individually measured IFO phases is formed in post analysis given by
\begin{equation}\label{eq:Phi_af}
    \Phi_{af} = \frac{2k_K}{k_{Rb}+k_K}\Phi_{Rb} - \frac{2k_{Rb}}{k_{Rb}+k_K}\Phi_K.
\end{equation}
This combination is insensitive to any common acceleration of the two species with respect to the reference mirror, but fully sensitive to a differential acceleration, and thus e.g.~a putative UFF violating signal. The IFOs are operated simultaneously ensuring that any motion of the reference mirror is indeed common to both species. The pre-factors in Eq.~\eqref{eq:Phi_af} are close to 1 ($\simeq 1\pm 0.008$) and for the derivation of the requirements below they will be approximated by 1, unless otherwise stated. Note that forming the combination \eqref{eq:Phi_af} requires some, not very stringent, knowledge of the $k_i$, treated in more detail in Sec. \ref{sec:laser_freq}.

\subsubsection{Non-interferometric measurements}
\label{sec.non.interf}

The other operation mode is a ``quantum expansion mode'', where the atoms are cooled to degeneracy, delta-kick collimated, left to expand freely and imaged at variable times. This mode is used for the test of quantum mechanics described in section \ref{sec.QM} and also for characterisation of atom parameters (temperature, position/velocity distribution and centre, etc.). Non-interferometric tests of collapse models are based on high precision measurement of the position variance of the atomic condensate.

\begin{table}
    \centering
    \begin{tabular}{c|c|c}
        Quantity & Value & Comment 	\\
        \hline  
       $\delta\sigma_0$, $\delta\sigma_t$ & $\leq1$ $\mu$m & Uncertainty of position width after $\mathcal{N}$ runs \\
       $\delta\sigma_{v,0}$ & $\leq0.02\,\mu$m/s & Uncertainty of velocity width after $\mathcal{N}$ runs \\
       $K_1$ & $\leq 10^{-3}$ s$^{-1}$ & 1-body loss coeff. (see \eqref{Nt}) for $N=10^6$ atoms \\
       $\tilde{K}$ & $\leq 10^{-15}$ s$^{-1}$ & 3-body loss coeff. (see \eqref{Nt}) for $N=10^6$ atoms \\
       $\delta v$ & $\leq 2\,$mm/s & drift during  imaging ($\sim500 \mu$s)\\ 
       $\sqrt{S_{a}(f)}$ & $\leq0.1\,\frac{{\rm m\,}{\rm s}^{-2}}{\sqrt{{\rm Hz}}}$ & vibrations during imaging ($\sim 500 \mu$s)\\ 
        \hline
    \end{tabular}
    \caption{\it Requirements from the test of quantum mechanics.}
    \label{tab:no_IFO_req}
\end{table}

\subsection{Requirements for the Quantum Mechanics test}
Testing and possibly quantifying the action of the collapse noise down to the GRW values requires the characterization of the initial preparation of the atomic cloud and that of the actions on the system of all the external noise sources to a total measurement inaccuracy of the  $\mu$m after a free evolution of 50\,s. Table \ref{tab:no_IFO_req} summarizes the related requirements.
To optimise the measurement, the Schr\"odinger contribution to the position variance should be minimized. This can be achieved by reducing the initial momentum variance of the atomic cloud, which is -- in the reference frame of the cloud -- proportional to its kinetic energy. Therefore, one needs to complement evaporative cooling techniques with delta-kick collimation ones. Ground-based experiments have successfully used the latter~\cite{Kovachy:2015ab}: an atom cloud, pre-cooled to nK through evaporative cooling, is initially trapped, and left to evolve freely for 1.1\,s. A harmonic trap is switched on for 35\,ms after which the cloud is left to evolve freely for a second time until the atom's position measurement is performed. The delta-kick imprints a force which is proportional to the displacement covered during the first free evolution, i.e. to the momentum $\sim \sqrt{\text{energy}}$ of the atom, thus slowing down the particles. In this way one can realize expansions of around 10\,pK in magnetic or optical traps~\cite{Kovachy2015,Deppner2021}.

The uncertainties in preparing and characterizing the initial state will enter in the total position variance $\sigma_t^2$ through the quantum-mechanical contribution: $\sigma_{\text{\tiny QM},t}^2=\sigma_0^2+\sigma_{v,0}^2t^2$, where $\sigma_0^2$ and $\sigma_{v,0}^2$ are  the position and velocity variances at time $t=0$ respectively.  The error on the initial position variance can be directly determined with a measurement at $t=0$, and corresponds to a requirement below the $\mu$m. Typically, $\sigma_{v,0}^2$ is determined through a measurement of $\sigma_{t}^2$ after some fixed time $t$, however this will interfere with determination of the collapse induced contribution. Nevertheless, since the latter is independent from the atom mass [cf.~Eq.~\eqref{eq.csl.position}], one can exploit the different variance evolution of the K and Rb clouds to quantify  the initial velocity variance and the corresponding errors, assuming the same position variances at $t=0$ and that they thermalise at the same temperature. This imposes a requirement on the error on $\sigma_{v,0}$ of the order of $2\times 10^{-8}\,$m/s.

Since the collapse noise adds to the standard noise sources, the best way to distinguish it is to minimize all the other noise sources. 
The major source of noise in STE-QUEST is atom loss, which needs to be minimized since it corresponds to a reduction of the signal-to-noise ratio when performing the measurement.  The principal contributors to atom loss are the one- and three-body recombination processes. The former corresponds to collisions of the BEC atoms with those of the background thermal cloud, whereas the latter accounts for inelastic collisions leading to molecule formation. These two processes can be characterized by the reductions of the number of atoms $N_t$ in the BEC that they induce~\cite{laburthe2003observation}:
\begin{equation}\label{Nt}
\frac{\operatorname{d}\! N_t}{\operatorname{d}\!t}=-K_1 N_t-\tilde{K}N_t^3,
\end{equation}
where $\tilde{K}=K_3/((2\pi)^3 3^{3/2} \sigma_t^6)$ with $K_3$ being the condensate three-body loss coefficient and we have assumed that the cloud is spherically symmetric with position variance $\sigma_t$~\cite{lepoutre2016production}.
Assuming  initially $N_0= 10^6$ atoms, in order to ensure 
an atom loss not larger than 10\% over 50\,s, one needs $\tilde{K}< 10^{-15}$\,s$^{-1}$  and $K_1< 10^{-3}$\,s$^{-1}$. On the other hand, working with $N_0=10^4$ atoms would require only $\tilde K<10^{-11}$\,s$^{-1}$ and $K_1<10^{-3}\,$s$^{-1}$. 
It is pivotal for the success of the STE-QUEST test of collapse models to reach sufficiently low values for $K_1$ and $\tilde K$ or, alternatively, to work with a non-negligibly smaller number of atoms. Employing $10^4$ atoms in place of $10^6$  imposes to run the experiment 100 times more to achieve the same statistical error, which is determined by $\sigma_{\text{\tiny QM},t}/\sqrt{N_t\times \mathcal N}$, where $\mathcal N$ is the number of independent runs of the experiment. For Rb, one has $\sigma_{\text{\tiny QM},t=50\,{\rm s}}=1.6\,$mm that requires $N_t\times \mathcal N\sim2.5\times 10^6$, while for K, $\sigma_{\text{\tiny QM},t=50\,{\rm s}}=2.3\,$mm requiring $N_t\times \mathcal N\sim5.3\times 10^6$.

Finally, another point to be addressed is the final measurement of the position variance. We consider the use of fluorescence detection technique which needs detection time of $\tau=500\,\mu$s ~\cite{Rocco_2014}. During this time, one needs to account for 1) the drift of the measurement apparatus with respect to the atom cloud, and 2) the vibrations acting on the measurement apparatus (since the atom cloud is freely falling, it is vibration free). The requirements on 1) are given by a drift smaller than the required accuracy of 1\,$\mu$m over 500\,$\mu$s giving a maximum drift of $2$\,mm/s. The requirement on 2) translates in a constrain on the vibrational noise $S_{a}(f)$ through
\begin{equation}
      \sigma^2(\tau)=\frac{1}{(2\pi)^2}\int \operatorname{d}\!f\,\frac{\left[\tau^2 f^2+\tfrac1{\pi^2} \sin ^2(\pi \tau f)-\tfrac1\pi \tau f \sin(2\pi f \tau)\right]}{f^4}\times S_{a}(f),
\end{equation}
and we want $\sigma(\tau) \leq 1\, \mu$m over $\tau=500\,\mu$s. Assuming a flat vibrational spectrum, one finds the constrain $\sqrt{S_{a}}\leq 10^{-1}\,$m/s$^2$/$\sqrt{{\rm Hz}}$, which is far above the vibration levels on board a S/C like STE-QUEST.

\subsection{Requirements for the UFF/LLI tests and dark matter searches}\label{sec:UFFrequirements}

As described in the previous sections the fundamental measurement for UFF/LLI-tests as well as DM searches is the differential acceleration of the two atomic species. Differences come mainly from the model dependent coupling to the composition of $^{87}$Rb and $^{41}$K, the altitude dependence and the phenomenology of the expected signal (modulation frequencies and oscillations). As a consequence the science requirements are derived with respect to the $\eta \leq 10^{-17}$ objective of the UFF test, but apply equally well to the other interferometric science objectives (LLI and DM). 

\subsubsection{Fundamental noise limit (quantum noise)}\label{sec:shot_noise}

For atom interferometric sensors the atom shot-noise (standard quantum noise) gives the ultimate limit that can be reached in the E\"{o}tv\"{o}s ratio determination, and yields the maximal achievable sensitivity to a potential violation signal, assuming that systematic and other stochastic errors can be kept below this level. That
fundamental noise source can in principle be reduced using quantum-noise reduction schemes like squeezing and entanglement, as already proposed~\cite{Geiger2018, Salvi2018,Szigeti2020,Corgier2021,Corgier2022} and implemented in some ground experiments~\cite{Greeve2021,Kasevich2022} for atoms in spatial superpositions. Whilst that may be an option for future missions, we consider the involved technology not sufficiently mature for STE-QUEST. Then the standard quantum noise per measurement cycle is given by the number of atoms in each species and the contrast and scale factors by 
\begin{equation}
  \sigma_{\Delta a}= \left(\sum_{i=K,Rb}(k_i T_i^2C_i\sqrt{N_i})^{-2}\right)^{1/2} \simeq \, 4.4 \times 10^{-14}\,\, \rm m/s^2  \,.
\end{equation}

Since atom shot noise is uncorrelated from shot to shot, it may be averaged down over many repeated cycles. For a space-borne mission on a circular orbit, where the satellite is kept inertial, the averaging over $n$ measurements gives a sensitivity to the Eötvös ratio of  
\begin{equation}\label{eq:S/N}
\sigma_\eta = \frac {\sigma_{\Delta a}\sqrt{2}} {g_0 \sqrt{n}}\,,
\end{equation}
where $g_0$ is the amplitude of the local gravitational potential gradient, and the $\sqrt{2}$ factor arises from the varying projection of the gravitational acceleration $\mathbf{g}$ onto the sensitive axis~\cite{Loriani2020}. Aiming for a target uncertainty of $\sigma_{\eta} \leq 1\times 10^{-17}$, with parameters as given in Table \ref{tab:params}, requires integration for  $\sim 20$~months. This is within the 24~months available science time (see Tab. \ref{tab:ref_mission}) leaving some time for additional/auxiliary measurements if necessary and specific measurements for other science objectives (e.g. sect. \ref{sec.non.interf}).

\subsubsection{Acceleration noise requirements}\label{sec:acc_noise}
\begin{wrapfigure}{R}{0.5\textwidth}
\includegraphics[width=0.48\textwidth]{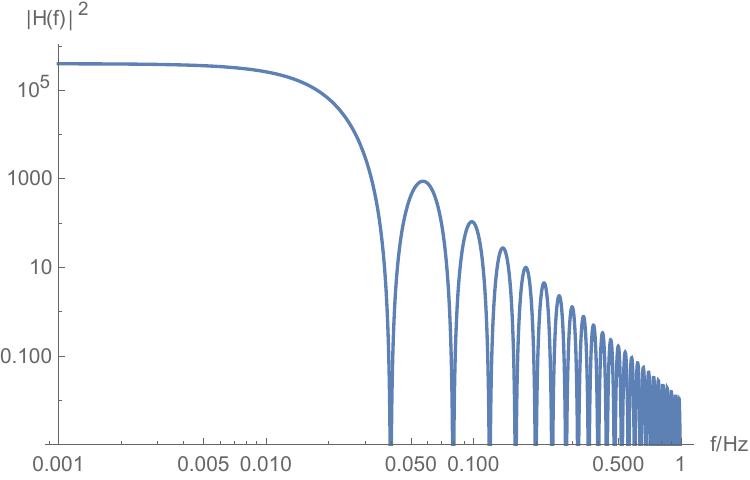}
\caption{\it Transfer function $|\frac{4}{(2\pi f)^2}\sin^2(\pi f T)|^2$.}
\label{fig:SingleH}
\end{wrapfigure}
Although the acceleration free combination defined in Eq.~\eqref{eq:Phi_af} is insensitive to accelerations of the reference mirror with respect to the atoms, there is still a residual requirement on the motion of the mirror coming from the fact that each individual IFO needs to be operated near half fringe to be maximally phase sensitive. 

\noindent Following the approach of Ref. ~\cite{Cheinet2008}, we use the single IFO sensitivity function to quantify the noise from residual mirror accelerations with an upper bound of\footnote{For a deviation of $\pi/10$ from mid fringe the slope (and thus noise) changes by $\leq$~5\%, which we consider negligible.} $\sigma_{\Phi_i} \leq \pi/10$, which provides the constraint
\begin{equation}\label{eq:Sa_constraint}
\int_{2\pi/T_c}^\infty \frac{S_{a}(\omega)}{\omega^4} \left|4\, k_{eff}\sin ^2(\omega T/2)\right|^2 d\omega \leq \left(\frac{\pi}{10}\right)^2    \,.
\end{equation}
Assuming white acceleration noise and $T_c=60$~s this gives a constraint of roughly $\sqrt{S_{a}(f)} \leq 4.0\times 10^{-10} \, \mathrm{m/s}^2/\sqrt{\mathrm{Hz}}$ in a band of approx. 0.01-0.5~Hz (at higher frequencies the transfer function has dropped steeply, see Fig.~\ref{fig:SingleH}, and the noise no longer contributes significantly). At low frequencies, such that $\omega T \leq 1$, Eq.~\eqref{eq:Sa_constraint} reduces to the variance of the ``usual'' IFO phase $k_\text{eff} \langle a\rangle T^2$, where $\langle a\rangle$ is the average value of $a$ when averaged over an interval of $2T$. The slowly varying $\langle a\rangle$ can be approximated by a low-order local polynomial and handled by feed forward from previous measurements onto the laser frequency in order to stay at mid fringe. For the feed-forward to work the additional condition is that $\langle a\rangle$ varies sufficiently little between cycles. In practice, that means $\Delta\Phi = k_\text{eff} T^2\langle\dot{a}\rangle T_c \leq \pi/10$, where $\Delta\Phi$ is the average change of IFO phase from one cycle to the next. Note that this procedure also caters for possible small acceleration biases stemming e.g. from the classical accelerometer used for the drag free and attitude control system (DFACS).

\subsubsection{Rotation noise requirements}\label{sec:rot_noise}
Contrary to accelerations, rotations of the S/C and reference mirror with respect to a local inertial frame are not suppressed in the ``acceleration free'' combination $\Phi_{af}$ defined in \eqref{eq:Phi_af}. They contribute to noise and systematics in that combination. Using the approach of Ref. ~\cite{Cheinet2008} we derive the sensitivity function of $\Phi_{af}$ to small angular motions 
\begin{eqnarray}\label{eq:Hrot2}
|H_{\theta}(\omega)|^2 &=& \left(2k_{eff}\right)^2\left[(\Delta v^yT)^2\sin^2(\omega T)\right.
+ \left. 4(\Delta y + \Delta v^y T)^2 \sin^4(\omega T/2)\right] \,,
\end{eqnarray}
valid for the case where the angular deviations $\theta \ll 1$ and with the sensitive axis of the ATI along $z$. We want the associated noise to be smaller than the quantum shot noise of Sect. \ref{sec:shot_noise}, which implies 
\begin{equation}\label{eq:rot_PSD_int}
\int_{2\pi/T_c}^\infty \frac{S_{\ddot{\theta}}(\omega)}{\omega^4} |H_{\theta}(\omega)|^2  d\omega \leq \frac{2}{N} \, .
\end{equation} 
This also ensures that the noise of each IFO is below $\pi/10$ required to stay at mid-fringe.

\begin{wrapfigure}{R}{0.5\textwidth}
\includegraphics[width=0.48\textwidth]{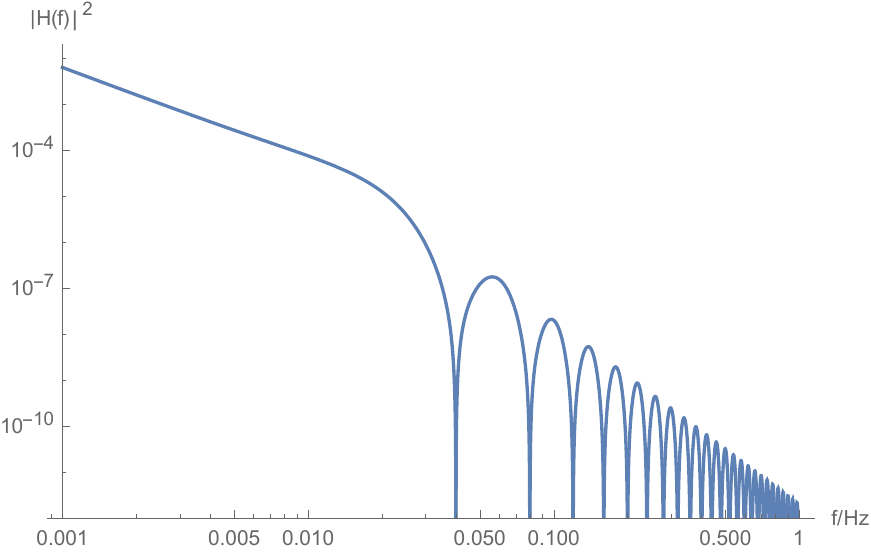}
\caption{\it Transfer function $|H_{\theta}(f)|^2/(2\pi f)^4$ of Eq.~\eqref{eq:Hrot2} in units of $k_\text{eff}^2$ with parameters as given in Tab. \ref{tab:params}.}
\label{fig:Hrot}
\end{wrapfigure}

Assuming white angular acceleration noise, this gives a constraint of roughly $\sqrt{S_{\ddot{\theta}}(f)} \leq 3.2\times 10^{-7} \, \mathrm{rad/s}^2/\sqrt{\mathrm{Hz}}$ in a band of approximately 0.01-0.5~Hz (at higher frequencies the transfer function has dropped steeply, see Fig.~\ref{fig:Hrot}, and the noise no longer contributes significantly). At low frequencies, such that $\omega T \leq 1$, Eq.~\eqref{eq:rot_PSD_int} reduces to the variance of the ``usual" Coriolis phase (coming from the differential Sagnac effect due to the small IFO areas created by $v^y$), given by $2k_{eff} \Delta v^y\langle \Omega\rangle T^2$, where $\langle \Omega\rangle$ is the average value of $\dot{\theta}$ when averaged over an interval of $2T$. The slowly varying $\langle\Omega\rangle$ is then a systematic effect and is treated in Sec.~\ref{sec:acc_rot_sys}. But, additional noise constraints come from the requirements to stay at half fringe for each shot and from the non zero spread in velocity of the atoms that will also generate a shot to shot noise through coupling to $\langle \Omega\rangle$, which then has to stay below atomic shot noise.  

The requirement to stay on half fringe is handled the same way as for accelerations (see Sect.~\ref{sec:acc_noise}), i.e. one can feed forward from previous measurements, which then implies that $\Delta\Phi = 2 k_\text{eff} T^2 v^y\langle\dot{\Omega}\rangle T_c \leq \pi/10$, where $\Delta\Phi$ is the average change from one cycle to the next. For the parameters of Table \ref{tab:params}, that corresponds to a requirement of $\langle\dot{\Omega}\rangle \leq 1.3\times 10^{-7}$~rad/s$^2$.

Each atom is prepared at finite temperature leading to a velocity  uncertainty at every shot of $\delta v = \sigma_v/\sqrt{N}$, which will translate into shot-to-shot phase noise of $2k_{eff} \delta v\langle \Omega\rangle T^2$. This additional phase noise has to be smaller than the shot noise $1/\sqrt{N}$. For the parameters of Tab. \ref{tab:params} that corresponds to a requirement of $\langle {\Omega}\rangle \leq 5.4\times 10^{-7}$~rad/s.

\subsubsection{Other noise requirements}
Other noise requirements come from self-gravity (through coupling to the spread in initial position/velocity of the atoms), thermal radiation and magnetic field fluctuations. They are treated in the respective sections on systematic effects (see Sections \ref{sec:self_grav}, \ref{sec:BBR} and \ref{sec:B-field}), as the more stringent constraints on the relevant parameters come from the systematics rather than the noise requirement.

\subsubsection{De-correlation of systematic effects}\label{sec:de-corr}
As briefly described in Sec. \ref{sec:ref_mission}, the signal we will be looking for is periodic at frequency $f_{sig}=f_{orb}$, but with irregular well controlled phase ``jumps'' of $\approx 10^\circ$ (details TBD). 
Let's consider a parasitic systematic effect at frequency $f_{p}$ of amplitude $A_{p}$; the impact of this perturbation on the searched signal can be reduced by several means (potentially cumulative):
\begin{itemize}
    \item If $f_{p}$ differs from $f_{sig}$, the perturbation can be de-correlated from our science signal provided that $|f_p-f_{sig}| > 1/T_{sc}$. Consequently, for a periodic effect of amplitude $A_{p}$ at frequency $f_{p}$ we will only consider the amplitude $A_{sys}$ of the component at $f_{sig}$ of that effect.
    \item The effect $A_{sys}$ is further reduced by a likely phase mismatch and the phase ``jumps'' mentioned above, which will be present in our signal but unlikely to be fully present in the systematic effect. The impact of most of the systematic effects will be considerably reduced in that way. Only a few perturbations, related to the direction of the Earth viewed from the satellite (e.g. thermal effects due to Earth radiations), will not be de-correlated and will have to be studied in more detail. Concerning the perturbations for which this technique is efficient, we will assume in the following an attenuation of $A_{sys}$ by a factor $10^{3}$. Indeed, we suppose that the attenuation could be limited by the imperfect control of the parameters of the modulation (angle, timing, etc\dots).
    \item Finally, if the systematic effect can be modelled (possibly with unknown parameters), its impact at $f_{sig}$ can be efficiently corrected provided this effect has also significant components at other frequencies different from $f_{sig}$, allowing the fitting of the model parameters to the data. A prime example of this is the effect of gravity gradients in MICROSCOPE whose amplitude $A_{sys}$ at $f_{sig}$ could be reduced by more than $10^{7}$ with respect to $A_{p}$ (at $f_p=2f_{sig}$) by fitting the model parameters (test-mass miscenterings) to the data~\cite{Touboul2017}, although the actual (unfitted) component $A_{sys}$ was only $\sim 10^{3}$ times smaller than $A_{p}$. 
\end{itemize}

\subsubsection{Systematics from accelerations and rotations}\label{sec:acc_rot_sys}
One of the main systematic effects may arise from residual accelerations and rotations of the S/C and/or the reference mirror. 

As explained in section \ref{sec:Instr_op_param}, the acceleration free combination \eqref{eq:Phi_af} is immune to accelerations that are common to the two species (e.g. S/C motion), and thus there are no hard constraints on accelerations at the signal frequency $f_{orb}$.

This is not the case for S/C (mirror) rotations with respect to the local inertial frame 
as it couples to the differential velocity $\Delta v^y$ of the two species leading to a differential phase shift $2k_{eff} \Delta v^y \Omega(t) T^2$, which has to be smaller than the one expected from the signal $k_{eff}\eta g_0 T^2$. More precisely, the component of the effect that is fully correlated with the modulated signal has to be smaller than that. Taking into account the $10^{3}$ factor due to the phase modulation (see Sec. \ref{sec:de-corr}) the corresponding requirement is $\Omega_{orb} \leq 3.3\times 10^{-7}$~rad/s, where $\Omega_{orb}$ is the component of $\Omega$ at orbital frequency.

\subsubsection{Laser frequency}\label{sec:laser_freq}
The acceleration free combination \eqref{eq:Phi_af} relies on knowledge of the wave vectors $k_i$ ($i=K,Rb)$. However, that knowledge will be limited, and the actual wave vectors will be $k_i+\delta k_i(t)$. The corresponding leading order error term is then
\begin{equation}
\delta \Phi_{af}(t) = a(t)T^2 (\delta k_{Rb}(t) - \delta k_K(t)) \,,
\end{equation}
where $a(t)$ is a residual acceleration of the S/C (the ATI mirror to be precise). The component of $\delta \Phi_{af}(t)$ that is at $f_{orb}$ and synchronous (including phase modulation) with our putative signal needs to be smaller than $\sim \eta g_0 k T^2$.

As the S/C is drag-free controlled $a(t)$ will be small, essentially driven by the bias and slow drifts of the classical accelerometer. Furthermore, at low frequency ($< 1/T_c$), the DFACS could be served to the actual measurement of each individual IFO (or their common mode combination) making the residual $a(t)$ even smaller, ultimately limited by shot noise of a single IFO per cycle ($\approx 3\times 10^{-14}\, \rm m/s^2$). Taking some margin to account for DFACS imperfections we assume $a(t) \leq 10^{-10}\, \rm m/s^2$ which gives $(\delta k_{Rb}(t) - \delta k_K(t))/k \leq 6.6\times 10^{-7}$. 

Note that this requirement applies to variations at $f_{orb}$ that are synchronous with the signal (incl. phase modulations) so there is ample margin in these requirements on $\delta k_i(t)$ as well as $a(t)$. In fact we expect the laser frequency knowledge to be dominated (at a similar $10^{-6}$ level) by gravity gradient cancellation (see Sec. \ref{sec:GGC}). 

\subsubsection{Earth's gravity gradient}\label{sec:GGC}
The Earth's gravity gradients (GG, the second derivatives of the gravitational potential) give rise to differential accelerations of two freely falling bodies (classical or atoms) of order $GM\Delta r/a^3$, where $a$ is the distance from the Earth's center and $\Delta r$ the separation of the two bodies. For a mission like e.g. MICROSCOPE, the effect is ``huge'', of order $3\times 10^{-12}\, g_0$ ($\Delta r \approx 20\, \mu\rm m$), when compared to the measurement objective of $10^{-15}\, g_0$. For free-fall tests of UFF on the ground, e.g.~\cite{Asenbaum2020}, the situation is even worse.

In space-tests of UFF the way to control and remove the GG effect is by now well established. Because the GG is a rank-2 tensor its effect on a circular orbit varies, to leading order, at twice the frequency $f_{sig}$ of the expected signal~~\cite{Touboul2017}. Components at the signal frequency are due to the deviation from spherical symmetry of the Earth's potential and residual eccentricity of the orbit, both about $10^{3}$ smaller than the leading order GG term at $2 f_{sig}$. The GG signal at $f_{sig}$ is then efficiently removed by modeling it, fitting the model parameters (e.g. $\Delta r$) at the main frequency $2 f_{sig}$, and thus removing the effect at $f_{sig}$. More precisely, the UFF-violating parameter $\eta$ and the GG parameters are fitted simultaneously to the data, and decorrelate to a large extent. For the MICROSCOPE mission this procedure allowed reducing the GG effect to a level compatible with a $10^{-18}$ test of UFF even in a single measurement session of 119 orbits (see Tab. III of Ref. ~\cite{Touboul2017, struckmann2024platform}).

The same methods can be applied to Atom-interferometric tests of UFF like STE-QUEST, however with two caveats. Firstly, the position and velocity spread of the atomic clouds ($\sigma_r, \sigma_v$) leads to a fundamental limit of the possible knowledge of $\Delta r$. Secondly, that spread and the resulting different trajectories, lead to a loss in contrast of the IFO and a resulting increase in phase noise. Recently a technique for circumventing both these difficulties has been proposed theoretically~\cite{Roura2017} and realized shortly afterwards in several ground experiments~\cite{Asenbaum2020,Overstreet2018,Damico2017}. That technique, called Gravity Gradient Cancellation (GGC) relies on changing ${\bf k}_{eff}$ of some of the laser pulses by a controlled amount, thus creating effectively an additional phase shift proportional to ${\bf \delta k}_{eff} r$ that is linear in $r$ but equal and opposite to the GG effect (also linear in $r$). The GGC technique is limited only by the knowledge of the local GG that needs to be compensated and by the capacity of laser control.

Applying GGC to satellite missions in an STE-QUEST-like scenario has been studied in detail in Ref. ~\cite{Loriani2020}, demonstrating reduction of the GG effect to below $10^{-17} g_0$ after only a few days of averaging, and likely even better if one uses appropriate modeling and fitting (see the MICROSCOPE example above and Sec. \ref{sec:de-corr}). This is the main reason for being able to relax the stringent (nm and 0.3~nm/s) requirements on $\Delta r$ and $\Delta v$ in the M4 proposal to the more comfortable $\mu$m and 0.1~$\mu$m/s in M7 (see Tab. \ref{tab:params}). It comes at the price of some additional complexity of the laser system and of the reflecting mirror as both need to be used to introduce changes in ${\bf k}_{eff}$ of order $10^{-4}$ in relative value~\cite{Loriani2020}. We note that these requirements on the initial positioning were satisfied by more than one order of magnitude in single species operations in reference~\cite{gaaloul2022} aboard the International Space Station. 

The GG-effect in STE-QUEST is of order $ 10^{-12}\, \rm m/s^{2}$ (1400~km altitude, $\Delta r = 1\, \mu$m, $\Delta v = 0.1\, \mu$m/s), which we want to reduce using GGC by 2-3 orders of magnitude (see~\cite{Loriani2020}) before further reduction by fitting and de-correlation. This imposes, roughly, a requirement of $10^{-3} - 10^{-2}$ on the knowledge of the local gravity gradient, the laser frequency shift and the angular change of the reference mirror. The mirror requires angular changes of order $10^{-4}$~rad between pulses (being static during the pulses), whilst still satisfying the requirements in angular stability given in sections \ref{sec:acc_rot_sys} and \ref{sec:acc_noise}. As an example the PAAM (Point-Ahead Angle Mechanism) mirror of the LISA mission is being developed for an angular stability of $<$~10~nrad/$\sqrt{\rm Hz}$ whilst maintaining a longitudinal stability $<$~1~pm/$\sqrt{\rm Hz}$, both of which are within the requirements of Tab. \ref{tab:platf_heritage} and Sec. \ref{sec:acc_rot_sys}. The laser frequency needs to be modified by $10^{-4} \nu_0$ ($\nu_0$ is the nominal frequency) with an uncertainty of $10^{-7}\nu_0 - 10^{-6}\nu_0$, and the knowledge of the gravity gradient at $10^{-4}$ is compatible with orbit knowledge and geopotential models. 

\subsubsection{Self gravity}\label{sec:self_grav}
We expand the gravitational potential of the satellite around a point corresponding to the nominal position $z_0$ of the Centres of Mass (CoM) of the BECs. For simplicity we do this only along the sensitive axis of the experiment, where the effect is largest, i.e. we write $V_{SG}(z)=\sum_0^\infty c_{SG}^{(n)}z^n$ similarly to the description of Ref. ~\cite{struckmann2024platform}. 

\noindent The $n=2$ coefficient is a local gravity gradient (GG) which can be compensated the same way as the Earth's GG, (see Sec. \ref{sec:GGC}), we thus require it to be no larger than that. For the rest, we calculate the maximum allowed value and uncertainty of each coefficient using the parameters of Sec. \ref{sec:Instr_op_param}. The uncertainty in the differential acceleration measured in the ATI, using the perturbative approach of Ref. ~\cite{ufrecht2020}, should not exceed $\eta g_0$  as detailed in Ref. ~\cite{struckmann2024platform}. 
\begin{wraptable}{tbr}{0.5\textwidth}
    \centering
    \begin{tabular}{c|c|c|c|c}
        n & $c_{SG}^{(n)}$ & $\delta c_{SG}^{(n)}$ & C=0.95 & unit	\\
        \hline 
        2 & $8.5\times 10^{-7}$ & $3.4\times 10^{-4}$ & $6.5\times 10^{-7}$ & s$^{-2}$ \\
        3 & $2.3\times 10^{18}$ & $2.1\times 10^{-8}$ & $5.8\times 10^{-5}$ &  m$^{-1}$s$^{-2}$ \\
        4 & $5.1\times 10^{-3}$ & $1.4\times 10^{-3}$ & $1.6\times 10^{-6}$ & m$^{-2}$s$^{-2}$ \\
        5 & $8.8\times 10^{2}$ & $6.4\times 10^{-8}$ & $1.3\times 10^{-4}$ & m$^{-3}$s$^{-2}$ \\
        6 & $1.5\times 10^{-2}$ & $4.0\times 10^{-3}$ & $4\times 10^{-6}$ & m$^{-4}$s$^{-2}$ \\
        7 & $1.3\times 10^{3}$ & $2.0\times 10^{-5}$ & $3\times 10^{-4}$ & m$^{-5}$s$^{-2}$ \\
        \hline
    \end{tabular}
    \caption{\it Self gravity requirements in terms of the maximum allowed coefficients $c_{SG}^{(n)}$ and uncertainties $\delta c_{SG}^{(n)}$ at DC following the approach in Ref. ~\cite{struckmann2024platform}.}
    \label{tab:self_grav}
\end{wraptable}
The same formalism is used to constrain the potential gradients induced by black body radiation (Sec.~\ref{sec:BBR}) and magnetic fields (Sec.~\ref{sec:B-field}). We take into account not only the direct differential effect of the potential, but also the associated effects due to $\Delta r, \Delta v, \sigma_r, \sigma_v$. 
We note that the corresponding effects are mainly at DC so should de-correlate from the expected signal to a large extent. We estimate the component at the signal frequency $f_{orb}$ to be driven by orbital temperature variations of $\leq 1$~K (max. value using MICROSCOPE thermal model) and a thermal expansion coefficient of the S/C of order $10^{-5}$~/K, leading to a reduction factor of $10^5$ with respect to the DC effect. Finally we apply the additional $10^{3}$ factor catering for the phase modulation (see Sec. \ref{sec:de-corr}). The requirements quoted in Table~\ref{tab:self_grav} are then the maximum allowed DC self-gravity coefficients and their uncertainties when taking into account the reduction factors above.  
Note that it should be relatively straightforward to model and correct these effects to some extent from onboard temperature measurements and a S/C thermo-elastic model, but we, conservatively, do not take that into account here. 
We have verified that the IFO phase noise introduced by the dependence of these terms on the initial velocity and position distribution of the atoms ($\sigma_r, \sigma_v$ of Table ~\ref{tab:params}) stays well below the atom shot noise for all values of Table ~\ref{tab:self_grav}. The only exception is the $n=2$ term which needs to be reduced by about two orders of magnitude using GGC, as already discussed. Similarly, we calculated the loss in IFO contrast for either of these terms. The corresponding constraints for the contrast to stay $\geq 0.95$ are also given in Table ~\ref{tab:self_grav}, but without the reduction factors due to signal modulation, as those do not affect the contrast loss. Note that, as one would expect contrast loss is mainly driven by even terms, which leads to more stringent requirements on those.
\subsubsection{Black body radiation}\label{sec:BBR}

\begin{wraptable}{tbr}{0.45\textwidth}
    \centering
    \vspace{-5mm}
    \begin{tabular}{c|c|c|c}
        n & $t_{\rm BBR}^{(n)}$ & $\delta t_{\rm BBR}^{(n)}$ & unit	\\
        \hline 
        1 & - & $2.5\times 10^{-4}$ & K.m$^{-1}$ \\
        2 & $6.1\times 10$ & $1.7 \times 10$ & K.m$^{-2}$ \\
        3 & $3.4\times 10^7$ & $6.8\times 10^{-4}$ & K.m$^{-3}$ \\
        4 & $2.1\times 10^2$ & $5.6\times 10$ & K.m$^{-4}$ \\
        5 & $2.9\times 10^7$ & $2.3\times 10^{-3}$ & K.m$^{-5}$ \\
        6 & $5.5\times 10^2$ & $1.5\times 10^2$ & K.m$^{-6}$ \\
        7 & $4.7\times 10^7$ & $7.4\times 10^{-3}$ & K.m$^{-7}$ \\
        \hline
    \end{tabular}
    \caption{\it Temperature gradient requirements in terms of the maximum allowed components at $f_{orb}$ of the coefficients $t_{\rm BBR}^{(n)}$ and uncertainties $\delta t_{\rm BBR}^{(n)}$, following the approach in Ref. ~\cite{struckmann2024platform}. An average temperature $t_{\rm BBR}^{(0)}=283$\,K with uncertainty $\delta t_{\rm BBR}^{(0)} = 1$\,mK and a gradient of $t_{\rm BBR}^{(1)} = 5$\,mK/m has been assumed here~\cite{Touboul2017}. Numerical values have been obtained for a  static polarizability of the atom: $\alpha_{Rb}=2\pi\hbar\times 0.0794 \times 10^{-4} $\,Hz.V$^{-1}$m$^2$ and $\alpha_K=\alpha_{Rb}/1.1$.}
    \label{tab:BBR}
\end{wraptable}

The effect of thermal radiation on the IFO measurement~\cite{Haslinger2018} is given in terms of spurious acceleration by
\begin{equation}\label{eq:BBR}
a_{BBR,i}=\frac{2\alpha_i \sigma }{m_ic \epsilon _0}\dfrac{\partial T_{\rm tube}^4(z)}{\partial z}\, ,
\end{equation}
where $\alpha_i$ is the static polarizability of atomic species $i$, $\sigma$ the Stephan-Boltzmann constant, $\epsilon_0$ the vacuum permittivity and $T_{\rm tube}(z)$ the temperature inside the vacuum tube at position $z$ along the sensitive axis.

To calculate the effect we expand $T_{\rm tube}(z)$ around a point corresponding to the nominal position $z_0$ of the Centers of Mass (CoM) of the BECs. We write $T_{\rm tube}(z)=\sum_0^\infty t^{(n)}_{BBR}z^n$. The calculations are then carried out following the same approach as for Sec. \ref{sec:self_grav}. 

For the systematic effect at orbital frequency, Table~\ref{tab:BBR} summaries the requirements up to order 7. Contrary to Sec. \ref{sec:self_grav} we state directly the requirements for the time varying part of those coefficients at orbital frequency (rather than the static DC part), which is likely to be significantly smaller than the DC part, but by an amount that is hard to quantify in the absence of a thermal model. However, as for the self-gravity, we do apply the $10^{3}$ reduction factor catering for the phase modulation (see Sec. \ref{sec:de-corr}). Note that it should be relatively straightforward to model and correct these effects to some extent from onboard temperature measurements and a S/C thermal model, but we, conservatively, do not take that into account here. 

We have checked that, with the constraints given in Table~\ref{tab:BBR}, the contrast and phase noise are not affected.

To compare these requirements to MICROSCOPE heritage, we have analyzed the data of temperature sensors positioned at the two ends and the middle of the MICROSCOPE payload along the sensitive axis (max. separation 0.159~m), for session 218 (119 orbits). The average temperature was 283~K with variations around that value of $<10^{-4}$~K. At frequencies above about 1~mHz the measurements were dominated by the temperature sensor white noise with a PSD of about $3.3\times 10^{-4} \, \rm K^2 Hz^{-1}$~\cite{Bars2019}. Below that frequency slow temperature drifts become visible. In the differential data of the sensors only the measurement noise remains. We fitted the amplitudes of oscillations at all relevant frequencies to that data ($f_{orb}, f_{spin}, f_{spin}\pm f_{orb}$), finding only measurement noise dominated gradients in the low $10^{-4}$~K/m and $10^{-3}$~K/m$^2$, which are worst case estimates as they are sensor noise dominated. 

\subsubsection{Magnetic fields}\label{sec:B-field}

The ATI is operated in $m_F=0$ states of $Rb$ and $K$ and is thus first order insensitive to magnetic effects. Then the effect of magnetic fields on the IFO measurement is given in terms of spurious acceleration due to the second-order Zeeman effect~\cite{Vanier1989} by
\begin{equation}\label{eq:MF}
a_{B,i}=\frac{\pi\hbar \chi_i}{m_i} \dfrac{\partial B_{\rm tube}^2(z)}{\partial z}\, ,
\end{equation}
where $\chi_i$ is the second-order Zeeman coefficient of atomic species $i$ and $B_{\rm tube}(z)$ is the magnetic field inside the vacuum tube at position $z$ along the sensitive axis. 
From ~\cite{Vanier1989} we have $\chi_{Rb} = 575.14\times 10^8\, \rm Hz/T^2$ and $\chi_{K} = 15460\times 10^8\, \rm Hz/T^2$. We evaluate the effect of the magnetic field gradients the same way as in the previous sections, i.e. we expand the magnetic field in a series expansion $B_{\rm tube}(z)=\sum_0^\infty b_{\rm B}^{(n)}z^n$ and calculate constraints on the coefficients $b_{\rm B}^{(n)}$. The results are presented in Tab. \ref{tab:2orderZeeman} up to order 7. 

\begin{wraptable}[16]{h}{0.45\textwidth}
    \centering
    \begin{tabular}{c|c|c|c}
        n & $b_{\rm B}^{(n)}$ & $\delta b_{\rm B}^{(n)}$ & unit	\\
        \hline 
        1 & - & $2.2\times 10^{-11}$ & T.m$^{-1}$ \\
        2 & $1.6\times 10^{-5}$ & $4.5\times 10^{-6}$ & T.m$^{-2}$ \\
        3 & $3.0$ & $1.1\times 10^{-10}$ & T.m$^{-3}$ \\
        4 & $4.1\times 10^{-5}$ & $1.1\times 10^{-5}$ & T.m$^{-4}$ \\
        5 & $4.6$ & $4.0\times 10^{-10}$ & T.m$^{-5}$ \\
        6 & $1.0\times 10^{-4}$ & $2.8\times 10^{-5}$ & T.m$^{-6}$ \\
        7 & $8.3$ & $1.3\times 10^{-9}$ & T.m$^{-7}$ \\
        \hline
    \end{tabular}
    \caption{\it Magnetic field gradient requirements in terms of the maximum allowed components at $f_{orb}$ of the coefficients $b_{\rm B}^{(n)}$, following the approach in Ref. ~\cite{struckmann2024platform}. An average field of $b_{\rm B}^{(0)}=100$\,nT with uncertainty $\delta b_{\rm B}^{(0)} = 50$\,pT and a gradient of $b_{\rm B}^{(1)} = 6$\,nT/m has been assumed here.}
    \label{tab:2orderZeeman}
\end{wraptable}
\noindent Contrary to Sec. \ref{sec:self_grav} we state directly the requirements for the time varying part of those coefficients at orbital frequency (rather than the static DC part), which is likely to be significantly smaller than the DC part, but by an amount that is hard to quantify in the absence of a detailed model of the magnetic shields and on orbit $B$-field variations. However, as for the self-gravity, we do apply the $10^{3}$ reduction factor catering for the phase modulation (see Sec. \ref{sec:de-corr}). We remark that the main time variation of $B^2(t)$ will be at $2f_{orb}$ because of the dipolar nature of the Earth's magnetic field, and thus decorrelate well from the EP-violating signal at $f_{orb}$. Furthermore modelling and fitting the main term at $2f_{orb}$ should allow reducing the effect at $f_{orb}$ by a large amount, as for the Earth's gravity gradient effect (see sect. \ref{sec:GGC} and \ref{sec:de-corr}). Knowing additionally, that control of magnetic field gradients below the nT/m level~\cite{Wolf2006a} are achieved on 30~cm scales on the ground (in a much more perturbed magnetic environment) and at a few nT/m over larger (8~m) scales~\cite{Wodey2020}, the requirements here on the $f_{orb}$ component should be achievable as one does expect it to be less than the DC value by at least two orders of magnitude.

\subsubsection{Wave-front aberrations}
Any deviations from a planar wave front lead to shifts that vary over the spatial extent of the atomic cloud. Following~\cite{Louchet2011,Schkolnik2015}, a parabolic curvature of the wave front leads to a bias acceleration of $a_\text{wf} = \sigma_v^2/R$, where $R$ is the radius of the curvature and $\sigma_v^2=k_B T_{\rm at}/m$ the effective expansion rate of the atomic ensemble. The resulting differential acceleration $\Delta a_\text{wf}$ should be below our target precision of $\eta g_0$. This requires $|\sigma_{v,\text{Rb}}^2 - \sigma_{v,\text{K}}^2|/R < \eta g_0$. Following~\cite{Corgier2020}, the relative differential expansion rate $\Delta\sigma_v/\sigma_v = 2|\sigma_{v,\text{Rb}} - \sigma_{v,\text{K}}|/(\sigma_{v,\text{Rb}} + \sigma_{v,\text{K}})$ can be reduced to $< 10^{-3}$ by tuning the timing of the dual-DKC. With this, it follows that the curvature must satisfy $R > 5 \times 10^{-4} (\sigma_{v,\text{Rb}} + \sigma_{v,\text{K}})^2 / (\eta g_0)$. Inserting the mission parameters\footnote{In Tab. \ref{tab:params}, for simplicity, we assumed equal temperatures (10~pK) for the two species, which leads to different $\sigma_v$ because of the mass difference. In practice the temperatures will be tuned for identical $\sigma_v$.} stated in Table \ref{tab:params} yields $R > 43.7$~km. The wave front curvature can be related to the mirror radius $r_m$ and peak-to-valley figure $\Delta z$ via $R = (\Delta z^2 + r_m^2)/(2\Delta z)$. The maximum BEC size (in terms of the Thomas-Fermi radius) during the interferometer is given by $R_{\rm TF,K}(2T) = 6.1$~mm, so one can estimate $r_m = 10$~mm, leaving a constraint on the peak-to-valley figure of $\Delta z < \lambda/700$ where $\lambda = 780$~nm. This is well within reach of mirror polishing and coating technology for large (35-55~cm) mirrors~\cite{Pinard2017}, and even more so for the small ($\sim$~cm) surfaces required here. Furthermore a non-zero curvature above that level will lead to a static shift in the differential measurement, which will decorrelate from the signal at $f_{orb}$ to a large extent, and even more when taking into account the phase modulation. Coupling to $f_{orb}$ is expected to be thermal (mirror deformations, laser intensities) or magnetic (DKC sequence) both of which should be small, but hard to estimate in the absence of a thermal and magnetic model. But either way, we do not foresee any difficulties to control wave-front aberrations below the required level, even with a large degradation (factor $>100$) of $R, \Delta \sigma_v/v, \Delta z$.


\subsubsection{Mean field effects}
An imbalance in the atoms' density in the two interferometer induces a phase shift due to a non-zero differential mean-field energy~\cite{Debs2011,HenselT.2021}. This can be treated as a statistical and a systematic uncertainty.

The statistical treatment sets a constraint on the BEC's size. Given a shot-noise limited accuracy for the first beam splitting ratio, the uncertainty reads
\begin{equation}
    \sigma_{\Delta\phi_{\rm MF}} = \frac{2T}{\hbar}\sqrt{N_{\rm at, Rb}\left(\frac{15 g_{\rm int, Rb}}{14 \pi R_{\rm TF,Rb}^3} \right)^2 + N_{\rm at, K}\left(\frac{15 g_{\rm int, K}}{14 \pi R_{\rm TF,K}^3} \right)^2}
\end{equation}
where the BEC's size is assumed to be constant and the interaction strength is defined as $g_{\text{int},i} = 4\pi \hbar^2 a_{\text{sc}, i}/m_{i}$ for $i=\text{Rb},\text{K}$ and the s-wave scattering length $a_{\text{sc, Rb}} = 98\, a_0$ and $a_{\text{sc, K}} = 60\, a_0$ where $a_0$ is the Bohr radius. This uncertainty is required to be below quantum shot noise. Since the interaction strengths of Rubidium and Potassium are approximately equal ($g_{\rm int,K}/g_{\rm int, Rb} \approx 1.3$), we can solve the requirement $\sigma_{\Delta\phi_{\rm MF}} < \sqrt{2}/\sqrt{N}$ for a common minimum required Thomas-Fermi radius $R_{\rm TF,min}$. Inserting the numbers of Table \ref{tab:params}, yields $R_{\rm TF, min} \approx 1.33$~mm or $\sigma_{r,\rm min} = R_{\rm TF,min}/\sqrt{7} \approx 500$~$\mu$m.

An imbalance of the first beam splitter, e.g., due to a finite fidelity, gives rise to a systematic uncertainty~\cite{Fitzek2020}. Suppose, after the beam splitter, we find $N/2 + \delta N/2$ atoms in the upper and $N/2 - \delta N/2$ atoms in the lower arm. Then, this imbalance induces an uncertainty in the phase given by
\begin{equation}
    \sigma_{\phi_{\rm MF},i}=\frac{2T}{\hbar}\frac{15 g_{\text{int},i}}{14 \pi R_{\rm TF}^3}\left( \left( \frac{N}{2} + \frac{\delta N}{2}\right) - \left( \frac{N}{2} - \frac{\delta N}{2} \right) \right)
\end{equation}
for $i=\text{Rb},\text{K}$. Setting the requirement $\sqrt{\sigma_{\phi_{\rm MF},\text{Rb}}^2 + \sigma_{\phi_{\rm MF},\text{K}}^2} < \eta g_0 k_{\rm eff}T^2$ yields a maximum allowed imbalance of $\delta N_{\rm max}/N = 9.48 \times 10^{-7}$ for the values in Table \ref{tab:params}. However, one does not expect that imbalance and the resulting phase shift to be correlated with the EP violating signal at $f_{orb}$ or it's phase modulation. The main cause of a variation in $\delta N(t)/N$ is likely a variation in the laser relative intensity of the first beam splitter pulse, which we estimate to be at most about $10^{-4}$ at $f_{orb}$. The phase modulation is expected to further decrease the part of $\delta N(t)$ that is correlated with our signal to below the required $10^{-6}$ in $\delta N(t)/N$.




\subsection{Platform stability requirements and heritage of previous missions}\label{sec:DFACS_summary}
Here we summarize the requirements on residual accelerations and rotations of the S/C from the sections above and compare them to the known performance of previous ``low inertial noise'' platforms: MICROSCOPE and LISA-Pathfinder (LPF). We also consider the performance of past (GRACE-FO) or future (NGGM) accelerometers to estimate the expected DFACS (Drag Free and Attitude Control System) performance, bearing in mind that at high frequency (larger than the typical DFACS bandwidth $\approx 0.01$~Hz) one can act on the laser frequency\footnote{For instance, an acceleration of $10^{-9}$~m/s$^2$ @ 0.1~Hz leads to a position change of the mirror by $\delta x \simeq 2.5$~nm, which can be compensated by changing the frequency of one of the laser pulses by $\delta f = f_0\delta x/x_0 \simeq 1.4$~MHz, where $f_0$ is the nominal laser frequency and $x_0$ the nominal distance between the atoms and the mirror ($x_0 \approx 70$~cm).} or mirror orientation rather than the S/C itself.

Table \ref{tab:platf_heritage} summarizes the constraints on S/C accelerations and rotations and compares them to previous and near future missions or their specifications.

\begin{table}[h!]\label{tab:SC_acc_rot}
\centering
\begin{tabular}{|m{1.5 cm}|m{3.1cm}|m{4cm}|m{2cm}|m{2cm}|m{2cm}|} 
\hline
{\bf Quantity} & \textbf{Constraint} & \textbf{Comment} & \textbf{$\mu$SCOPE} & \textbf{LPF} & \textbf{GRACE-FO} \\
\hline 
$\sqrt{S_a(f)}$ & $4.0\times 10^{-10}$ m/s$^2$/$\sqrt{{\rm Hz}}$ in [0.01:0.5]~Hz & From (\ref{eq:Sa_constraint}) assuming white noise & OK in (\ref{eq:Sa_constraint})$^{(1)}$ & OK in (\ref{eq:Sa_constraint})$^{(1)}$ & OK \\
\hline 
$\langle\dot{a}\rangle_{2T}$ & $2.5\times 10^{-13}$ m/s$^3$ & $\langle\dot{a}\rangle_{2T}=$ average (over 2T) of $\dot{a}$, cf. text after (\ref{eq:Sa_constraint}). & OK$^{(2)}$ & (OK)$^{(3)}$ & (OK)$^{(4)}$ \\
\hline  
$\Omega_{orb}$ & $3.3\times 10^{-7}$~rad/s $^{(5)}$ & Amplitude of component of $\Omega$ at orbital frequency cf. Sec. \ref{sec:acc_rot_sys} & OK & OK & - \\
\hline
$\sqrt{S_{\dot{\Omega}}(f)}$ & $3.2\times 10^{-7}$ rad/s$^2$/$\sqrt{{\rm Hz}}$ in [0.01:0.5]~Hz & From (\ref{eq:rot_PSD_int}) assuming white noise & OK$^{(6)}$ & OK & - \\
\hline
$\langle\Omega\rangle_{2T}$ & $5.4\times 10^{-7}$ rad/s & cf. Sec. \ref{sec:rot_noise} & OK & OK & - \\
\hline
$\langle\dot{\Omega}\rangle_{2T}$ & $1.3\times 10^{-7}$ rad/s$^2$ & cf. Sec. \ref{sec:rot_noise} & OK & OK & - \\
\hline

\end{tabular}
\caption{Requirements on S/C accelerations and attitude. Superscripts in brackets refer to the notes in the text. Note that the $\langle \dots \rangle_{2T}$ constraints apply to variations at frequencies such that $2\pi f T < 1$.}
\label{tab:platf_heritage}
\end{table}

\noindent Notes on Table \ref{tab:platf_heritage}:
\begin{enumerate}
\item OK when integrating real noise in Eq. (\ref{eq:Sa_constraint}), ie. without assuming white noise.
\item OK but with little margin. But can be improved by better feed forward strategy (rather than simply propagating the value of the previous cycle).
\item Marginally OK ($5.9\times 10^{-13}$~m/s$^3$). Limited by out of loop noise, which could be reduced by acting on laser frequency, which allows increasing loop bandwidth. And can be improved by better feed forward strategy.
\item Marginally OK ($3.8\times 10^{-13}$~m/s$^3$). Limited by detector noise. But can be improved by better feed forward strategy.
\item Assuming $10^3$ suppression by modulating the phase of the EP-violating signal c.f. Sect. \ref{sec:de-corr}.
\item OK for session 218. Marginally OK for session 216.
\end{enumerate}


For MICROSCOPE we use in-flight data of sessions and 216 (inertial) and 218 (after removing the constant spin). The attitude files provide directly S/C attitude at 4~Hz sampling. The S/C DFACS used one of the test masses as a sensor, then the differential acceleration file (4~Hz sampling) between the two test masses provides the ``out of loop'' measurement of residual accelerations. The MICROSCOPE spectral densities are obtained directly from the flight data. Evaluating Eq.~\eqref{eq:Sa_constraint} we obtain 0.14~rad, which satisfies the $\pi/10$ requirement, but with relatively little margin. The $\langle\dots\rangle_{2T}$ are evaluated from the MICROSCOPE data by simulating the STE-QUEST sequence: 
\begin{itemize}
\item We calculate $\langle \Omega\rangle_{2T}$ for every interval $2T$ out of $T_c$ (i.e. every 50~s followed by 10~s dead-time). We obtain $\langle \Omega\rangle_{2T} = (-0.4 \pm 9.2)\times 10^{-7}\, \rm rad/s$ for session 216 (median of the three axis) and $\langle \Omega\rangle_{2T} = (0.00016 \pm 3.0)\times 10^{-7}\, \rm rad/s$ for session 218 (median of the three axis).
\item For $\langle \dot{a} \rangle_{2T}$ and $\langle \dot{\Omega} \rangle_{2T}$ we use the same procedure to calculate $\langle a\rangle_{2T}$, $\langle \Omega\rangle_{2T}$ and then calculate the difference between successive values to simulate a simplistic feed forward procedure. We obtain $\langle \dot{a} \rangle_{2T} = (-0.012 \pm 6.65)\times 10^{-14}\,\rm m/s^3$ (session 218), and $\langle \dot{\Omega}\rangle = (0.0006 \pm 1.28\times 10^{-8})\,\rm rad/s^2$ (median of the three axis of session 216, with similar values for 218).
\end{itemize}
Finally we fit amplitude and phase of a sine at orbital frequency to the rotation data to obtain $\Omega_{orb} = 5.17\times 10^{-8}\, \rm rad/s$ for session 216 and $\Omega_{orb} = 1.4\times 10^{-9}\, \rm rad/s$ for 218 (median of the three axes), which are well below the requirement.

For LPF we use the results reported in Figs. 6,7 of Ref. ~\cite{Armano2019}. We integrate the PSD according to eqs. \eqref{eq:Sa_constraint} and \eqref{eq:rot_PSD_int}. For the $\langle\dots\rangle_{2T}$ we integrate the PSD in a $[0:1/{2T}]$ band. We find that LPF satisfies all rotational requirements with orders of magnitude margin. However, the acceleration requirements are only marginally satisfied. For \eqref{eq:Sa_constraint} we obtain $0.15\, \rm rad$, which satisfies the $\pi/10$ requirement, but with little margin. For $\langle\dot{a}\rangle_{2T}$ we obtain $5.9\times 10^{-13}\, \rm m/s^3$, which is about a factor 2 larger than our requirements, but limited by out of loop noise on LPF, which could be reduced by acting on laser frequency, which allows increasing the loop bandwidth. It could also be handled by a better feed forward strategy.

For GRACE-FO, we use the measured performance of the onboard accelerometers as reported in Fig.~2 of Ref. ~\cite{Christophe2015}. For \eqref{eq:Sa_constraint} we obtain $0.03\, \rm rad$, which satisfies the $\pi/10$ requirement. For $\langle\dot{a}\rangle_{2T}$ we obtain $3.8\times 10^{-13}\, \rm m/s^3$, which is about a factor 1.5 larger than our requirements, limited by detector noise. It could also be handled by a better feed forward strategy, and is likely to be less for the next generation ONERA accelerometers (see below).

In conclusion, we remark that the DFACS and accelerometer performance of either of the heritage missions is sufficient for STE-QUEST, but with little margin in some cases, for which we have provided some suggestions for improvement. Let us also add that accelerometers under development by ONERA for the next generation gravity mission (NGGM) are expected~\cite{Rodrigues2022} to perform about one order of magnitude better than those on GRACE-FO, and also have improved sensitivity to angular accelerations at about $10^{-9}$~rad/s$^2$, which is comparable to LPF performance.


\subsubsection{Orbit determination requirements}
Similarly to MICROSCOPE, orbit determination requirements are not very stringent, because the measurements are local differential measurements and thus orbit errors only play a role via perturbing effects from external factors. The main driver is then the error in the gravity gradient coming from an error in position $r$, which leads to an incorrect estimation of the corresponding acceleration, which we want to be below $\eta g_0$. Assuming a simple spherical potential we then have as a requirement
\begin{equation}
    \delta r \leq \frac{\eta g_0 r^4}{3 GM \Delta r} \approx 200\,\rm m \,.
\end{equation}
Using an onboard GNSS receiver this requirement should pose no difficulties. Furthermore it is highly pessimistic as we did not take into account GGC (see section \ref{sec:GGC}), and more importantly ignored any de-correlation and phase mismatch between the combined effect of orbit errors, GG and our signal (see Sec. \ref{sec:de-corr}).

\section{Scientific instrument}

\subsection{Payload}\label{section:payload}
The payload consists of a dual species atom interferometer (ATI) which compares the free evolution of matter waves of ultra-cold potassium $^{41}$K and rubidium $^{87}$Rb atoms. The differential acceleration between the two samples is continuously measured over the spacecraft orbit. 

\begin{figure}[ht]
\centering
\includegraphics[width=0.7\textwidth]{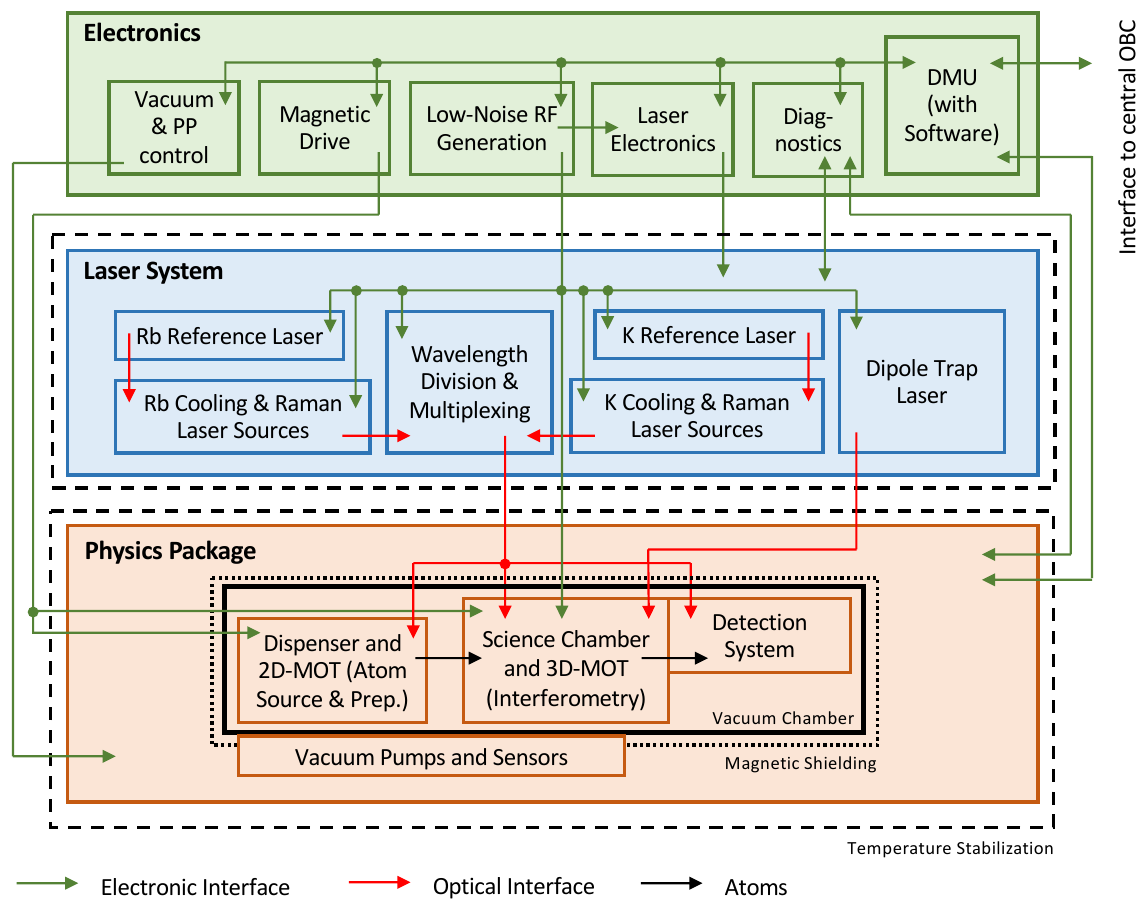}
\caption{Functional diagram of the atom interferometer. }
\label{fig:functional}
\end{figure}

The ATI consists of the following three subsystems: Physics Package (PP), Laser System (LS) and Electronics. They are detailed in the following sections, together with the software which is implemented within the Data Management Unit (DMU). The functional diagram of the ATI is shown in figure \ref{fig:functional}. 

\subsubsection{Physics Package}

The Physics Package denotes the subsystem in which the atoms are captured, cooled, coherently manipulated, and detected during an experimental sequence (see ref.~\cite{Schuldt2015} for the setup proposed in M3). It interfaces with the Electronics Package for driving electronics inside the physics package, e.g. coils, and recording data, as well as with the laser system via optical fibres providing light fields for manipulating the atoms, and the satellite bus for mounting and dissipating heat.

The planned experiments require an ultra-high vacuum system to be maintained by an ion-getter pump and a passive pump, and consists of several chambers. Inside the science chamber, an atom chip provides magnetic and radio frequency fields for capturing and evaporating atoms. The science chamber is connected to the pump system, and a 2D$^+$-MOT chamber providing a cold atomic beam to load the 3D-MOT on the atom chip. It additionally features tubes to accommodate the baseline of 130\,cm required for the atom interferometer. On one end, a viewport enables flashing in the beam splitting light pulses, on the opposing side, a steerable, intra-vacuum mirror retro reflects these light fields. Both the 2D$^+$-MOT and main chamber feature multiple viewports for application of light fields for cooling and preparation as well as reading out the signal of the atoms. The 2D$^+$-MOT chamber itself is connected to two reservoirs housing a sample of $^{87}$Rb or $^{41}$K, to supply a background vapor for the 2D$^+$-MOT which can be adjusted by heating. A differential pumping stage prevents contamination of the main chamber. The vacuum chambers are manufactured from titanium with anti-reflection coated N-BK7 viewports.

Multiple beam shaping optics and sets of coils are directly mounted to the vacuum chambers. The optics shapes the laser beams for cooling, preparation, coherent manipulation, with collimated beam diameters depending on their function, typically one to few centimeters, and the crossed optical dipole trap, with a focus above the atom chip. Additional sets of optics are implemented for detection with cameras. The coils provide magnetic fields to operate the 2D$^+$-MOT, 3D-MOT and magnetic trapping in conjunction with the atom chip, as well as the offset field for interferometry. One pair at the main chamber supports magnetic field generation of up to 80\,G for tuning the $^{41}$K-$^{87}$Rb inter-species Feshbach resonance.

Multiple additional sensors are integrated in the Physics Package, both for housekeeping, analysis and optimization of the atom interferometer. Photo diodes read out the laser power supplied to the physics package, thermal sensors record the temperature, and magnetometers track the switching and strength of the magnetic fields.

The vacuum chamber including the peripherals, but excluding the pump section, is housed inside a multi-layer magnetic shield. This shield suppresses external stray fields which could harm the preparation of the atom and the interferometry, and contains the magnetic fields generated inside the Physics Package.

\subsubsection{Laser system}
\label{sec:LaserSysSummary}
\begin{figure}[ht]
\centering
\vspace{0.5cm}
\includegraphics[width=0.8\textwidth]{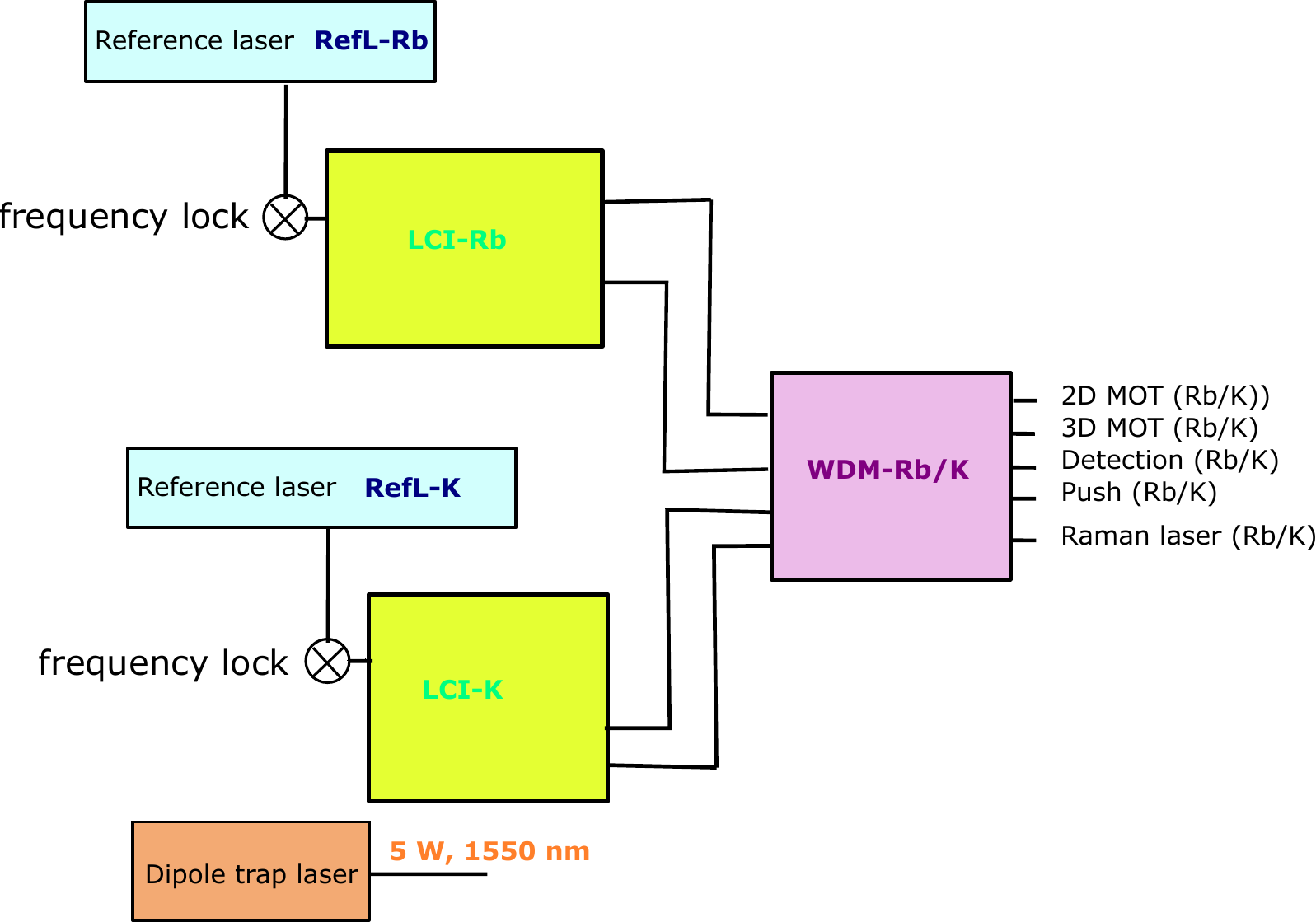}
\caption{Block diagram of the laser system. For each atomic species, a reference laser is composed of a master laser source servo locked on a Rubidium (respectively Potassium cell) by absorption spectroscopy (Reference Laser, RefL). The laser sources for cooling and interferometry (LCI) are then frequency locked on the reference laser with a beat note unit. The two laser sources at different wavelengths (780\,nm for Rubidium and 767\,nm for Potassium) are then combined in a wavelength division multiplexing unit (WDM Rb/K) which provides all the outputs corresponding to the different functions of the laser system (2D MOT, 3D MOT, Raman/Bragg, detection and push). A separate laser is dedicated to the dipole trap. }
\label{fig:LaserBlockDiagramSTEQUEST}
\end{figure}

The laser system is based on telecom technology and frequency doubling, inherited from industrial and research activities led since 15 years (see the heritage described in section \ref{sec:LaserHeritage}). Fig. \ref{fig:LaserBlockDiagramSTEQUEST} presents the block diagram of the full laser system. The subsystems for Rubidium and Potassium are identical on a technology and architecture point of view. Because of the wavelength difference (780\,nm and 767\,nm), it is not possible to use a single laser source for both species due to the different phase matching of the PPLN waveguide and the limited tunable optical frequency range of the laser source. 

The reference laser (RefL) includes a narrow linewidth laser diode (External Cavity Laser Diode), an all fibered frequency doubling module (PPLN waveguide), and a laser reference unit (Rubidium or Potassium cell, photodiode, discrete optics). Each laser for cooling and diffraction (Raman or Bragg) is frequency locked on the reference laser using a beat note. They include two phase locked laser diodes to supply the two frequencies required for laser cooling and two photon transitions. Each laser source is amplified in an Erbium Doped Fiber Amplifier (EDFA) and doubled in the PPLN waveguide. They can be turned off independently, using a fibered AOM situated between the EDFA and the PPLN. 

Since the atom source for the same species are produced simultaneously at the same location, an additional dichroic micro-optical bench (WDM-Rb/K) recombines the two wavelengths 767\,nm and 780\,nm to inject the two laser lights in the same optical fiber for each function: 2D MOT, 3D MOT, Detection, Push, Raman/Bragg. 

The dipole trap laser is a dedicated subsystem including a fibered laser source at telecom wavelength, two fibered AOMs and two mechanical shutters to control independently the two arms of the far detuned dipole trap. 

Note that for gravity gradient cancellation, it will be necessary to shift the laser frequency by typically 100\,GHz  before the $\pi$-pulse of the atom interferometer. This function will require a tunable laser source and additional electro-optical modulators and will require a dedicated study.

\subsubsection{Electronics}

\begin{figure}[ht]
\centering
\includegraphics[width=1\textwidth]{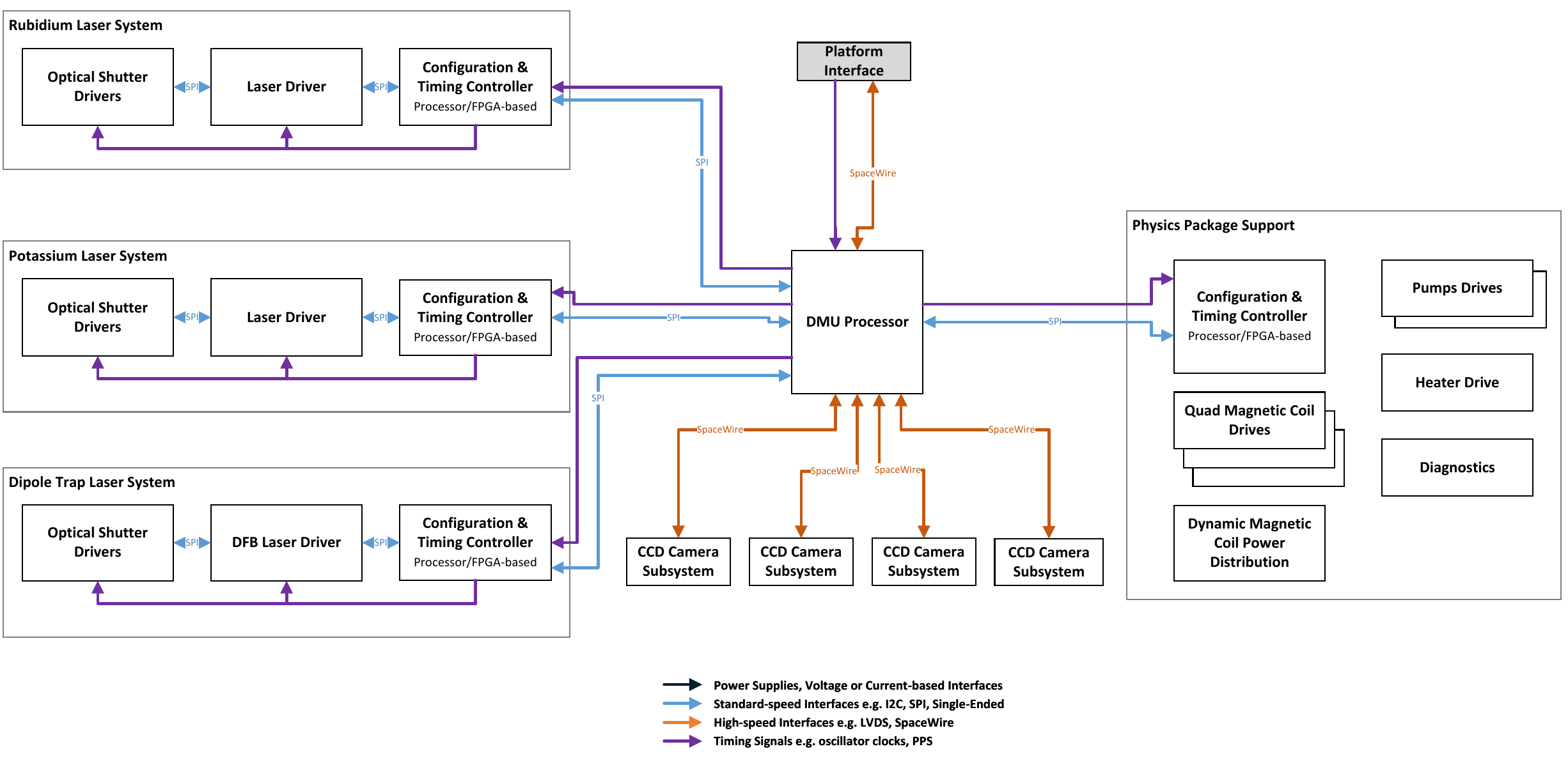}
\caption{Electronics System Diagram}
\label{fig:ElectronicsSystemDiagramSTEQUEST}
\end{figure}

The electronic functionality required to operate the STE-QUEST payload is substantial and is reliant on a range of targeted high-performance subsystems running specific functions.  This modularized approach provides programmatic advantages during development and testing and allows functionality to be readily duplicated across the payload. Figure \ref{fig:ElectronicsSystemDiagramSTEQUEST} shows a block diagram representing the payload control electronics and all of its modules.\\

\begin{figure}[ht]
\centering
\vspace{-0.5cm}
\includegraphics[width=1\textwidth]{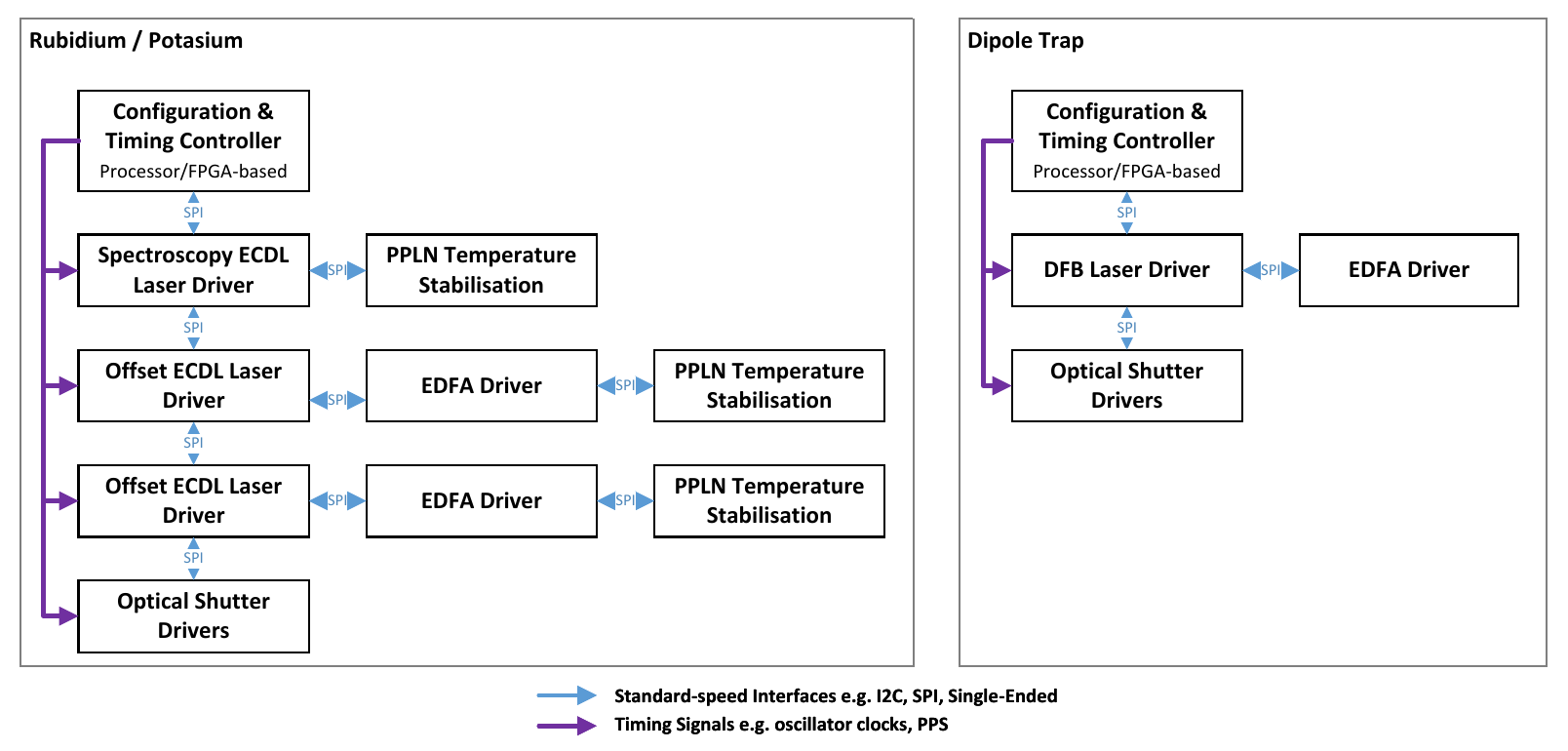}
\vspace{-0.8cm}
\caption{Block diagram representing the Laser Control Electronics for the Rubidium and Potassium (left) and Dipole Trap (right) laser systems. }
\label{fig:ElectronicsRbKLaserControlSTEQUEST}
\end{figure}

\noindent\textbf{Laser Control System}\\
Many of the subsystems rely on a core current drive circuit, alongside a high-performance temperature sense circuit with bi-directional TEC current drive to allow for temperature control and stabilization.  The circuit topology is shared across these subsystems, however the exact parameters of these circuits can be adapted to meet specific requirements, such as maximum allowable current range, or to trade-off performance parameters such as noise and bandwidth. \\
The Spectroscopy-Locked ECDL driver provides the core current driver and temperature stabilization for the Reference ECDL, whilst also incorporating the drive and sense circuitry for the laser reference unit and spectroscopy cells.  To ensure robust operation in the space-environment the Spectroscopy ECDL laser driver is required to independently scan and identify the appropriate atomic transitions to lock the reference frequency to, and all control algorithms are implemented in the digital-domain to maintain flexibility. \\
In a similar manner to the Spectroscopy-Locked ECDL, the Offset-Locked ECDL driver provides the core current driver and temperature stabilization for the cooling and diffraction lasers.  The Offset ECDL driver contains a frequency-measurement circuit to determine the beatnote frequency between the reference and relevant cooling/diffraction beams whilst an internal FPGA-based control loop is used to maintain the laser frequency at the desired setpoint. These units are combined with the PPLN temperature controller, EDFA drivers and DFB laser driver to form the control electronics for the laser system outlined in Section \ref{sec:LaserSysSummary}. In Figure \ref{fig:ElectronicsRbKLaserControlSTEQUEST} the block diagram of such a control system is shown. From a hardware perspective, there is little difference in the control electronics required between the Rubidium and Potassium-based interferometers, and broad similarities with the dipole trap.  Whilst each system requires its own dedicated electronic hardware in the payload, a modularized approach offers flexibility and efficiencies during development, assembly and test.\\
\\
\textbf{Detection System}\\
The design of the detection system is driven by the performance requirements, in terms of pixel noise and dynamic range, as well as the need to capture two images in a short succession on the same imaging plane.  This has driven the detector selection towards a frame-transfer CCD and in-turn the readout electronics towards high-performance systems seen in other space imaging or spectroscopy applications.  This contrasts with other systems using CMOS-based sensors, which potentially offer a simplification of control electronics, but do not meet the requirements as they are currently envisaged.\\
To achieve the required performance in a flight-suitable implementation, the CCD readout electronics contains all the associated digitization circuits, bias generation and readout clock sequencing.  A high-speed data link is provided to the DMU where image storage and processing shall take place. \\ \\
\textbf{DMU and Physics Package Support}\\
Alongside the core laser and detection systems electronics there are a significant number of support functions required to operate the physics package and overall payload operation sequence. These include functions such as the magnetic coil drivers, getter pump control and overall payload power distribution. \\
In addition, the DMU is required to bring everything together to provide data storage, processing capability and the overall spacecraft interfaces. A significant requirement is the timing resolution and accuracy of events across the payload during the science sequence.  This is likely to drive the overall design topology towards a distributed FPGA-based solution for timing and synchronization where configuration parameters are stored locally within subsystems, and high-resolution synchronization pulses are issued from the DMU to initiate experiment sequence transitions.\\ 
\subsubsection{Software}

Developing software for the DMU implies using embedded techniques. The specific system requirements, not only software but also electronic ones, tie the software architecture to the hardware necessities. The system works in a time-constrained environment and performance is one of the most important goals.\\

\noindent\textbf{Architecture and design}\\

The Atom Interferometer (ATI) software for Command, Control and Data Processing consists of two separated subsystems: 
\begin{itemize}
    \item {the one that runs on the S/C Computer, also called OnBoard Computer (OBC). It is known as OnBoard Software, being the Experiment Operations and Control Software}
    \item{the ICU Software running on the DMU}
    
\end{itemize}

The ICU Software is split into the Boot (or Basic) Software (BSW), in charge of the initialization and troubleshooting of the DMU, and the Application Software (ASW), based on RTEMS~\cite{RTEMS} operating system, implementing the required science computations, control and data management for the rest of the ATI electronics units.

\textit{BSW} is stored in the PROM, providing the minimum functionality necessary to:
\begin{itemize}

\item{assess and report on the overall DMU hardware health status}
\item{establish a reliable communication link with the OBC, implementing an adequate subset of the Packet Utilization Standard (PUS) protocols (ECSS-E-70-41A)}
\item{check and provide access to RAM and EEPROM memory (where the ASW shall be stored)}
\item{allow remote patching of Application Software }
\end{itemize}

\textit{ASW} is an extension of the BSW. ASW’s functionality can be summarized in three main tasks:

\begin{enumerate}
    \item {Handling of the ATI subsystems
    \begin{itemize}
        \item {reroute tele commands (TC) from OBC to ATI subsystems}
        \item{reroute telemetry (TLM, housekeeping) from ATI subsystems to OBC}
        \item{power management}
        \item{data acquisition rerouted to OBC for scientific purposes}
    \end{itemize}}
    
    \item {Computation of science data
    \begin{itemize}
        \item {controlling experiment sequences, like parameter optimization and sensor pictures processing}
        \item{controlling ATI subsystems}
        \item{packing and sending to OBC}
    \end{itemize}}

    \item {System monitoring, including health status and Onboard Monitoring Function, the standard service specified by CCSDS}
\end{enumerate}

\begin{figure}[hbt]
\centering
\includegraphics[width=0.9\textwidth]{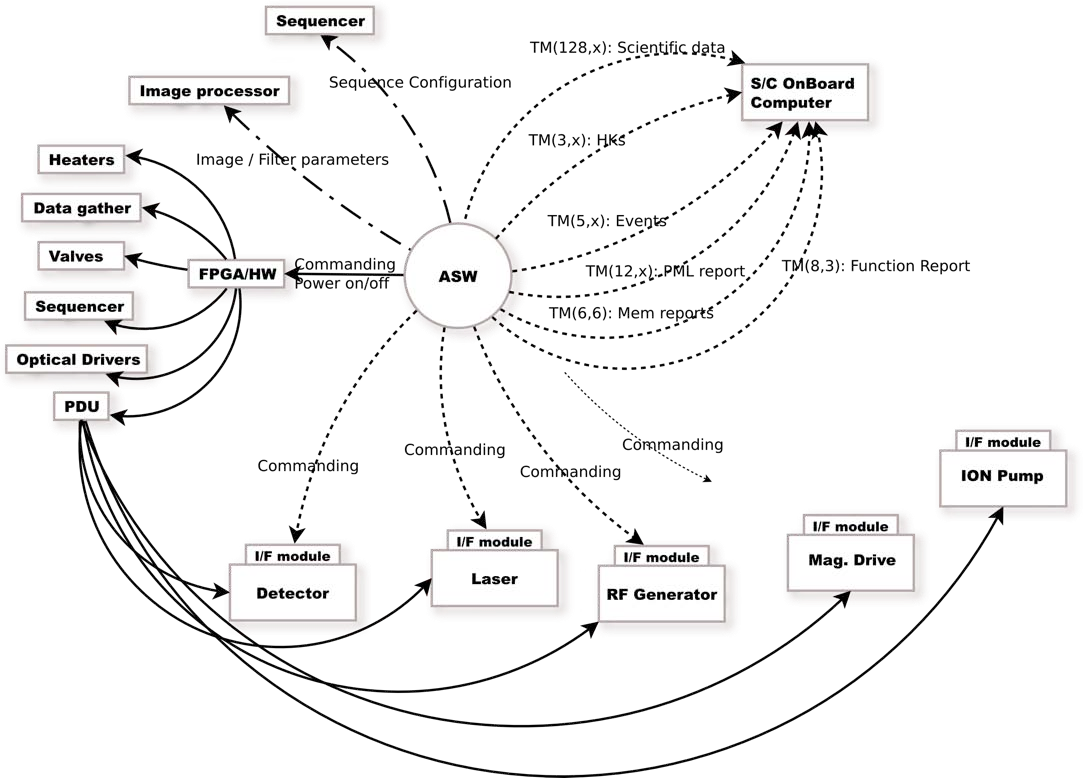}
\caption{ATI software output context diagram (Application Software).}
\label{fig:ASW_ctx_out}
\end{figure}

The output context diagram for the ASW is shown in Figure \ref{fig:ASW_ctx_out}. 

The main behavior required for ATI software from experimental and science point of view, apart from standard housekeeping and monitoring, can be grouped in 3 main blocks:

\begin{itemize}
    \item { \textit{Experiment management}: Atom interferometry experiments require simultaneous actions on several devices. The timings involved in the experiment sequences are considered a hard and critical requirement; this means that the whole software (running under an up to 85 MHz CPU (TBC)) must be able to manage critical time sequences (steps around $\mu$s and changes about nanoseconds). Such strict timing requirements, despite using a real-time operating system, are assumed to be unreachable in strict terms of software. The best approach is to implement the experiment sequence using dedicated hardware electronics (FPGA), able to satisfy the timing requirements, and let the management and control of the parameters definitions to the software.}
    
    \item { \textit{Parameter optimization}: The parameters in the experiment sequences must be very precise in order to produce best results. These parameters need to be computed using some function parameter optimization technique. As the processing power of the DMU is moderate, a hybrid approach is foreseen: a preliminary optimization must be performed on ground, leaving flight software to only further improve them.}
    
    \item { \textit{Image processing}: CCD sensors produce images that ATI software must manage for later being sent to ground, depending on communications availability, and mass storage present in DMU. Besides, some on-board processing is needed, to apply filters and fit algorithms to extract useful information from the images. The preliminary approach is to use a dedicated FPGA to implement the filters, and use the DMU software as a co-CPU to produce final image data to download.
    }
\end{itemize}

\noindent\textbf{Conventions, procedures, standards and quality} \\

Regarding design methodology, the chosen one is the Ward-Mellor method~\cite{WardMellor}, based on the well-known Yourdon structured analysis. It provides extensions taking into account real-time needs, and completely covers the needs for the whole DMU software development. 
More heavyweight methodologies, like object-oriented developments using UML or RUP, are not justified for real-time applications with low-level and not very complex architecture like ours.

Documentation and code is traced and versioned using appropriate tools for control versioning, issue tracking and requirement management as it is mandatory for Software  Engineering, following the Configuration Management standards (ECSS-M-ST-40C).

The main programming language to be used for developing the DMU SW is C. For BSW, some parts may be directly developed using the CPU assembly language. These languages are best suited due to the need of access to underlying hardware at low level, and in order to ensure that the size and CPU consumption of the resulting applications is well within budget.

Quality assurance is an integral process enclosing all stages of the DMU software development life cycle. It
relies on the ESA Space Software Engineering and Software Product Assurance standards (ECSS–E–ST-40C, ECSS-Q-ST-80C),
tailoring them at first stage of the software design phase.

Software testing and validation also follows the tailoring regarding quality. All the software produced is unit-tested by the same developer team and validated by external institutions/companies. Official test and validation
software campaigns are planned prior to reach Qualification Review meeting and Acceptance Review.

\subsection{Heritage}

The STE-QUEST atom interferometer as detailed in chapter \ref{section:payload} can rely on extensive heritage in ground based experiments, as well as developments for space (including dedicated demonstrators towards space applications and atom interferometers, operated on a sounding rocket and in 0-g-flights). In the frame of the ESA Cosmic Vision program (M3), the STE-QUEST satellite test of the equivalence principle was down-selected for a phase A study~\cite{STE-QUEST_Yellow_book,Aguilera2014}. The outcome of this study validated the main concepts for such an operation with a dual-condensed source of Rb isotopes testing the UFF at the $2 \times 10^{-15}$ level. Main technological limitations which have been identified at that time have been overcome, mainly thanks to the developments of national programs in France and Germany.


The heritage for the different subsystems is detailed in the following.

\subsubsection{Physics package}

The Physics Package for STE-QUEST M7 is a modified version of the payload anticipated for M3~\cite{Aguilera2014,Schuldt2015} / M4. Required changes for M7 accommodate the increased free-fall time and baseline of the atom interferometer, higher atomic flux of well collimated $^{87}$Rb and $^{41}$K ensembles, detection capabilities for the additional scientific goal, as well as the means for a tighter control of error sources to $10^{-17}$. The Physics Package benefits from the heritage of various microgravity activities, ICE~\cite{Condon2019,barrett2016,Geiger2011} onboard a zero-g Airbus, QUANTUS~\cite{Deppner2021,Rudolph2015,vanZoest2010} and PRIMUS~\cite{Vogt2020,Kulas2017} in the drop tower in Bremen, the sounding rocket mission MAIUS~\cite{Lachmann2021,becker2018} activities, and NASA's Cold Atom Lab (CAL)~\cite{aveline2020,elliott2018} on the ISS (International Space Station). In addition, the MAIUS collaboration planned~\cite{Piest2021} and conducted a second sounding rocket experiment, and the BECCAL (Bose-Einstein Condensate and Cold Atom Laboratory) collaboration is preparing a multi-user multi-purpose facility for atom optics and atom interferometry on the ISS~\cite{Frye2021}.

QUANTUS~\cite{Deppner2021,Rudolph2015,vanZoest2010} and MAIUS~\cite{Lachmann2021,becker2018} successfully operated $^{87}$Rb-BEC experiments utilizing ultra-high vacuum systems with 2D-MOTs, atom chips and peripherals similar as planned for STE-QUEST in the drop tower in Bremen and onboard of a sounding rocket, respectively. Recently, an upgraded version of the physics package in MAIUS showed the capability for dual-species Rb-K BEC generation on ground~\cite{Piest2021}. Designed for operating an optical dipole trap rather than an atom chip, the physics package of the ICE project supported dual-species Rb-K atom interferometry on a plane~\cite{barrett2016}, as well as evaporation of $^{87}$Rb in an optical dipole trap in a µg simulator~\cite{Condon2019}, and the physics package of the PRIMUS project enabled evaporation of $^{87}$Rb in an optical dipole trap in the drop tower in Bremen~\cite{Vogt2020,Kulas2017}. CAL~\cite{aveline2020,elliott2018} followed a different approach for the vacuum chamber by implementing a glass cell, but also relies on an atom chip for BEC generation. The physics packages of the aforementioned experiments relied on (multi-layer) magnetic shields to suppress the impact of external magnetic stray fields. BECCAL~\cite{Frye2021} builds on the heritage of QUANTUS, MAIUS, and CAL, implementing a modified design from QUANTUS and MAIUS, and including a box-shaped three-layer magnetic shield and a tip-tilt stage for rotation compensation.

Summarizing, core technology and functions required for the physics package of STE-QUEST were pioneered by various payloads operated in µg which serve as a solid basis for the adaptation in size, shape, and performance.

\subsubsection{Laser system}
\label{sec:LaserHeritage}

Telecom lasers are robust solutions for future space missions such as STE-QUEST.
Using second harmonic generation to generate near infrared light (767--780$\,$nm), this technology applies for Rubidium and Potassium.
Many of these fiber-coupled sources are already Telcordia qualified, satisfying demanding specifications in terms of vibrations, shocks, temperature variations and lifetime.
Moreover, these commercial products constitute a large catalog of highly reliable and/or redundant components.
This includes narrow linewidth laser diodes suitable for atom interferometry, phase/intensity modulators to simplify architectures and to provide fast tunable laser systems,  acousto-optical modulators (AOMs), Erbium Doped Fibered Amplifiers (EDFAs), and fast photodiodes.
Radiation hardness has also been tested for a number of these components and some of them are now space qualified.
The most critical component is the EDFA but specific irradiation hardened doped fibers have been developed to tackle this issue.
The frequency doubling stage is a robust fibered PPLN waveguide, providing high efficiency and delivering high optical power at the output of a mono-mode optical fiber (typically 500 mW). High power (1W at 780 nm) frequency doubling modules are now available if required.
Telecom lasers are present in the first commercial atom gravimeters~\cite{menoret_gravity_2018}, experiments in microgravity~\cite{menoret_dual_2011}, compact navigation devices~\cite{battelier_development_2016}, and atomic based gravitational waves antennas~\cite{sabulsky_fibered_2020}.

During the M3 STE QUEST phase A, a CNES study had been led by SODERN on a reference laser based on telecom technology and frequency doubling. It was the first study including a realistic architecture including an evaluation of the thermal aspects. More recently, a CNES study has been led by the French company iXblue (muQuanS) to develop a demonstrator of an all fibered laser system. This prototype allows to produce all the functions required for atom cooling and interferometry, using a single fiber output such as an atom gravimeter. Even if this architecture is not adapted to STE QUEST, it includes all the components and subsystems required for our mission. Test campaigns on the subsystems and the full laser system were led to increase their TRL, and especially included tests in shocks/vibrations, irradiation and heating under vacuum. Finally the performances of the laser system were maintained at a good level: frequency servo-lock, polarization, preservation of the optical power, efficient thermal dissipation under vacuum for a typical working of the instrument and a good preservation of the frequency doubling efficiency. 

The dipole trap can be produced using a telecom laser, which has the strong advantage to share the same technology as for the cooling and Raman/Bragg laser (LCI) except for a higher power required for the fibered amplifier (EDFA).  Fast ($\approx 1$ s) Bose Einstein Condensation were demonstrated on ground for Rubidium~\cite{Clement2009} and Potassium~\cite{Salomon2014}. More recently, similar approaches were adapted to produce a BEC in 1 second in microgravity on the 0g simulator in Bordeaux~\cite{Condon2019}, and a full optical BEC onboard the 0g plane has been demonstrated during the flight campaign of March 2022.

\subsubsection{Electronics}
Extensive work has taken place to begin the initial developments of the electronics hardware.  Whilst use of ground-based electronics systems has been used across the consortium to develop the functional requirements for STE-QUEST --- these are typically unsuitable to take through to high-reliability flight operation in the space environment.   

A number of previous programmes have therefore sought to re-build these functional capabilities using designs and component choices that have been designed with the space environment in mind.  The most notable have been MCLAREN~\cite{MACLAREN2017} which produced a wide-ranging preliminary system design for an atom interferometer alongside prototypes of laser drive, spectroscopy-lock, offset-lock, RF generation and magnetic coil drive circuitry.  This project also focused on the development on synchronization and timing distribution to achieve ~10\,ns accuracy of events between subsystems. 

This has been further developed by CASPA Accelerometer~\cite{Ferreras2022} and GSTP High Stability Laser which initially developed and characterized an integrated laser drive and temperature stabilization circuitry, and further developed autonomous spectroscopy and offset-lock capability into an integrated module.  

Whilst less focus has been placed on prior development of the DMU or payload support hardware it is envisaged these present a lower-level of technical risk, and typically rely on processors or hardware that are not uncommon in the flight environment. 

CCD-based detection systems have extensive heritage in the flight environments (e.g. within the NASA missions SDO, GOES), with a track record for a range of approaches tailored to mission requirements including Digital Correlated Double Sampling techniques or dedicated ASIC-based solutions to optimize size, weight and power. 

\subsubsection{Software}

The control software in the STE-QUEST DMU will build upon a strong heritage on the LISA Pathfinder experience, being the same team involved in both missions. 

We expect a further detailed analysis on the system functionalities during the phase A study, which will set a more clear development path. However, in the current design stage, we assume that the STE-QUEST DMU software needs will be similar to those of an Instrument Control Unit for an M mission. In that sense, the use of PUS standard (ECSS-E-ST-70-41C) and the know-how developed during LPF will be a clear advantage. 

The embedded architecture and the need to have an RTOS is also an LPF heritage that we consider of potential interest for the current proposal. In terms of communications, STE-QUEST shares with LPF the need of a rigorous timing, for which we foresee that previous experience of the software development team in synced communications will be an asset. In LPF, the implementation was based on MIL-BUS (MIL-STD-1553), a know-how that can be certainly of interest for the current proposal. 

Finally, it is worth noting the software team experience in terms of software testing campaigns, which we consider a key heritage. ESA acceptance requirements in terms of software testing campaigns are tough to meet and the associated effort should not be underestimated. Based on this heritage, software testing will be considered a key aspect of the software development from the initial phases of the design and its impact will be carefully evaluated during the different design phases of the mission.

\subsection{Required budgets and interface requirements}
\begin{figure}[ht]
\centering
\includegraphics[width=1\textwidth]{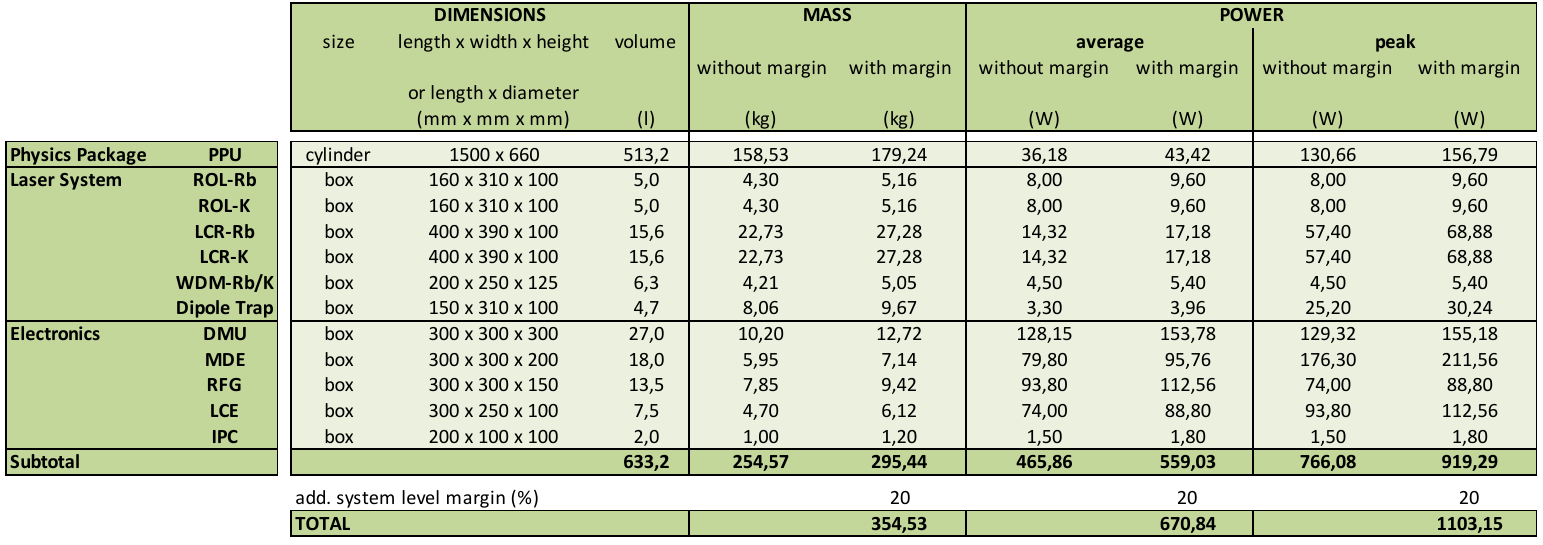}
\caption{Budget overview of the atom interferometer payload. The total power is delivered to the electronics box via one single interface from the spacecraft. Physics Package and Laser System obtain their power via an interface to the electronics box. The given power values for the three subsystems is the power dissipated therein.}
\label{fig:budgets}
\end{figure}

Within the M3 phase A study, a detailed breakdown of the atom interferometer payload down to the component level has been worked out. The payload has been grouped to physical boxes with corresponding budgets. The budget requirements concerning volume, mass and power have been updated to the M7 scenario, see figure \ref{fig:budgets}. Including component and additional system level margins, the overall payload is assessed to have a mass of about 355\,kg, an average power consumption of 670\,W and a peak power consumption of 1100\,W. Data rates have been evaluated to be $<110$\,kbps and assessed to be feasible. 


\subsection{Development plan and model philosophy}
\label{sec:dev_plan}

The overall technology readiness of the relevant subsystems is given in Table \ref{tab:TRLSubsystems}. It includes the assessment of the TRL by today, where TRL 4 refers to commercially available technologies and to technologies demonstrated in laboratory environment. While technical feasibility and implementation is demonstrated, background work with respect to the specific STE-QUEST mission requirements might still be required. TRL 4--5 refers to technologies which already have been developed with respect to space applications but have not yet been environmentally tested according to the specific STE-QUEST mission requirements. According to the development plan detailed in the following, all subsystems will achieve TRL 5 or TRL 6 by the end of Phase A (2026). 
A detailed list at the component level can be found in Appendix II on page \pageref{appendix:TRL}.

\begin{table}[h]\footnotesize
\hspace{-5mm}
\begin{center}
\hspace{-5mm}
\begin{tabular}{|c|c|c|c|c|c|}
\hline 
{\bf Subsystem} & {\bf TRL 2022} & {\bf TRL 2026} \\
\hline\hline
Atomic Source & 4 (anti-straylight coating: 3--4; atom chip: 3) & 5/6 \\\hline
Interferometer Optics / Beam Splitters & 4 (tip-tilt stage: 3--4) & 5/6\\\hline
Detection System & 4--5 (quantum mechanics test objective: 3)& 5/6\\\hline
Vacuum System (incl. Pumps) & 4--5 & 5/6\\\hline
Magnetic Shielding & 4--5 & 5/6 \\\hline
Laser System & 4--5 (dichroic filter: 3--4) & 5/6\\\hline
Dipole Trap Laser & 4 & 5/6\\\hline
DMU & 3--4 & 5/6 \\\hline
Control Electronics & 3--4 & 5/6\\\hline
RF-Generation & 3--4 & 5/6\\\hline
\hline
\end{tabular}
\caption{\it Overview on the technology readiness level for the ATI subsystems where components with TRL$<$4 are explicitly mentioned. Space-grade anti-straylight coatings are available, however their use at the vacuum level required by STE-QUEST needs to be proven; atom chip technology has been demonstrated in various ground-based experiments and on sounding rocket, but needs to be adapted for the STE-QUEST atom source; the tip-tilt stage needs functional and performance tests, including vacuum compatibility at the required level; the detection system for the additional science goal (test of quantum mechanics) requires a dedicated study to first validate the design and then function and performance; the dichroic filter has been demonstrated onboard the 0g plane, but needs to be tested according to the STE-QUEST environmental requirements; electronics technology has been demonstrated in laboratory environment as well as in $\mu$g environments and sounding rockets, but needs to be adapted for the STE-QUEST mission environment.} \label{tab:TRLSubsystems}
\end{center}
\end{table}

\subsubsection{Physics package}

Due to the preceding µg experiments~\cite{aveline2020,Condon2019,barrett2016,Vogt2020,Lachmann2021,Deppner2021}, a substantial part of the technology and components for the physics package of STE-QUEST M7 was tested in compact and robust devices, and subjected to shocks or vibrations, indicating a TRL of 4 or 4+. They require a mission specific delta qualification, e.g. for the vibration loads and shocks, thermal cycling and radiation hardness, to reach TRL 5. It is implicitly assumed, that changing size or shape of the vacuum system or optics does not decrease the TRL. Low TRL items at 3 or 3-4 are the anti-straylight coating inside the vacuum chamber~\cite{Vovrosh2020}, the atom chip with adapted design to support higher atoms numbers, the tip-tilt stage / retroreflector actuator for rotation compensation~\cite{Frye2021}, and the detection system~\cite{Rocco_2014} for the additional science goal (test of quantum mechanics). The anti-straylight coating and the tip-tilt stage are expected to simply require a functional and performance test, including vacuum compatibility at the required level, to reach TRL 4 and then be subject to the delta qualification. Atom chip designs implemented in existing experiments~\cite{Deppner2021,Lachmann2021,becker2018,Rudolph2015} require an evaluation for the possibility to accommodate an increased atom number. Depending on the outcome of the evaluation, design updates may be necessary, followed by functional and performance tests. The detection system for the additional science goal requires a dedicated study to validate the design approach followed by functional and performance tests. Beyond reported and ongoing experiments in microgravity, further activities are planned utilizing physics packages with relevant technology for STE-QUEST M7, a follow-up sounding rocket mission with Rb and K (MAIUS-2)~\cite{Piest2021}, developments towards a pathfinder mission with Rb interferometry (CARIOQA)~\cite{Alonso2022,carioqa-moriond2022,carioqa-pmp-approved2022}, and the BECCAL mission (Bose-Einstein Condensate and Cold Atom Laboratory) for atom optics and atom interferometry on the International Space Station~\cite{Frye2021}.

\subsubsection{Laser system}

The TRL of the laser system components in 2022 (cf. Table \ref{tab:TRL_LS} in the annex) is based on the heritage described in section \ref{sec:LaserHeritage}. The high TRL for the laser system comes from the CNES demonstrator study led by iXblue/muquans. The development plan to reach TRL 5/6 in 2026 mainly consists in performing complementary tests with the relevant conditions for STE QUEST, especially in terms of life time and radiation hardness. Between 2023 and 2027,
an engineering model (EM) of a laser system for an atom accelerometer dedicated to a pathfinder mission (CARIOQA) will be led by CNES and industrial partners. Despite the fact that the CARIOQA system will be more simple, this EM will be based on the very same technology. The required optical power for the dipole trap is higher than for cooling and Raman/Bragg. This is why it leads to a lower TRL. A dedicated development is required to validate a high-power version of the EDFA. Moreover, an improvement of the double species atom source in dipole traps in term of flux and atom number is required and will be demonstrated in studies by the scientific team on the the microgravity platforms (0g plane, 0g simulator in Bordeaux, QUANTUS/MAIUS, CAL/BECCAL).  
The dichroic filter has been tested in microgravity onboard the 0g plane, allowing the production of simultaneous Rubidium/Potassium atom interferometers. Nevertheless relevant environment tests are still to be done. More specifically, this component has a natural sensitivity to temperature. Similarly to what is done for laser diodes and PPLN waveguides, a dedicated package to control the temperature of the filter will be developed.

\subsubsection{Electronics}
Atom interferometry experiments have all utilized a range of electronics, demonstrating the viability of all aspects of the electronic technology.  The overall maturity of electronics technology could therefore be considered as TRL 4; however it is important to recognize further design work is required to ensure suitability for the flight environment leading to the general assessment for the electronics of TRL 3-4. 

Table \ref{tab:TRL_electronics} in the annex summarizes the TRL of the electronics system components in 2022 and their expected value at the end of a potential Phase A in 2026. The development of electronics and control subsystem for the STE-QUEST payload, in a manner in that enables progression to reach the higher technology readiness levels later in the programme, is considered a key task of the payload development.   

During the Phase A study it is envisaged that development work will be required across subsystems to further inform the electronics architecture and design approach required to meet the performance requirements.  This process will also identify any components or subsystems where there is an increased risk of the environmental factors impacting performance (e.g. temperature, radiation) such that suitable mitigation or screening can be initiated earlier in the study. 

A breadboard of the payload control electronics shall be developed to enable the performance characterization of the system in a representative manner against the STE-QUEST objectives, thus bringing a full payload electronics to TRL 4. 

Targeted environmental testing will enable potentially sensitive areas of the payload control to be verified in a relevant environment, bringing all aspects of the payload electronics to TRL 5/6 by the end of the Phase A study.

The Diagnostics package will include high precision environment sensors on-board the satellite. Among them, temperature, magnetic fields and magnetic field gradients are potential noise sources for the STE-QUEST instruments that require precision monitoring. Heritage from LISA Pathfinder Diagnostics Subsystem~\cite{Armano2019a,Armano2020}, with similar performance requirements, will be an advantage for the mission in this aspect. 

\subsubsection{Software}

The initial phases of the DMU control software development requires the consolidation of requirements. This implies, at the same time, the definition of communications with the rest of subsystems and OBC. The requirements and interfaces —and the associated ICDs— will determine the final DMU needs in terms of CPU, RAM, and communications.

Once the requirements are established, the next phase is the application design. Since this design phase will run in parallel with the hardware development, we do not expect to have real hardware until later stages of the mission development. Hence, we foresee the need of CPU emulators or CPU development boards in the initial phases of the software development. These will enable the development of the main structure of the application as well as to start building the software framework to be used later with the real hardware (EM/EQM). 

Given the constraints in timing and synchronization in STE-QUEST, a thorough assessment of compilers and RTOS is expected during this initial phase. This will set the bases of a sound task scheduler upon which the team will build the final software. In these early stages, software development requires as well the definition of the full testing framework, that will be implemented at later stages. However, the testing platform is crucial and for that needs to be part of the definition phase. As an example, in LPF, the testing platform took five times more lines of code than the flight software itself. Test campaigns will have a crucial role in the development, and are key part of the acceptance in the milestones of the mission. The test framework may also have a crucial role in terms of EGSE needs, as it may be used not only to test the DMU and its software, but also for the subsystems to test their interaction with the DMU.

Once STE-QUEST enters in implementation phase, the software development will run in parallel to the hardware development with software releases synchronized with the sequential reviews of the project. Each of these software releases will implement further functionalities, in accordance with the development plan and the planned instrument test reviews.

\subsubsection{Model philosophy}
The ATI development foresees the technology development bread-boarding activities as detailed above during Phase A/B1. An Engineering and Qualification Model (EQM) of the complete ATI will be realized in Phase B2/C, undergoing the full set of qualification level environmental and functional testing. EQM performance testing might be carried out in the Einstein Elevator in Hannover. In parallel, a Structural Thermal Model (STM) is realized to simulate and qualify the thermal and structural properties of the instrument. In Phase D, the Flight Model (FM) will be realized. It is planned to adapt the EQM and to make it functionally identical to the FM, in order to serve as a GTB (Ground Test Bed). The GTB remains on ground during the operation of the payload in orbit. It serves as a monitoring system and will be used to test software updates, track the source of potential faults occurring during flight, and test remedies. The model philosophy will be revisited during phase A to ensure that the EQM $\to$ GTB strategy is realistic and fulfills the main functional requirements for the GTB. 

\section{Mission configuration and profile} \label{sec:config}

\subsection{Mission profile}\label{sec:miss-profile}

The mission configuration and profile are the result of optimization for the main science objectives of testing the equivalence principle and searching for dark matter. For the third science objective (test of quantum mechanics) the choice of orbit is not critical and it will not be a driver for the considerations here. The general philosophy is to find an orbit that allows reaching the target of $\eta \leq 1\times 10^{-17}$ within a maximum of 3 years total mission duration, whilst minimizing perturbations and systematic effects, like residual S/C accelerations and rotations, gravity gradients, thermal effects, magnetic effects, etc \dots

\subsubsection{Orbit optimization trade-off}
\begin{figure}[t]
\centering
\includegraphics[width=0.75\textwidth]{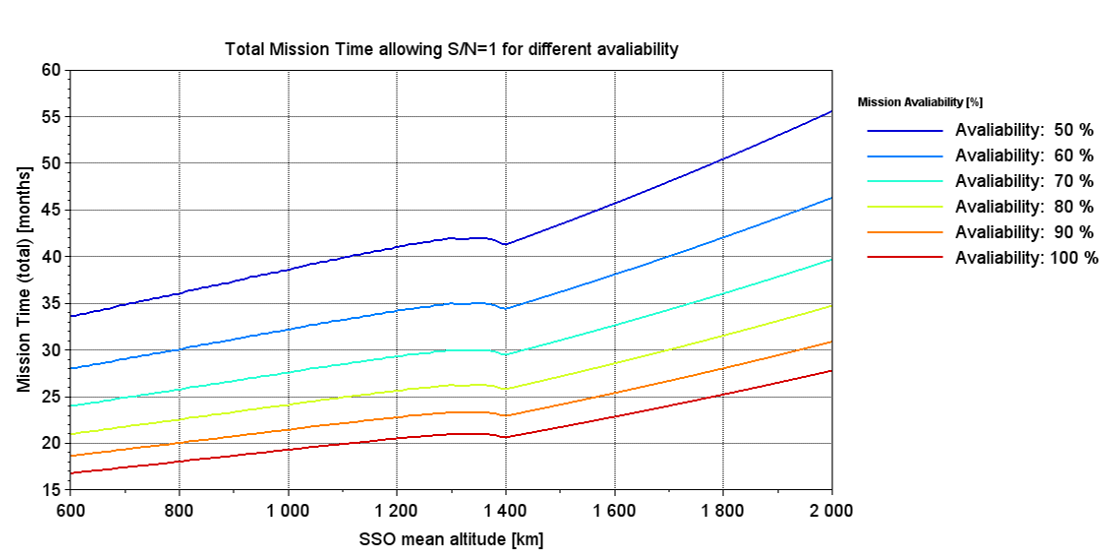}
\caption{\it Mission duration (after commissioning) to reach $\eta = 10^{-17}$ for different availabilities.}
\label{fig:Orbit-SN}
\end{figure}

\begin{wraptable}{h}{0.55\textwidth}
        \centering
        \begin{tabular}{c|c}
            {\bf high} & {\bf low}\\
			\hline \\ [-1.5ex]
            Eclipses & Radiation\\
            DFACS & End of Life\\
           GG (Sec. \ref{sec:GGC}) & S/N (mission duration)\\
            Thermal (Sec. \ref{sec:BBR}) & \\
            Magnetic (Sec. \ref{sec:B-field}) & \\
            \hline
        \end{tabular}
        \caption{\it Orbit altitude trade-off drivers.}
        \label{tab:altitude}
    \end{wraptable}
Given the experience with MICROSCOPE, a trade-off study was conducted by CNES engineers and members of the STE-QUEST core team, many of which were heavily involved in MICROSCOPE. The baseline choice was a sun synchronous circular Earth orbit (as for MICROSCOPE) that could be reached by direct injection from VEGA-C given the total satellite wet mass of $\sim$1187~kg (see Sec.~\ref{sec:S/C}). The main trade-off objective was then to find the optimum altitude. Tab. \ref{tab:altitude} summarizes the main trade-off drivers.
\noindent The expected signal is proportional to the local derivative of the gravitational field $g_0$ and thus increases with lower altitude, whilst the noise is dominated by atomic shot noise and is independent of altitude (see Eq.~\ref{eq:S/N})). 
\begin{wraptable}[12]{tbr}{0.35\textwidth}
    \centering
    \begin{tabular}{c|c}
        \hline
        SMA & 7798.1~km\\
        Eccentricity & 0.0009789\\
       Inclination & 101.6 deg\\
        Arg. of perigee & 90.0 deg\\
        RAAN & 190.3 deg \\
        mean anomaly & -90.0 deg \\
        Perigee altitude & 1412.4 km \\
        Apogee altitude & 1427.6 km \\
        Keplerian period & 114.2 min \\
        \hline 
    \end{tabular}
    \caption{\it Orbit parameters.}
    \label{tab:orbit}
\end{wraptable}
However, at altitudes below 1400~km SSO (6~h, 18~h) orbits experience eclipse seasons (e.g. $\approx$55 days/year at 1000~km), when the satellite passes in the Earth's shadow and the resulting thermal fluctuations make science operations impossible, and may lead to thermal instabilities even after the eclipse season. 
Figure \ref{fig:Orbit-SN} shows the total mission duration required to reach the $\eta \leq 1\times 10^{-17}$ target for different availabilities and taking into account eclipse seasons. Adding 6 months for commissioning, altitudes up to 1700~km are compatible with an overall mission duration of 3~years (at 80\% availability).
Also many other drivers favor high altitude: Drag free and attitude control (DFACS) are mainly driven by residual atmospheric drag and magnetic torques on the spacecraft, both decrease with altitude. Thermal perturbations on the instrument from the Earth albedo, magnetic perturbations from the Earth's magnetic field and gravity gradient effects also decrease. One would then tend for a relatively high orbit, but that is limited by end of life disposal and by radiations. 
The former is not very critical (see Sec. ~\ref{sec:EoL}), but the latter is limiting, because the radiation dose increases rapidly with altitude as one enters the radiation belt (increase by a factor $\sim$4 when going from 1000~km to 1400~km, see Sec.~\ref{sec:radiations}).

As a consequence of this study the baseline orbit chosen for STE-QUEST is at the minimum altitude that avoids eclipses (1400~km ), whilst at the same time keeping radiations as low as possible. The reference orbit is a quasi-circular sun synchronous orbit with a mean local time of ascending node of 6h (or 18h). Table \ref{tab:orbit} summarizes the corresponding orbit parameters. However, lower altitude choices (e.g. 1000~km) are possible and compatible with the mission objectives, if in the course of more detailed phase A studies the radiation environment turns out to be critical for S/C or payload.

\noindent The orbit is not maintained but the drag free sub system will compensate for the small residual air drag. Other perturbations will have a small effect on the other parameters. The orbit parameters will be optimized in the frame of mission analysis activities.

\subsubsection{Attitude and operational mode}
Similarly to MICROSCOPE and LPF, the STE-QUEST S/C will be operated in drag-free mode with actively controlled attitude using a hybridization of several sensors (classical accelerometer, star-trackers, gyroscopes). Additionally, if necessary, the main instrument (ATI) can be used for low frequency ($<$1/$T_c$), acceleration control along the sensitive axis at high accuracy (low drift). The requirements on the DFACS are summarized in Sec. \ref{sec:DFACS_summary} and can be met with the by now ``standard'' cold gas $\mu$N thrusters.

The satellite operates in inertial mode leaving the orientation of the sensitive axis of the instrument unchanged in an inertial frame. This leads to a modulation of the expected signal at orbital frequency. As mentioned earlier, for further de-correlation from systematic effects we will modify the orientation of the sensitive axis by irregular (every 50 orbits on average) rotations of $\approx 10^\circ$ in the orbital plane, leading to an additional phase modulation of the expected signal \footnote{The optimal modulation sequence is yet to be determined, the sequence described here is a first estimate.}. Preliminary estimates indicate that the DFACS cold gas consumption for a 3 year mission lifetime is of order 35~kg at an altitude of 1400~km. It slowly increases with decreasing altitude as the main driver (as in MICROSCOPE) is the coupling of the S/C magnetic moment to the Earth's magnetic field. For example, at 1000~km the consumption is estimated to be 40~kg. The estimates were obtained using the MICROSCOPE data adding 100\% margin to account for the unknown magnetic moment of the STE-QUEST S/C and adding another 20\% margin to account for complementary tests of DFACS and systematics. The regular maneuvers to re-orient the S/C every 50 orbits (on average) will take about 800~s each and cost a total of about 0.5~kg additional cold-gas, which is negligible with respect to the DFACS consumption.

\subsubsection{Radiation}\label{sec:radiations}
\begin{figure}[t]
\centering
\includegraphics[width=0.48\textwidth]{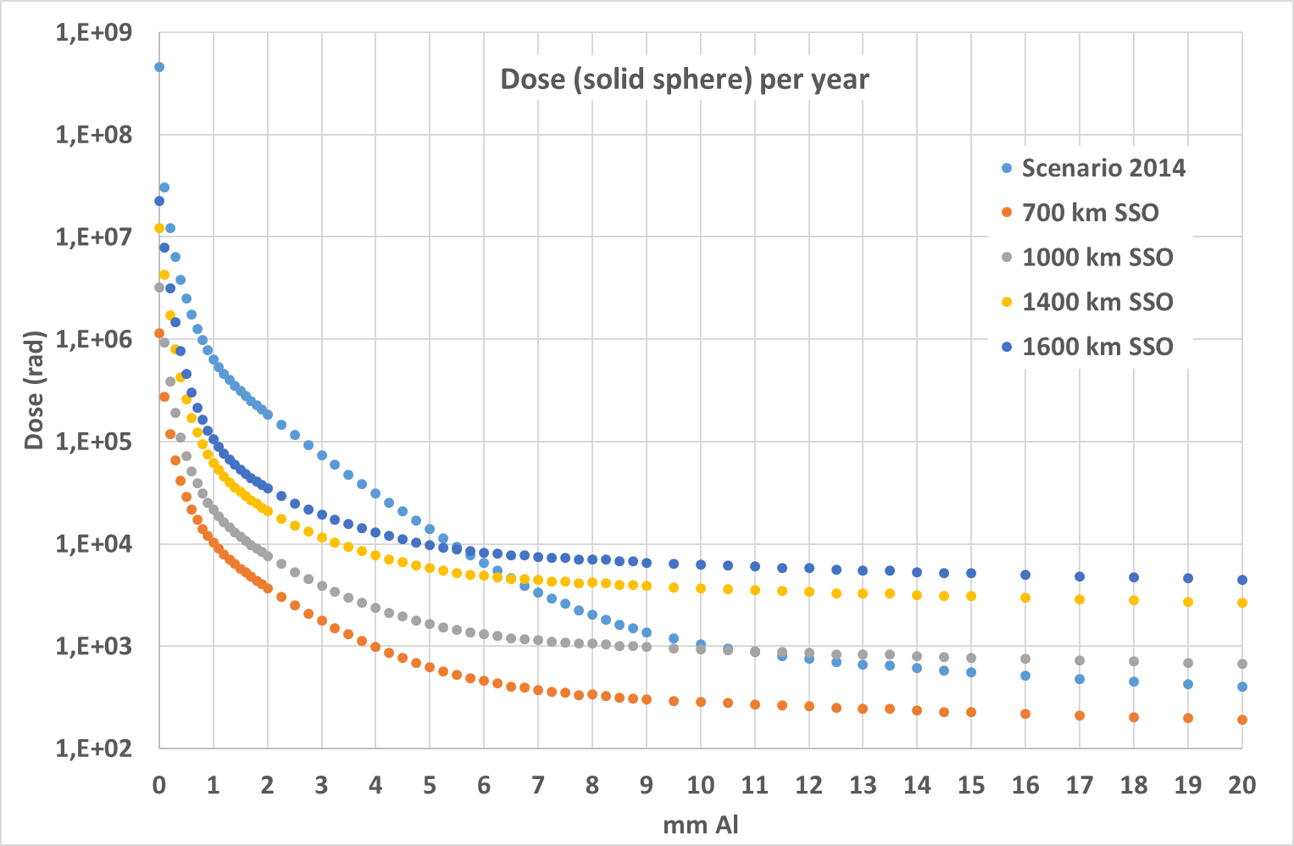}
\hspace{0.05cm}
\includegraphics[width=0.48\textwidth]{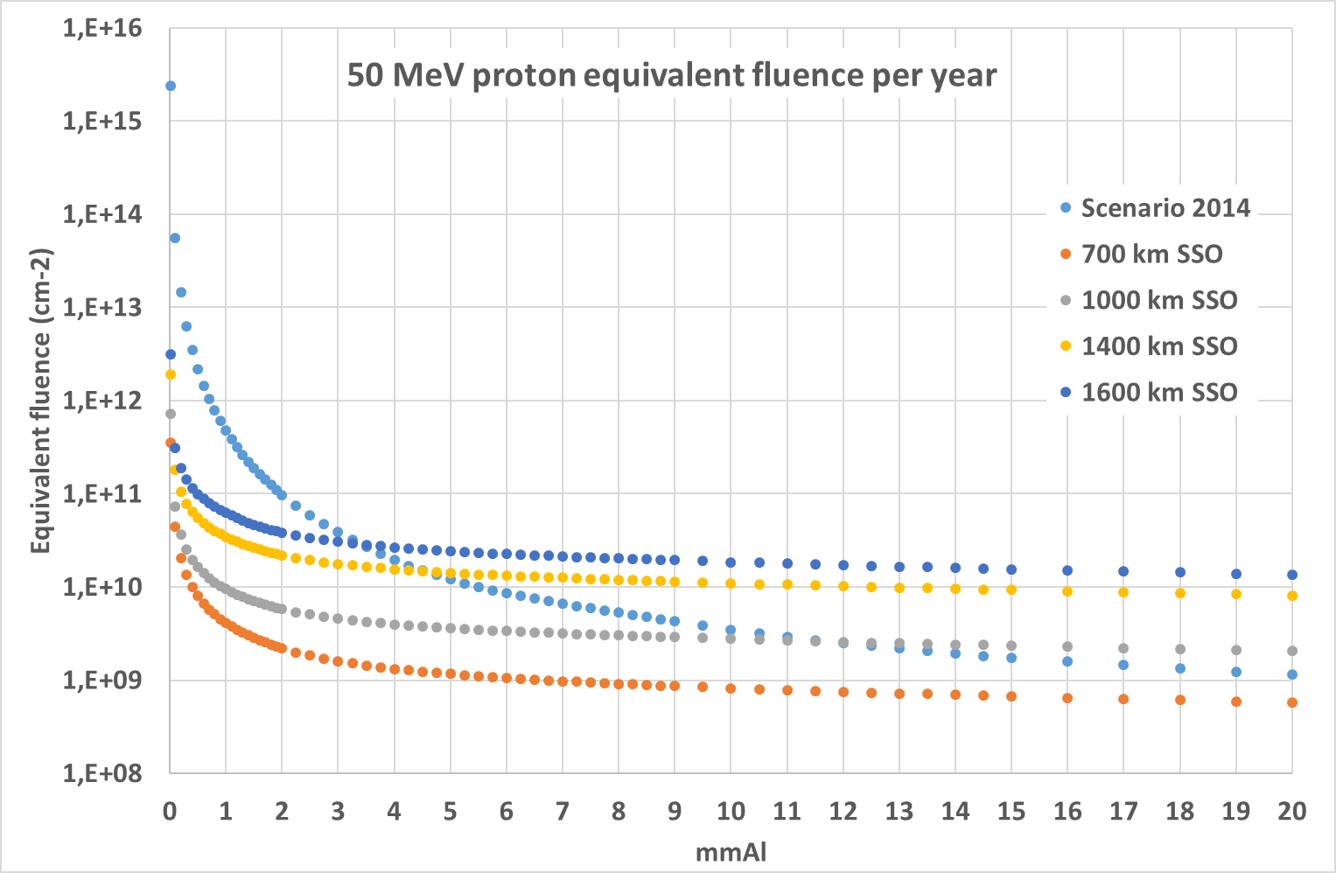}\\
\vspace{0.3cm}
\caption{\it Radiation levels in different altitude SSO orbits and the M4 (2014) HEO scenario as a function of Al shielding thickness.
}
\label{fig:radiations}
\end{figure}

A comparative radiation analysis was carried out for SSO orbits at four different altitudes (700~km - 1600~km). The results were also compared to the M4 STE-QUEST mission profile (2014 scenario). They are presented in Fig. \ref{fig:radiations}.

As expected, radiation doses increase rapidly with altitude (factor $>$10 between 700~km and 1400~km). Whilst the present baseline orbit at 1400~km has about a factor 4 stronger radiation than the alternative 1000~km one and up to an order of magnitude more than the M4 (2014) orbit, we still consider this a reasonable option, the reason being the relatively short mission duration (3 years) and the experience of JASON 1 and 2 satellites in very similar orbits (1340~km). Those satellites were equipped with dosimeters, mounted on the inner face of the outer panel, that measured radiation doses of 2659~rad/yr and remained in operation for $>$10~years (for an initially planned 5~yr extended mission duration). If the same satellites were flown in an STE-QUEST orbit the corresponding dose would be 2285~rad/yr in the 2014 scenario, 2967~rad/yr in the M7 1400~km option and 774~rad in the M7 1000~km option. So whilst the later allows a very calm radiation environment, the 1400~km radiation levels do not seem prohibitive.

\subsubsection{End of life aspects}\label{sec:EoL}
For end of life disposal from a 1400~km SSO orbit there are two fundamental options. One is a re-entry, the other a Hohmann transfer to an altitude $>$2000~km, which would also comply with current space debris regulation. Whilst the latter option is a little less costly in terms of propellant, it is ethically less responsible and also has additional complications (two successive $\Delta v$). We thus opt for a controlled re-entry. This requires a $\Delta v$ of about 360~m/s (incl. 5\% margin). Assuming a specific impulse (ISP) of $\approx$~300~s for the solid propulsion booster, the maneuver corresponds to about 130~kg propellant for STE-QUEST, which is accounted for in the overall STE-QUEST mass budget. For comparison, for a 1000~km orbit that mass decreases to about 98~kg, a difference which is not considered critical.

\subsubsection{Launch}
Given the chosen orbit (1400~km SSO) and the overall S/C mass (1187~kg) a launch and direct orbit injection with VEGA-C is a well adapted option, with a mass margin of about 200~kg~\cite{ESA2022}. If in the course of the phase A study the VEGA-C launch turns out to be problematic a lowering of the orbit could be envisaged.

\subsubsection{Summary}
In summary the STE-QUEST mission profile is an SSO orbit with the minimum altitude that allows eclipse free operation i.e. 1400~km. The mission duration to reach the science objectives in that case is about 32~months including 6 months commissioning and 80\% availability for science. This leaves a comfortable 4 months margin with respect to the overall 3 year mission duration, that will be used for additional measurements or tests, checks, etc. In case that the more detailed phase A study concludes that such an orbit is problematic, a lower orbit (e.g. 1000~km) would also satisfy the scientific requirements, but may cause more complications in terms of thermal and magnetic effects. However if additional S/C and payload design/qualification requirements for radiation hardness (and associated extra cost) or launcher incompatibilities turn out to be critical, such a lower orbit is a perfectly possible fallback option.

\subsection{Spacecraft design}\label{sec:S/C}
%


This section outlines the spacecraft design strategy and provides an overview of a reference spacecraft architecture which can be considered for the present scenario, for further elaboration and more detailed assessments in upcoming study phases.

While mission requirements driven by the science objectives are discussed in previous sections, the most stringent requirements which drive the S/C design are in general terms associated with the need to provide a stable environment at the instrument in a LEO profile, in the presence of disturbances resulting from e.g. residual drag, gravity gradients, but also Earth radiation  magnetic fields, and eclipses and occultation (in case of a $<1400$~km orbit choice).

The key characteristics of the reference architecture elaborated at this level are indicated in Tab. \ref{tab:S/C}. The orbit choice has been discussed in the previous section. A 1400~km SSO orbit was chosen, subject to further optimization in the 1000-1400~km range. The selection of a SSO profile, among other advantages, guarantees a more favorable thermal environment through maintaining a constant relative geometry to the Sun, and also allows for a simplified design approach, e.g. avoiding any mechanisms to change the orientation of the solar array (see below).
\begin{table}[h!]
        \centering
        \begin{tabular}{l|m{12 cm}}
           	\hline
           	\multicolumn{2}{c}{\bf Mission scenario}	\\
			\hline
            Mission profile & SSO, 1400~km, subject to further optimization (range 1000-1400~km)\\
            Mission duration & 3 years nominal \\
            Science mode & Inertial attitude (experiment run autonomously, remotely controlled from ground) \\
            \hline
            \multicolumn{2}{c}{\bf Reference architecture}	\\
			\hline
			Attitude/orbit control & 3-axis control, drag-free (science mode) \\
			- Sensors & STRs, IMU/Fiber optic gyro, accelerometer assembly (e.g. GRACE-FO, NGGM, $\mu$-STAR)\\
			- Actuators & Micro-Propulsion System (MPS) based on linear cold gas thrusters \\
			Power subsystem & 30V unreg., final solar-array sizing dependent on power budget consolidation \\
			Thermal control & Primarily passive (on platform level), plus limited use of heaters \\
			Structure & Panels with an aluminium honeycomb structure, CFRP face sheets \\
			End of Life & Solid fuel propulsion for controlled re-entry \\
            \hline
            \multicolumn{2}{c}{\bf Instrument accommodation/resources}	\\
			\hline
			Instrument type & Dual-species ($^{87}$Rb-$^{41}$K) atom interferometer, GNSS receiver (dual-band GPS) \\
			Accommodation & PLM at inner central part of the S/C with instrument core (physics package) isostatically mounted, and (ideally) co-located at the S/C CoM.\\
			Resources & Resources (in particular, mass and power) for the payload, as derived from instrument budgets (incl. maturity, and 20\% additional margin):\\
			 & - Instrument mass (total): 355 kg \\
			 & - Instrument (average) power demand: 671 W \\
            \hline
        \end{tabular}
        \caption{\it Summary of the reference S/C architecture considered in this phase.}
        \label{tab:S/C}
    \end{table}

In science mode, the S/C is nominally operated in inertial mode.  As the atom interferometer poses stringent requirements on attitude stability and non-gravitational accelerations during science measurements, the platform would be operated drag-free-controlled. As primary sensors for the S/C active control, i.e. attitude and translational control, the platform accommodates star trackers, inertial measurement unit (IMU), and an accelerometer assembly for drag-free control capabilities; a Micro-Propulsion System (MPS) operated in science mode can be based on e.g. MICROSCOPE/GAIA/LPF heritage (employing e.g. a set of micro-proportional cold-gas thrusters for S/C pointing and translational control in science mode).

As to the spacecraft configuration, different options have been considered and conceptually studied at this level, and will have to be further assessed and traded-off in upcoming phases. In particular, major design drivers for a possible configuration turn out to result from the instrument accommodation, decoupling and thermo-mechanical stability.

A reference conceptual design elaborated here (on the basis of heritage from previous assessment studies e.g. M3, and other previous feasibility studies such as HYPER which in particular adopted a LPF-like S/C configuration) consists of a service module (SVM) with platform equipment and propellant tanks, and a payload module (PLM) which accommodates the instrument core, isostatically mounted at the inner central part of the spacecraft. The physics package (1500$\times$660 mm cylinder) is ideally co-located at the CoM of the spacecraft to minimize by design the impact of residual disturbances resulting from coupling terms associated with the S/C residual dynamics.

\begin{figure}[t]
\centering
\includegraphics[width=0.41\textwidth]{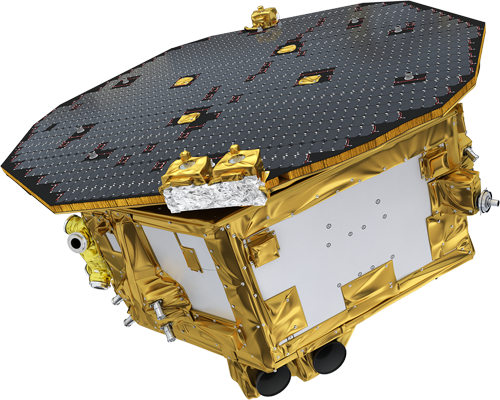}
\hspace{0.05cm}
\includegraphics[width=0.55\textwidth]{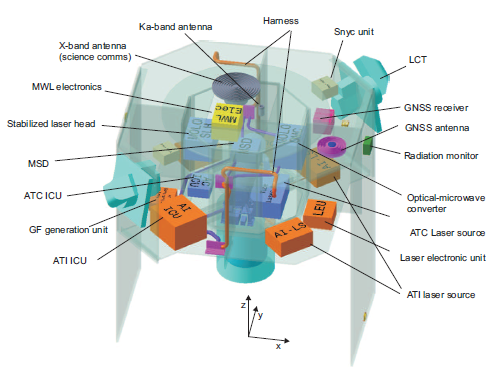}\\
\vspace{0.3cm}
\caption{\it Heritage for S/C design: LISA-Pathfinder (credit: ESA) and STE-QUEST M3. Only for heritage reference, e.g. many of the payload elements from M3 are not on STE-QUEST M7 (LCT, MWL, ATC, Stabilized laser, Ka-band antenna, \dots).
}
\label{fig:SC-heritage}
\end{figure}

The spacecraft, here outlined essentially at conceptual level, and subject to further consolidation in upcoming study phases, can be based on an octagonal shape with a diameter of $\sim$2.1 m and a height of the order $\sim$1~m. These dimensions are based on LPF heritage (2.1$\times$0.85 m) but will be adapted in phase A to further optimize payload accommodation and decoupling. 
The PLM, as mentioned, is accommodated in the well protected central region of the spacecraft, which places the instrument core (physics package) close to the center-of-mass (CoM) of the S/C and therefore minimizes the coupling to rotational accelerations (whilst also providing optimal shielding against the doses of radiation accumulated over the mission duration). A GNSS receiver (dual-band GPS) is used to support the primary measurements with POD. Communication is through S-band (X-band for science data). In summary, the S/C design draws on heritage from the STE-QUEST M3 study and the LISA-Pathfinder design. Both are shown in Fig. \ref{fig:SC-heritage} for reference.

The thermal control of the S/C can exploit the advantages of a SSO profile; radiators can be primarily accommodated on the S/C panels with radiative surfaces facing deep space; if needed, as a result of detailed assessments, sufficient thermal conductivity between dissipating units and radiators could be provided through the use of heat-pipes which could be embedded in the instrument baseplates.

The SVM can be based on a structure made of aluminum honeycomb panels, and carbon-fiber reinforced polymer face sheets; its nominal dry mass (i.e. with maturity margins included on the various platform subsystems) is at this level estimated at approximately 497 kg. The payload nominal mass (i.e. with all margins, 20\% on subsystem level and 20\% on payload level, see Tab. \ref{fig:budgets}) is currently estimated at 355 kg. That leads, when we include propellant mass and additional 20\% system margin, to a total S/C mass (excluding launch adapter) at approximately 1187 kg. The top-level mass budget is summarized in Tab. \ref{tab:SC-mass-budget}.

Although the detailed design will have to be conducted in upcoming study phases, the concept is expected to be designed/engineered targeting compatibility with a launch on e.g. VEGA~C; while mission scenario, total launch mass, envelope are expected to be compatible, based on preliminary assessments conducted under present assumptions, loads and in particular mechanical properties (e.g. structural Eigen-frequencies) will have to be further analyzed for compatibility with the corresponding requirements. A few distinctive design features on spacecraft subsystem level are discussed in more detail in the following, subject to further assessment in successive phases.

\subsubsection{Thermal control subsystem (TCS)}
Compared to previous assessment studies (M3), the re-baselining of the payload (i.e. omission of atomic clock and optical link) and of the mission profile (i.e. selection of a SSO profile) has overarching implications also on the design of several subsystems (which pose now potentially less severe challenges compared to the M3 study).    

In previous assessments (M3), the payload dissipation (of a large amount of power in the range of approximately 2kW - M3 estimate for the entire payload at that time i.e. atom interferometer, atomic clock and optical link, including maturity and system margins) represented a major challenge for the thermal system which could only be met through dedicated heat-pipes transporting the heat from the protected accommodation region in the spacecraft center to the radiator panels. Also problematic was the fact that the baseline orbit was not sun-synchronous and featured a residual drift of the right-ascension of the ascending node (RAAN), which led to seasonally strongly variable thermal fluxes incident on the spacecraft from all sides.

Although the (detailed) design will have to be conducted in upcoming phases, exploiting e.g. the advantage of a SSO profile, the thermal control of the spacecraft can in principle be simplified (e.g. based primarily on passive thermal control techniques, with limited use of controlled heaters, at selected temperature reference points (TRPs), and at the interface with the instrument). Dissipating units, payload electronics and spacecraft OBC (OnBoard Computer) can be accommodated on a S/C panel with radiating surfaces towards deep space, in an effective configuration that fully exploits the advantages of a SSO, while MLI blankets cover the SVM panels, and a low-thermal-conductance interface structure supports the solar array. The thermal interface with the instrument still is expected to pose more severe challenges, which could nevertheless be addressed with a multi-layer insulation system (in a combination of passive insulation and active control techniques), with the outer layer of the thermal shielding actively controlled. The PLM, located at the center of the S/C, with the physics package at its inner central part, will have to be to a large extent decoupled, radiatively and conductively from the rest of the S/C. Low-conductance isostatic mounts will be designed to minimize conductive coupling as well as mechanical distortions.

\subsubsection{Electrical power subsystem (EPS)}

Similarly to the TCS, also the design of the electrical power subsystem (EPS) can take advantage from the re-baselined mission profile and experiment concept. 
In previous assessments (M3), the EPS design was driven by the highly variable orbit featuring a large number of eclipses, the high power demand of multiple instruments, and the satellite pointing strategy in addition to the required stability of the spacecraft power bus. That resulted in a design with 2 deployable solar arrays at a cant angle of 45 degrees, rotated around the spacecraft y-axis (that configuration, in combination with two yaw-flips per year, ensured a minimum solar flux of approximately 1 kW/m2 on the solar panels which generated a total power of 2.4 kW, including margins, much above the average payload consumption which was at 1.3 kW). 
With a SSO profile and only one instrument baselined (atom interferometer), power requirements (average power and peak power) are now less demanding (see Tab. \ref{fig:budgets}), the EPS design can be simplified, mechanisms and moving parts can be avoided, and a design with a body-mounted array (in a LPF-type configuration) will be explored, depending on the further consolidation of the power budgets in upcoming study phases and the final S/C dimensions.

\subsubsection{Mechanical subsystem, and assembly, integration, testing (AIT) aspects}
    
The spacecraft structure can be made almost entirely from panels with an aluminum honeycomb structure (40 mm thickness considered at this level) and carbon-fiber reinforced polymer (CFRP) face sheets, which is favorable from a mechanical and mass-savings perspective. Furthermore, the structure can be designed to optimize AIT and aspects of parallel integration (and functional testing of the atom interferometer on-ground). As an important feature in the integration process, payload and service module components are completely separated in their respective modules up to the final integration steps, when they are finally joined and their respective harness connected on easily accessible interface brackets. 
Another important aspect related to the structural design and instrument accommodation is to maintain a symmetrical configuration (cylindrical symmetry) in order to minimize by design orientation-dependent effects in the inertial-pointing SSO orbit, while at the same time constraining self-gravity at the instrument core. 

Although (structural) design aspects associated with the radiation environment are less critical compared to previous assessment studies - M3 (as a result of the selection of a LEO), nevertheless, dedicated provisions, shielding and protected harness routing are foreseen for sensitive units such as the laser unit, payload electronics, and in particular optical fibers, sensitive to radiation-induced degradation.

\subsubsection{Summary tables}

\begin{table}[h]
\begin{minipage}{0.4\textwidth}
        \centering
        \begin{tabular}{c|c}
          \hline
		    Service Module & 497~kg \\
		    Payload Module & 355 kg \\
		    System margin (20\%) & 170 kg \\
		    Propellant (solid fuel) & 130 kg \\
		    Propellant (cold gas) & 35 kg \\
		    \hline
		    S/C wet mass & 1187 kg \\
            \hline
        \end{tabular}
        \caption{\it Top level mass budget preliminary estimates. Wet mass includes all S/C provisions except launch adapter.}
        \label{tab:SC-mass-budget}
    \end{minipage}
    \hspace{0.1\textwidth}
    \begin{minipage}{0.4\textwidth}
        \centering
        \begin{tabular}{c|c}
          \hline
		    Service Module & 309~W \\
		    Payload Module & 671 W \\
		    Losses (PCDU, harness, 5\%) & 49 W \\
		    System margin (20\%) & 206 W\\
		    \hline
		   Total power demand & 1235 W  \\
            \hline
        \end{tabular}
        \caption{\it Top level average power budget preliminary estimates.}
        \label{tab:SC-power-budget}
    \end{minipage}
\end{table}
  
For completeness, in Tabs. \ref{tab:SC-mass-budget} and \ref{tab:SC-power-budget}, the mass and power budgets of a reference satellite configuration considered at this level are provided, with the specific values estimated for the service module and the payload module. 

The S/C total mass (wet mass, i.e. resulting from the S/C dry mass – SVM plus PLM – with the addition of 20\% system margin, propellant and incl. all S/C provisions/harness, but excluding the launch adapter) is estimated at 1187 kg. The power budget adds up to a current estimate (S/C total i.e. including both SVM and PLM, average, with unit margins and 20\% system level margin) at a level of $\sim$1.24~kW, subject to further consolidation in upcoming study phases.
    
Both budgets (mass and power) were estimated by Airbus Defence and Space (Friedrichshafen) for the reference architecture considered under present assumptions, based on the expected mass and power demand at equipment level. The power estimates were then cross-checked/compared with the M3 phase-A estimates (i.e. M3 power budget at perigee), removing all equipment no longer on-board, and obtaining good agreement.
Finally the estimated $<$110~kbps downlink data rate required for the science data should not pose any difficulties and leaves quite some margin in the S/X-band communication channels~\cite{ESA2022}.

\section{Management scheme}
\subsection{Management scheme overview}
\begin{wrapfigure}{L}{0.5\textwidth}
\includegraphics[width=0.49\textwidth]{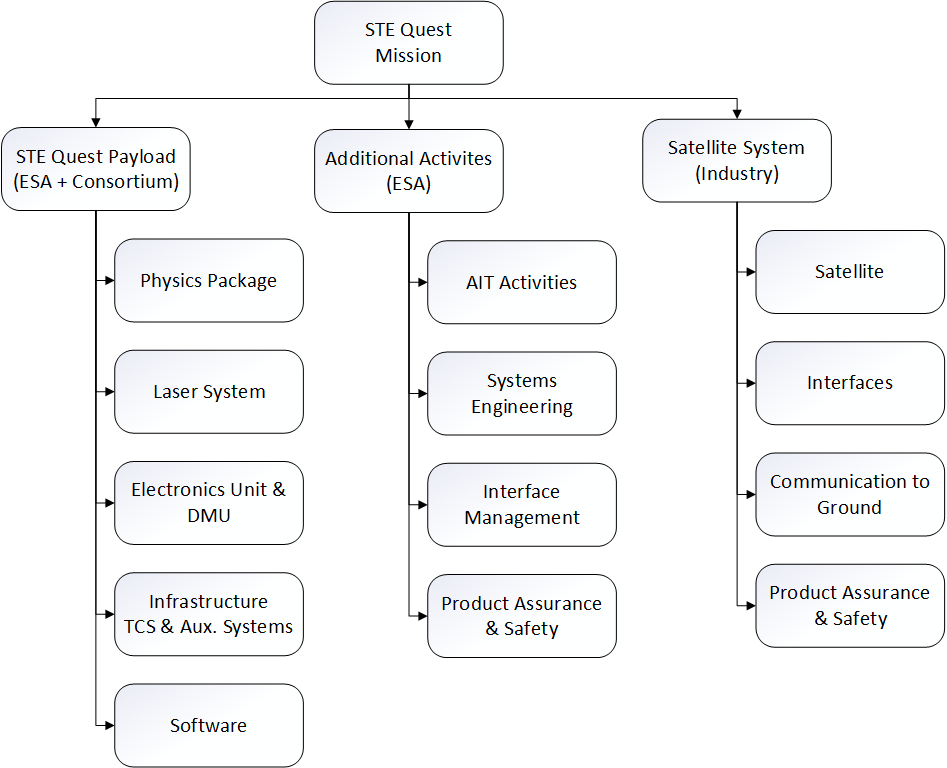}
\caption{\it Overview over the management structure and the relevant subsystems.}
\label{fig:mngt}
\end{wrapfigure}
On a top level, the STE-QUEST scientific mission can be divided into the payload part, under overall ESA responsibility but with sub-systems provided by the STE-QUEST consortium, and the satellite system, realized by an industrial partner that will be selected by ESA. Thus ESA will be in charge of overall system engineering and AIVT activities. The three main branches and their major constituents, such as the payload subsystems, ESA activities, and satellite components, are displayed in figure~\ref{fig:mngt}. 

In the following, the three branches are explained in greater detail:

The payload is divided into different subsystems. The main subsystems are the Physics package, the Laser system, and the Electronics unit. They are supported by the infrastructure and experiment software. These subsystems including the distribution of responsibilities are described in more detail in section~\ref{sec:payload}. The overall payload is managed centrally, with teams led by local managers to complete the efforts on the individual subsystems. The different subsystems, as explained in section~\ref{sec:payload} are not independent of one-another but require well defined interface control to work in unison, enabling the defined scientific goals. In consequence, regular interchange meetings and usage of joint systems engineering tools and file servers are necessary to ensure mission success. While the according systems engineering, product assurance, and interface control activities are managed by ESA, each subsystem defines a responsible for those areas within the payload. 

As outlined above, ESA coordinates the assembly, integration, and test activities including the overall interface control, product assurance, and systems engineering as well as development and delivery of STE-QUEST MOC and SOC. In addition to the interfaces between the payload subsystems, this includes the interfaces to the satellite and launcher, as well as the coordination of all product assurance and systems engineering activities. Due to the complexity of the payload each system and subsystem instates their representatives to interact closely with ESA and the other partners in the mission.

For these activities, ESA appoints a Project Manager, who implements and manages ESA’s responsibilities during the development and implementation phases, until launch and system commissioning.
The ESA Project Manager will be directly supported in the execution of the programme by the engineering, administrative, and project control staff of the ESA Project Office. The Project Manager is supported by the Project Scientist and Payload Manager, who oversee development of the mission throughout the different phases. The Project Office will hand over responsibility of the mission to the ESA Mission Manager after system commissioning.
The Mission Manager takes responsibility for spacecraft operations, the payload, and the ground segment, excluding the nationally funded IOCs and DPCs.
A Science Team will be appointed by ESA and, chaired by the Project Scientist, will develop the science strategy and guide science operations planning and execution.

Finally, the satellite itself is under the responsibility of an industrial partner. They design and build the satellite bus based on the requirements put forward by the payload during the definition phase. Consequently, the industrial partner is part of the milestone reviews and technical interchange meetings.
    
The distribution of the subsystems in the consortium and their organization can be viewed in figure~\ref{fig:consortium_organization}. This organizational chart displays the responsibilities for the payload subsystems.

\subsection{Work Breakdown Structure}

A top-level display of the work packages for STE-QUEST is outlined in figure~\ref{fig:wbs}. 
This mission is broken down in general activities, such as project management, overall systems engineering, interface management, product assurance, and risk management. 
Those are supported by the general activities for the payload, spacecraft, and launcher, for which individual system and subsystem management, engineering, and product assurance activities are set in place.

As outlined in the schedule (Fig.~\ref{fig:schedule} in Appendix III), the payload will be developed first on component and prototype level.
The subsystems of the engineering and qualification model (E(Q)M) and the flight model (FM) are the same. 
As it discussed above, component and bread board activities are set in place to increase the maturity prior to the preliminary design review and achieve TRL 5-6 in 2026. The E(Q)M and FM are fully integrated. In a later stage the usage of the E(Q)M as ground test bed is envisaged. 
Each of the prototyping activities and different models includes general activities, such as model management, systems engineering, and product assurance, and work packages for the different subsystems. 

\begin{figure}[ht]
\centering
\includegraphics[width=0.8\textwidth]{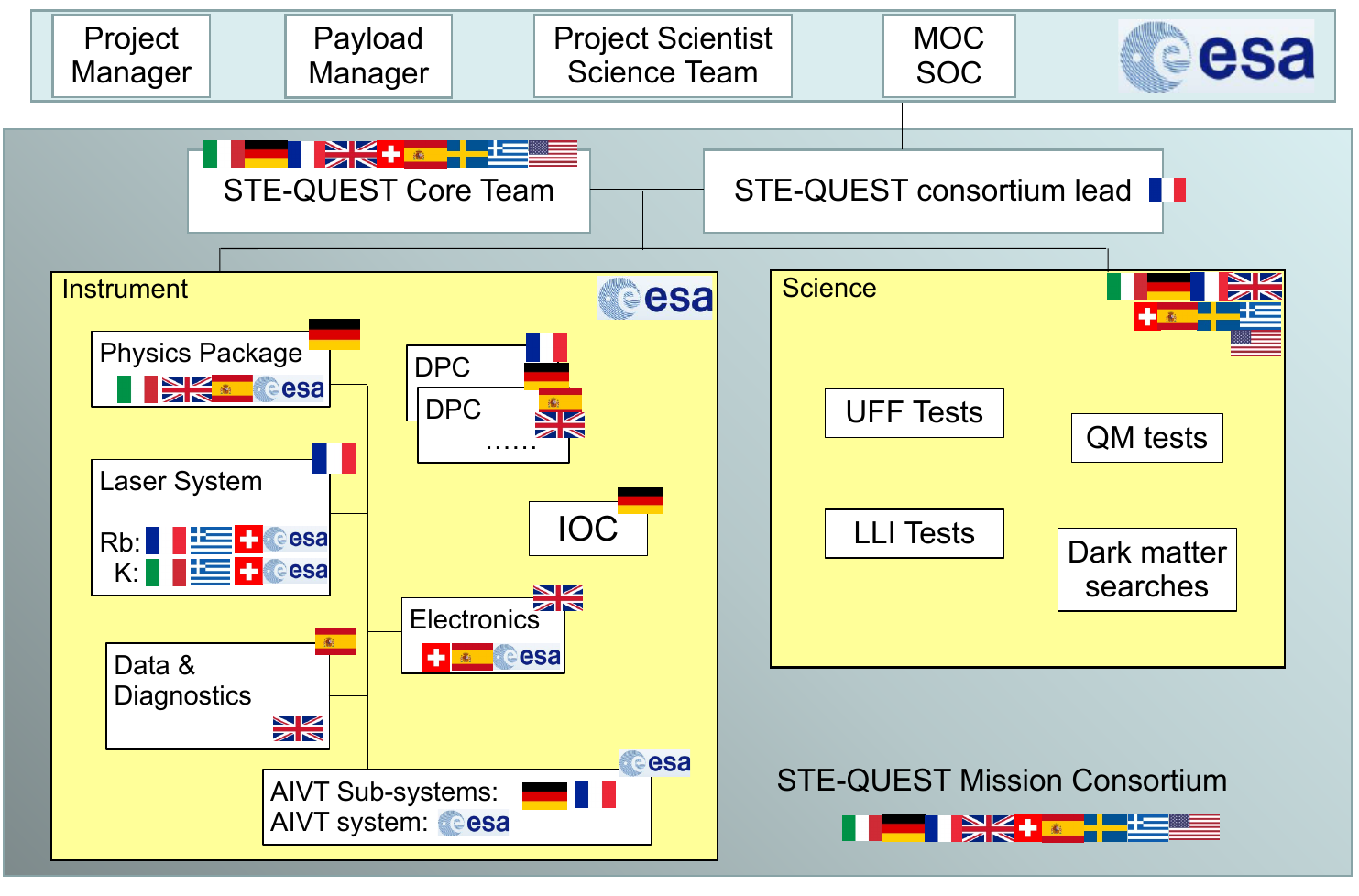}
\caption{Proposed STE-QUEST top level management structure. MOC: Mission Operation Centre, SOC: Science Operation Center, DPC: Data Processing Center, IOC: Instrument Operation Center, AIVT: Assembly Integration Validation and Testing.}
\label{fig:consortium_organization}
\end{figure}

\begin{figure}[ht]
\centering
\includegraphics[width=\textwidth]{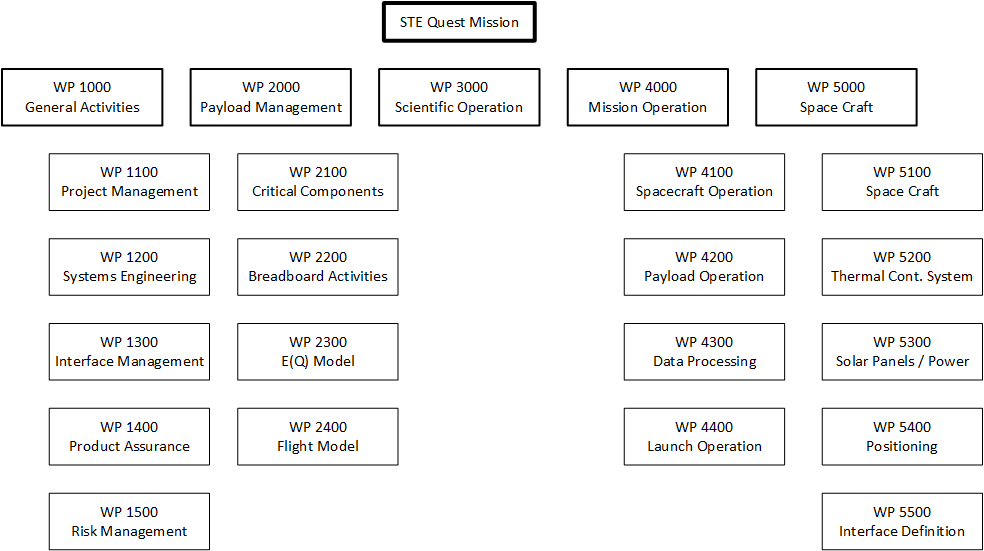}
\caption{An overview over the work breakdown structure.}
\label{fig:wbs}
\end{figure}

The responsibilities for the different work packages is sketched in figure~\ref{fig:consortium_organization}.

\subsection{Requirements Organization}

The first responsibility of the consortium is the definition of scientific requirements and their flow down to system and subsystem requirements. Those are the baseline for defining the requirements towards the satellite. The set of requirements is completed under the responsibility of ESA in close contact with the consortium, ensuring the consideration of both, scientific and engineering requirements.

The responsibilities for the requirements and subsystems are reflected in figure~\ref{fig:consortium_organization}. A mission consortium is formed by members of the scientific community, national agencies and ESA scientists (see section~\ref{sec:team}) that is responsible for the scientific and subsystem requirements including data analysis and preparation of experimental sequences. ESA is responsible for the interfaces between the payload and the satellite, which includes a cross check for necessary requirements to complete a meaningful satellite study. The resulting satellite requirements, covering internal systems, are the responsibility of the industrial partner.

The resulting requirements tree reflects the mission structure with requirements derived from scientific necessities, engineering limitations, and environmental specifications. By tracking inter-dependencies of the requirements, full transparency is ensured and impacts of changes or non-conformances can be traced. 


\subsection{Schedule and TRL}

Following the official kick-off of the project, the scientific requirements and top level system requirements are outlined, which lead to a preliminary design as a baseline for the satellite design study. During that three-year phase A study, the design of the planned payload is refined and adapted with regards to outer limitations and requirements set towards the payload. Once the initial study is completed, a preliminary design review will take place accounting for the necessity of possible changes or adaptations once the satellite design is chosen. Component level tests, breadboard activities and prototypes of subsystems are the basis of the payload engineering and qualification model (EQM). The EQM is deployed for qualification level environmental and functional verification. With successful completion of the EQM test campaigns, the critical design review (CDR) is held. This is then followed by the construction of the flight model (FM).  With these milestones, a potential launch in 2037 is would have been possible. 

In parallel to the payload development, the satellite is developed.
The current spacecraft design considerations are detailed in section~\ref{sec:S/C}.
Based on the results of the initial phase, the satellite design as well as the mission profile (see section~\ref{sec:miss-profile}) will be adapted to fit the mission needs. 

An example of a schedule is displayed in appendix~\ref{appendix:shedule}. 
It details the development of the payload and the satellite bus.

Following the above detailed model philosophy, environmental, and functional test campaigns, the technological maturity of the payload is increased during the project's run time.
The targeted technological readiness levels are outlined in section~\ref{sec:dev_plan}.
To reach TRL 5-6 in 2026, bread-boarding activities will be performed. 
This includes both functional and environmental tests on component and subsystem level. 
These activities are supplemented by prototypes of components which have been judged critical, have a low TRL, or deviate substantially from previously tested or flown hardware. 
These activities are performed at the beginning of the project and in parallel to the satellite study to ensure the validity of the preliminary design of STE-QUEST and the requirements towards the spacecraft. 
The timeline of these developments are shown in appendix~\ref{appendix:shedule} and the relevant milestones discussed in Table~\ref{tab:milestones}. 

The schedule does not show the ongoing interface control, product assurance, and system engineering activities necessary for the success of the mission.

\subsection{Milestone List}

In the following, the major milestones for STE-QUEST following the schedule in appendix~\ref{appendix:shedule} are displayed.
The milestones are separated along the project, with the first phase ending with the decision after phase A (initially expected to start in 2022) and the mission adoption marked after phase B1.
This review is preceded by a mission requirements review (M-RR), component prototypes and tests, and the satellite study. 
With those, the TRL of the individual components for STE-QUEST is increased according to the development plan, see section~\ref{sec:dev_plan}, to TRL 5 prior to the mission decision for critical components and to TRL 6 prior to mission adoption.

The second set of milestones describes the developments towards the preliminary design review, including the design development and additional prototyping activities as necessary. 
It ends with the preliminary design review. 
Afterwards, the first complete model, the engineering and qualification model (E(Q)M) is assembled, tested, and its functions verified. 
These activities have an impact on the design. 
Accordingly, this phase ends with the critical design review on payload (P-CDR), satellite (S-CDR), and mission level (M-CDR).
With the design being finalized, procurement of the flight model (FM) components start.  
Finally, the flight model is assembled, tested and integrated into the satellite. 
Following the mission flight acceptance review, the satellite is integrated into the launcher and the launch campaign prepared. 
After successful operation in orbit, the mission end of life review (M-EOL) is the final milestone within STE-QUEST.

The milestones will be supplied by necessary procurement and manufacturing readiness reviews as well as mandatory and key inspection points during the course of phase C. 

\begin{longtable}{lcc}
\hline 
{\bf Milestone } & {\bf Description} & {\bf Quarter} \\
\hline
KO & Kick Off & Q4 2022\\
M-RR & Mission Requirements Review & Q1 2023\\
S-D & Satellite Decision / Satellite Study Review & Q1 2025\\
P-C-TR & Payload Component Test Review & Q4 2025\\
M-D-A & Mission Selection after Phase A & Q2 2026 \\
\hline
P-BBM-TR & Payload Bread Board / Prototype Test Review & Q2 2029\\
M-D-B1 & Mission Adoption after Phase B1 & Q2 2029\\
\hline
P-PDR & Payload Preliminary Design Review & Q1 2030\\
S-PDR & Satellite Preliminary Design Review & Q1 2030\\
M-PDR & Mission Preliminary Design Review & Q1 2030\\
\hline
P-E-IRR & Payload E(Q)M Integration Readiness Review &  Q4 2031\\
P-E-TRR & Payload E(Q)M Test Readiness Review & Q3 2032\\
P-E-TR & Payload E(Q)M Test Review & Q4 2032\\
P-CDR & Payload Critical Design Review & Q4 2032\\
S-CDR & Satellite Critical Design Review & Q4 2032\\
M-CDR & Mission Critical Design Review & Q4 2032\\
\hline
P-F-IRR & Payload FM Integration Readiness Review &  Q3 2034\\
P-F-TRR & Payload FM Test Readiness Review & Q2 2035\\
P-F-TR & Payload FM Test Review & Q3 2035\\
S-D & Satellite Delivery & Q4 2035\\
P-SIRR & Payload Satellite Integration Readiness Review & Q4 2035\\
M-TRR & Mission Test Readiness Review & Q1 2036\\
M-FAR & Mission Flight Acceptance Review & Q4 2037\\
\hline
L-IRR & Launcher Integration Readiness Review & Q2 2037\\
M-FRR & Mission Flight Readiness Review & Q2 2037\\
M-L-C & Launch Campaign & Q3 2037\\
\hline
M-ORR & Mission Operation Readiness Review & Q3 2037\\
M-EOL & Mission End of Life Review & Q2 2040 \\
\hline
\caption{The milestones planned for the STE-QUEST M7 proposal.}\label{tab:milestones}
\end{longtable}

\subsection{Payload provision and responsibilities}
\label{sec:payload}

STE-QUEST is a mission with a single payload (the atom interferometer, ATI), with the satellite interface being essential, e.g., satellite self-gravity, drag-free, attitude control, etc. Some auxiliary payloads or systems (e.g. electrostatic accelerometer, GNSS receiver, de-orbiting system) are necessary. Since STE-QUEST is a mission with a single ``large, complex'' payload with sensitive interfaces to the spacecraft, ESA system engineering, AIVT and overall payload responsibility seems particularly adapted to this mission. This was already the case (for AIVT of the ATI) in the M4 proposal, and thus presents a moderate change with respect to the M4 version. The three payload subsystems, Laser system (LS), Physics package (PP), Electronics unit (EU), will be provided under Member-State responsibilities. International (NASA) collaboration is excluded at this phase due to the lack of a financial inter-agency agreement. This could be reconsidered at a later stage if this is changing. A close scientific contact with American scientists interested in STE-QUEST is maintained (see Section~\ref{sec:international}), as NASA was part of the M4 proposal. Table \ref{tab:payload} gives a distribution of payload subsystem contributions and responsibilities reflecting the outcome of discussions among the core team and with national agencies.

\subsubsection{Project scientist and science team}\label{sec:team}
The STE-QUEST Science Team (SST) monitors and advises the STE-QUEST Project/Operations Team on all
aspects affecting STE-QUEST scientific performance. The following key roles have been identified in the Science
Team:
\begin{itemize}
    \item The ESA Project Scientist (PS), representing the link between the Science Team and the STE-QUEST Project/Mission Operations Management in ESA
\item The STE-QUEST Consortium Lead (SCL), the formal interface of the STE-QUEST consortium to ESA. The
SCL provides link between the STE-QUEST consortium the SST and ESA, ensuring that the performances of the
mission meet the science requirements.
\item The STE-QUEST core team includes scientists from different backgrounds covering all aspects of STE-QUEST science, and providing links to all
national agencies that participate in STE-QUEST payload development.
\item The Payload Managers, focal point for the science-related aspects as well as the scientific performance of the
STE-QUEST instrument/s. 
\item The Data Analysis Coordinator/s (DAC), responsible for the definition of scientific algorithms for data
analysis, mission products generation, and exploitation.
\end{itemize}

\subsubsection{Procurement}
The Study and Definition Phase will be conducted following ESA best practice either through
parallel competitive contracts or by choosing a single system prime through open competition. A single system
prime will be chosen through open competition after Mission Adoption for the Implementation Phase (B2/C/D). The
industrial structure will take into account the geographical distribution requirements. The industrial prime will deliver
the fully integrated system to ESA and be responsible for design, manufacturing, integrations, testing, and verification
of the spacecraft. ESA will control and monitor the activities.

The procurement process will be supported by according procurement and manufacturing readiness reviews to ensure that the items are in line with the requirements of the project. 

\subsection{Science and data management}

Data directly resulting from the STE-QUEST spacecraft and ground segment (raw and calibrated data) will be owned by ESA and are provided by ESA to the STE-QUEST Science Team for analysis and publication of the scientific results. Wherever possible we will adopt a full open-data approach.

The data handling falls into two different periods:  A first (embargo) period which will be followed by an open-access period.  

The embargo period will last for 1 year following the acquisition and full calibration/verification of the dataset. It will focus the data analysis  on the key mission objectives. During this period the access to the data will be limited to specific focus groups: 

\begin{itemize}
    \item Members of the Science Team, and their support teams
    \item Members of the STE-QUEST consortium Core Team
    \item Members of the Science Operation Centre
    \item Members of the Data Processing Centres
    \item Members of the Instrument Operations Centres
    \item Members of the Instrument Consortia
    \item Accredited external users
\end{itemize}

During the embargo period, all scientific publications require validation and express approval by the STE-QUEST Science Team. Of course, such a review shall not unduly withhold the publication and shall be carried out within a reasonable time.

All data shall be protected, distributed, stored and handled by ESA in accordance with the applicable data policy. Arrangements shall be made with the STE-QUEST users so that they are committed to: Expeditiously provide to ESA an analysis of the results obtained from the planned scientific investigations; Take all reasonable steps to make these results available to the scientific community, or alternatively, authorize ESA to do so, through publication in appropriate journals or other established channels as soon as possible and consistent with good scientific practice.  Requests by External users to participate in the data analysis  of the STE-QUEST data during the embargo period will apply through the proper channels of ESA. Upon positive evaluation of the data analysis proposal by ESA and the STE-QUEST Science Team, the responsible scientists will be given access to the complete STE-QUEST data or part of it for analysis and publication of the results, as appropriate.

In a second period, the data will be made publicly available.  Great care will be taken to ensure full, meaningful accessibility.   The STE-QUEST Science Team and consortium Core Team will coordinate an optimal utilization and exploitation plan of the STE-QUEST data and data products. We envisage to use the data also for outreach activities, where the general public will be encouraged to perform a guided form of data analysis. All published uses of the data shall cite its usage in a predefined fashion.

\subsection{Community, Outreach and Communication}

\subsubsection{Community Engagement}

STE-QUEST touches a very large community ranging from atomic physics, over quantum mechanics to relativity to cosmology and beyond. 
These communities are all involved in the development and design of the STE-QUEST mission. 
Its ultimate success will come from on the involvement of all of these communities in the exploitation of the research results.
One of the main instruments in achieving this is the STE-QUEST workshop series.

A first Community Workshop on Cold Atoms in Space~\cite{CommunityWorkshop} established a   community road-map and milestones to demonstrate the readiness of cold atom technologies in space, as proposed in the Voyage 2050 recommendations, and in synergy with EU programmes.  
A more focused first STE-QUEST workshop~\cite{STE-QUESTWORKSHOP}, held on May 17 and  18, 2022, 
explored the science opportunities offered by the STE-QUEST mission. This workshop brought together leading representatives of the cold atom, quantum mechanics, particle physics, astrophysics, cosmology, fundamental physics, geodesy and earth observation communities to participate in shaping the details of the science program and mission profile.



The workshop was instrumental in building a wide STE-QUEST consortium, embracing Cold Atom technology experts as well as prospective Users.  
In total, 299 people from 26 countries registered as participants in the Workshop. 
As anticipated for a Europe-based event and ESA targeted mission proposal, about 80\% of the registrations are from European countries. The largest contingents were from from Germany (72), the United Kingdom (48), Italy (32), and France (30). Greece (8) and Spain (7) were also well represented. 
There is also significant North American (24)  and Asian (10) participation. 
The geographic distribution of the European participants  match well the overall responsibility sharing that is outlined in this STE-QUEST mission proposal. 
In terms of self-declared research interests, the registered participants of the workshop shows an almost even split between implementation of the sensor (e.g. cold atoms 54\%) and the application of its results.
Space industry was also very well represented. 
The scientific user community represent a rather divers field covering several key areas of Fundamental Physics, Earth Observation and Industry. 
The participants display an excellent and diverse mix of expertise, building an outstanding basis for the wider STE-QUEST consortium, and will provide the backbone for long-term planning and the support needed to see the challenging STE-QUEST cold atom missions through to its successful completions.
For a  list of registered supporters of STE-QUEST see Appendix III on page \pageref{appendix:supporters}.

\subsubsection{International Contributions }\label{sec:international}

Participation from JPL/NASA is actively being discussed. There is significant expertise and space mission implementation experience available at NASA/JPL, thanks to the CAL and BECCAL ultra-cold atom experiments on the ISS, which will be of great benefit to STE-QUEST. There is also a strong interest from US scientists to participate. It is our understanding that NASA's financial support for individuals participating in non-NASA led missions will be through agreements with NASA's mission partners. Such a strategic partnership on the fundamental physics mission of STE-QUEST will be heavily dependent on the outcome of the decadal study currently underway~\cite{decadal}. The STE-QUEST mission concept was submitted to the decadal whitepaper call jointly with several American scientists co-authoring it~\cite{STE-QUEST-decadal}. The decadal report and its recommendations are expected to be published in the summer of 2023. Therefore,  US/NASA participation and support may be determined no earlier than 2024. US contributions/NASA support in hardware can provide a financial margin, and may provide additional flexibility in the event of a cost overrun or funding difficulties of one or several national agencies or ESA itself.

\subsubsection{Outreach and Communication}
In order to maximize this impact, STE-QUEST will include an active communication strategy towards the technically minded and the general public, 
addressing the diverse communities as a whole and individually through workshops and targeted publications in specialized journals.
Considerable effort will be placed on achieving a large geographic spread especially within Europe but also beyond.

STE-QUEST will place considerable emphasis on the engagement of stakeholders and the public at large.
The large spread of targeted science cases together with the novel measurement principles (Quantum Sensor) makes it an ideal vehicle for education. 
Our research topics of gravity, general relativity, quantum mechanics and dark matter stimulate great interest among the general public, particularly the young.
Many of the STE-QUEST community have considerable experience and affinities with these agendas and are well-placed to deliver significant societal impact.
We will take this opportunity to inspire, educate and engage with stakeholders and the public concerning the underlying quantum technology and fundamental science of the STE-QUEST program. To this purpose, we will implement, apply, and further develop the outreach resources and tools investigated within a research-oriented framework in the pilot project Quantum Technologies Education for Everyone (QUTE4E), conducted within the Quantum Education Coordination and Support Action of the the Quantum Flagship, whose consortium's partners have contributed to~\cite{QUTE4E1,QUTE4E2,QUTE4E3,QUTE4E4}.  
  
The STE-QUEST community will lead the development of a professionally-made public-facing web-page that will provide access to an up-to-date status of the project, photos, explanatory materials, and a list of outreach contacts. 
This strategy will be reinforced by an outreach campaign directly aimed at schools. 
It will provide an interconnected set of resources on the web page, combined with exhibitions, and interactive content, designed to be suited to diverse interested audiences such as students, teachers, general public, and even policy makers. 
The designed and produced resources will also serve to promote STEAMs (Science-Technology-Engineering-Arts-Mathematics), which is lately suffering in Europe from a low influx of talent. 
Similarly, considerable efforts will be directed towards improving the gender balance in STEAM.
We believe that STE-QUEST path towards this goal will be especially innovative, in view of the composition of the Core Team and the significant intertwining of basic fundamental science and advanced technology contents of STE-QUEST.
We will also set STE-QUEST up on social media to provide up-to-the-minute status reports and advertise events.
These activities will be underpinned by a continuing program of outreach presentations at Open Days, schools, other educational establishments and science societies by the STE-QUEST community.
In all these activities we will follow the principles and framework of Responsible Research and Innovation in all their six dimensions to engage into a dialogue with the public and help shape the direction of future technology research and developments, ensuring public acceptance and societal benefit. 
Special attention will be devoted to the aforementioned dimensions of public engagement, education, gender, and open access, also implementing the RRI guidelines for quantum technologies outreach produced by QUTE4E~\cite{QUTE4E_report}. 

\section{Costing}

\subsection{Payload costs}

 STE-QUEST is an integrated mission built essentially around one major payload element which is the ATI. Payload funding is covered by national agencies, with contributions from ESA. As no single national agency can carry the full ATI cost, this requires a clear distribution of the ATI in terms of sub and sub-sub systems with clear responsibilities and interfaces (see Fig.~ \ref{fig:consortium_organization}). The individual financial contributions of each national agency and of ESA are calculated from the detailed breakdown of costs (reflected at top level in Table~\ref{tab:subsystems-costs}) and tailored to fit the capabilities of all partners. Particular care was taken to keep interfaces and responsibilities clearly defined (e.g. sub-system AIVT of all models in the same country as the largest sub-system contributor). 
 The top level responsibilities are indicated in Table~\ref{tab:payload} with individual contributions and corresponding costs.

\begin{table}[H]
\centering
\begin{tabular}{cc}
\hline 
{\bf Work package} & {\bf Cost (k\euro{})} \\
\hline \hline
Project management & 2990 \\
System Engineering & 7540 \\
Product Assurance & 4810 \\
Demonstrator, BB Activities & 6725 \\
Assembly, Integration, Verification and Test & 18460 \\
ATI Engineering Model & 34779 \\
ATI Flight Model & 48517 \\
ATI Science Ground Segment & 8326 \\
Launch Service & 2600 \\
\hline
Total w/o margin & 134747 \\
\hline
20\% Margin & 26949.4 \\
\hline
{\bf Total} & {\bf 161696.4} \\
\hline
\end{tabular}
\caption{\it Top level cost breakdown of ATI and scientific data analysis}
\label{tab:subsystems-costs}
\end{table}

In the STE-QUEST M4 proposal the total ATI cost was estimated by the consortium to be 126~M\euro{}, with an ESA contribution estimated at the time to be 30~M\euro{}. In the present baseline, the total ATI payload is estimated to cost 161.6~M\euro{} with an assumed ESA contribution of about 48~M\euro{}. The cost was obtained by detailed analysis of the M4 payload elements (themselves resulting from the M3 phase-A study), modifications where necessary for the new mission profile and overall performance, and adding 30\% to account for 2022 economic conditions. The current baseline for ESA contribution includes the lead of the system engineering, AIVT, Product Assurance (PA) and main contributions to the laser system (dipole trap), electronics (RF synthesis) and physics package (magnetic shielding).
\begin{table}[H]
\hspace{-5mm}
\begin{center}
\hspace{-5mm}

\begin{tabular}{ccc}
\hline 
{\bf Country/Agency} & {\bf Subsystem contribution} & {\bf Cost contribution} \\
\hline
France & {\bf Laser system (LS) Rb, LS AIVT} & 26.9 M\euro \\
Germany & Atom chip source, {\bf Physics package (PP) AIVT} & 26.4
M\euro \\
Greece & LS contribution & 2.9 M\euro \\
Italy & {\bf Laser system (LS) K}, Vacuum system & 15.1 M\euro \\
Spain & {\bf Data \& Diagnostics subsystem}, software, DMU & 8.9 M\euro \\
Sweden & Science & 0.7 M\euro \\
Switzerland & LS, EU contribution & 6.8 M\euro \\
United Kingdom & {\bf Electronics Unit (EU)}, Detection system & 25.8 M\euro \\
ESA & {\bf ATI syst. eng. \& AIVT}, PP, LS, EU contribution & 48.1 M\euro \\
\hline
{\bf Total} &  & {\bf 161.6 M\euro } \\
\hline
\end{tabular}
\caption{\it Preliminary atom interferometer (ATI) payload contributions and responsibilities (boldface indicates overall subsystem or payload responsibility). The cost contributions include a 20\% margin.}\label{tab:payload}
\end{center}
\end{table}

\subsection{Overall ESA costs}

The present cost estimation by the consortium is based on the STE-QUEST M4 estimate conducted by ESA (debriefing note, 2015) which differed from that of the consortium by about +17~\%. The resulting ESA cost estimate was deemed too high for the stringent M4 boundary conditions (450~M\euro{} Cost at Completion, CaC). Table~\ref{tab:cost} shows the M4 ESA cost estimates as a reference and the present consortium cost estimates for M7. Significant cost savings have been taken into account like the Vega-C rather than Soyuz launch (saving $\approx$~30~M\euro), the de-scoped payload elements (MWL) and the simplified mission scenario (circular orbit) and ground segment in the absence of MWL ground stations and time/frequency infrastructure. 

\begin{table}[H]
\hspace{-5mm}
\begin{center}
\hspace{-5mm}
\begin{tabular}{cccc}
\hline 
& {\bf ESA M4 (M\euro)} & {\bf M7 lower bound (M\euro)} & {\bf M7 upper bound (M\euro)} \\
\hline
ESA project team & 56 & 58.2 & 58.2 \\
Industrial cost  & 224 & 220 & 260 \\
Payload contribution (ESA) & 55 & 48.1 & 48.1 \\
Mission Operations (MOC) & 43 & 48 & 48\\
Science Operations (SOC) & 27 & 15 & 15 \\
Launcher & 73 & 45.5 & 45.5 \\
Contingency (10\%) & 40 & 39 & 43 \\
\hline
{\bf Total} & {\bf 518} & {\bf 474} & {\bf 518} \\
\hline
\end{tabular}
\caption{\it M7 consortium estimate of the ESA CaC for the STE-QUEST proposal with the M4 ESA estimate as a reference. The contingency (10\%) excludes the payload contribution for which a 20\% margin is already applied. The lower and upper bounds reflect the range in costs of  industrial activities to build the S/C as estimated by Airbus Defence and Space.}\label{tab:cost}
\end{center}
\end{table}
Two scenarios are displayed in Table~\ref{tab:cost} corresponding to the lower and upper bounds in industrial costs of the S/C platform as estimated by Airbus Defence and Space specifically for the M7 mission profile and baseline. The ESA project team cost of M4 was reduced by 20\% to account for the absence of the MWL development. The mission operations (MOC) were adapted from M4 to a 3-year operation (instead of 3.5). Both were then increased by 30\% to reflect 2022 economic conditions. The Science Operations (SOC) were divided by half (no MWL and time/frequency ground infrastructure) compared to M4, adapted to a 3-year operation and also increased by 30\%. A contingency of 10\% was added on top of all costs excluding the payload contribution which is derived with 20\% margin already. Either estimate, with all margins and overheads included, fits well in the 550~M\euro{} cost envelope of the M7 call with a comfortable margin in case of cost overruns.
\\


\section{Summary}

In summary, STE-QUEST is a fundamental physics mission concept that tackles several of the most puzzling questions in modern physics: violation of the principles of General Relativity, the foundations of Quantum Physics and searches for Dark Matter. 

The mission concept has a maturity of more than a decade since it was first proposed as an M-mission candidate which allowed the core team to overcome critical technological, scientific and financial challenges.

The science case of STE-QUEST is attractive to quite a large, inter-disciplinary community of supporters worldwide and is occupying a central position in the cold atom roadmap for space offering clear synergies with Earth Observation and other quantum space technology missions.  

\clearpage

\section{Acknowledgments}
We thank Robert Ecoffet, Domenico Gerardi, Thierry Martin, Alain Robert, Noah Saks and the numerous colleagues who contributed to the past STE-QUEST proposals.

\newpage
\clearpage

\bibliographystyle{JHEP}
\bibliography{sample}

\clearpage

\section*{Appendix I: Acronyms \label{appendix:acronyms}}

\begin{longtable}{l l}
ACES & Atomic Clock Ensembles in Space \\
AIT & Assembly, Integration and Testing \\
AIVT & Assembly, Integration, Verification and Testing \\
ASIC & Application-Specific Integrated Circuit \\
ATI & Atom Interferometer \\
AOM & Acouto-Optic Modulator \\
ASW & Application Software \\
BEC & Bose-Einstein Condensate \\
BECCAL & Bose-Einstein Condensate and Cold Atom Laboratory\\
BSW & Boot Software \\
CaC & Cost at Completion \\
CAL & Cold Atom Laboratory \\
CCD & Charged-Coupled Device \\
CCSDS & Consultative Committee for Space Data Systems \\
CDF & Concurrent Design Facility \\
CFRB & Carbon Fibre Reinforced Polymer \\
CMOS & Complementary Metal-Oxide Semiconductor \\
CoM & Centre Of Mass \\
COTS & Commercial off-the-Shelf \\
CSL & Continuous Spontaneous Localization model \\
DFACS   &  Drag Free and Attitude Control System \\
DFB & Distributed Feedback Laser \\
DKC & Delta Kick Collimation \\
DM & Dark Matter \\
DMU & Data Management Unit \\
DP & Di\'osi-Penrose model \\
DPC & Data Processing Center \\
ECDL & External Cavity Diode Laser \\
ECSS & European Cooperation for Space Standardization \\
EDFA & Erbium Doped Fibred Amplifier \\
EEP & Einstein Equivalence Principle \\
EEPROM & Electrically Erasable Programmable Read-Only Memory \\
EPS & Electrical Power Subsystem \\
FPGA & Field Programmable Gate Array \\
FPR-AT & Fundamental Physics Roadmap - Advisory Team \\
GG & Gravity Gradient \\
GGC & Gravity Gradient Cancellation \\
GOES & NOAA/NASA's Geostationary Operational Environmental Satellites\\
CoM & Center of Mass \\
GNSS & Global Navigation Satellite System \\
GR & General Relativity \\
GRW  & Ghirardi-Rimini-Weber\\
GTB  & Ground Test Bed\\
HEO & Highly Elliptical Orbit \\
IFO & InterFerOmeter \\
IMU & Inertial Measurement Unit \\
IOC & Instrument Operation Center \\


LCI & Laser Cooling and Interferometry \\
LEO & Low Earth Orbit \\
LLI & Lorentz Local Invariance \\
LPF & LISA PathFinder \\
LS & Laser System \\MAIUS & Materiewelleninterferometrie unter Schwerelosigkeit \\
MOC & Mission Operation Centre \\
MOT & Magneto-Optical Trap \\
MPS & Micro-Propulsion System \\
MWL & Microwave Link \\
NGGM & Next Generation Gravity Mission \\
OBC & OnBoard Computer \\
PCDU & Power Conversion and Distribution Unit \\
PLM & PayLoad Module \\ 
PP & Physics Package \\
PPLN & Periodically Poled Lithium Niobate \\
PRIMUS & Präzisionsinterferometrie mit Materiewellen unter Schwerelosigkeit \\
PROM & Programmable Read-Only Memory \\
PSD & Power Spectral Density \\
PUS & Packet Utilization Standard \\
QUANTUS & Quantengase unter Schwerelosigkeit \\
RAM & Random-Access Memory \\
RefL & Reference Laser\\
RF & Radio Frequency \\
RTEMS & Real-Time Executive for Multiprocessor Systems \\
RTOS & Real-Time Operating System \\
RUP &  Rational Unified Process \\
S/C & Spacecraft \\
SDO & NASA's Solar Dynamic Observatory\\
SME & Lorentz violating Standard Model Extension \\
SOC & Science Operation Center \\ 
SSO & Sun Synchronous Orbit \\
STM & Structural Thermal Model \\
SVM & SerVice Module \\
TC & Telecommand \\
TCS & Thermal Control System \\
TEC & Thermo-Electric Cooling \\
TLM & Telemetry \\
TRL & Technology Readiness Level \\
TRP & Thermal Reference Point \\
UFF & Universality of Free Fall \\
ULDM & Ultra Light Dark Matter \\
UML & Unified Modeling Language \\
WEP & Weak Equivalence Principle \\
WDM-Rb/K & Wavelength Division Multiplexing unit\\
\label{tab:acronyms}
\end{longtable}

\clearpage

\section*{Appendix II: Detailed TRL assessment  \label{appendix:TRL}}

\begin{longtable}{|p{0.2\linewidth}|c|c|p{0.2\linewidth}|p{0.3\linewidth}|}
\hline 
{\bf Components} & {\bf TRL 2022} & {\bf TRL 2026} & {\bf Heritage} & {\bf Development plan}  \\
{\bf Physics Package}&&&&\\ \hline\hline

Science chamber & 3--4 & 5 & QUANTUS~\cite{Deppner2021,Rudolph2015}, MAIUS~\cite{Piest2021,Lachmann_2021,becker2018} & Design adaptaion (size, shape), STE-QUEST environmental tests \\\hline
- Anti-straylight coating & 3--4 & 5/6 & Ref.~\cite{Vovrosh2020} & Functional, performance, and vacuum test, STE-QUEST environmental tests \\ \hline
- Metal body, viewports, sealing technique & 4--5 & 5/6 & QUANTUS, MAIUS, ICE~\cite{barrett2016,Geiger2011}, PRIMUS~\cite{Vogt2020,Kulas2017} & STE-QUEST environmental tests\\\hline
Atom chip & 3 & 5/6 & QUANTUS, MAIUS & Dedicated evaluation of QUANTUS, MAIUS atom chips + adaptation, STE-QUEST environmental tests\\ \hline 
- Chip wire structures, mesoscopic wire structures, RF structures & 4--5  & 5/6 & QUANTUS, MAIUS & STE-QUEST environmental tests \\ \hline
2D-MOT chamber & 3--4 & 5 & QUANTUS, MAIUS & Design adaptation (size, shape), STE-QUEST environmental tests \\ \hline
- Metal body, viewports, sealing technique & 4--5 & 5/6 & QUANTUS, MAIUS & STE-QUEST environmental tests \\ \hline
Oven / revervoir for Rb, K & 4--5 & 5/6 & PHARAO / ACES~\cite{PHARAO_synthesis}, QUANTUS, MAIUS & STE-QUEST environmental tests \\ \hline
Valve (between reservoir and 2D-MOT chamber) & 4--5 & 5/6 & COTS part & Functional, performance, and vacuum test, STE-QUEST environmental tests \\ \hline
Beam shaping optics & 4--5 & 5/6 & QUANTUS, MAIUS, ICE, PRIMUS & STE-QUEST environmental tests \\\hline
Retroreflection mirror & 4 & 5/6 & COTS part & STE-QUEST environmental tests \\\hline
Tip-tilt stage (for retroreflection mirror) & 3--4 & 5/6 & BECCAL~\cite{Frye2021}, COTS part & Functional, performance, and vacuum test, STE-QUEST environmental tests, EO pathfinder mission \\
\hline
Ion getter pump (incl. magnetic shield) & 4--5 & 5/6 & QUANTUS, MAIUS, ICE, PRIMUS & STE-QUEST environmental tests\\ \hline
Passive getter pump & 4--5 & 5/6 & QUANTUS, MAIUS, ICE, PRIMUS & STE-QUEST environmental tests\\ \hline
Coils & 4--5 & 5/6 & QUANTUS, MAIUS, ICE, PRIMUS & STE-QUEST environmental tests\\ \hline
Cameras & 4--5 & 5/6 & QUANTUS, MAIUS, ICE, PRIMUS / COTS part & STE-QUEST environmental tests\\ \hline
- Detection system for additional science goal (camera + optics) & 3 & 5/6 & QUANTUS, MAIUS, ICE, PRIMUS & Dedicated study / development, functional, performance, STE-QUEST environmental tests\\ \hline
Magnetic shield & 4--5 & 5/6 & PHARAO / ACES, QUANTUS, MAIUS, ICE, PRIMUS, BECCAL & STE-QUEST environmental tests\\ \hline
\caption{\it Summary of the technology readiness level for the components of the physics package and development plan to reach TRL 5/6 in 2026 for all the components.} \label{tab:TRL_PP}
\end{longtable}

\begin{longtable}{|p{0.2\linewidth}|c|c|p{0.2\linewidth}|p{0.3\linewidth}|}
\hline 
{\bf Components} & {\bf TRL 2022} & {\bf TRL 2026} & {\bf Heritage} & {\bf Development plan}  \\
{\bf Laser System}&&&&\\ \hline\hline
External Cavity Laser Diode & 4--5 & 5/6 & CNES demonstrator &    STE QUEST environmental tests \\\hline
Telecom Optical isolator & 4--5 & 5/6 & CNES demonstrator & STE QUEST environmental tests \\\hline
Fibered splitter & 4--5 & 5/6 &  CNES demonstrator   & STE QUEST environmental tests\\\hline
Phase modulator & 5 & 5/6 & GRACE-FO & STE QUEST environmental tests\\ \hline
Erbium Doped Fiber Amplifier (EDFA)& 4--5  & 5/6 & CNES demonstrator& STE QUEST environmental tests \\ \hline
Telecom fibered AOM& 4--5 & 5/6 & CNES demonstrator & STE QUEST environmental tests \\ \hline
PPLN waveguide & 4--5 & 5/6 & CNES demonstrator & STE QUEST environmental tests \\ \hline
780 nm fibered AOM & 4--5 & 5/6 & CNES demonstrator & STE QUEST environmental tests \\ \hline
Shutter & 4--5 & 5/6 & CNES demonstrator & STE QUEST environmental tests \\ \hline
Micro-optical Bench (MOB) & 4--5 & 5/6 & CNES demonstrator & STE QUEST environmental tests \\\hline
Dichroic filter 767 nm/780 nm & 3--4 & 5/6 & ICE experiment & environmental tests \\
\hline
Laser Reference Unit & 4--5 & 5/6 & Pharao/ACES (Cs) & Adaptation to Rubidium and Potassium\\ \hline
High Power EDFA (Dipole trap) & 4 & 5/6 & CNES demonstrator & Adaptation to higher optical power\\ \hline
\caption{\it Summary of the technology readiness level for the components of the laser system and development plan to reach TRL 5/6 in 2026 for all the components.} \label{tab:TRL_LS}
\end{longtable}

\begin{longtable}{|p{0.2\linewidth}|c|c| p{0.5\linewidth}|}
\hline 
\bf{Components} & \bf{TRL 2022}&\bf{TRL 2026}&\bf{TRL Notes}\\
{\bf Electronics}&&&\\ \hline\hline
Spectroscopy ECDL Driver &
3-4 &
5 &
Flight-targeted laser driver, temperature stabilization and spectroscopy locking verified in laboratory environment. Further work required to characterize against STE-QUEST performance requirements and subsequent progression to relevant environment.\\ 
\hline

Offset ECDL Driver &
3-4 &
5 &
Flight-targeted laser driver, temperature stabilization and offset locking verified in laboratory environment. Further work required to characterize against STE-QUEST performance requirements and subsequent progression to relevant environment. \\
\hline

PPLN Temperature Controller &
3-4 &
5 &
Flight-targeted temperature stabilization of laser diode verified in laboratory environment.  Further work required to translate to PPLN, characterize against STE-QUEST performance requirements and subsequent progression to relevant environment.\\
\hline

EDFA Driver &
3-4 &
5 &
Flight-targeted low-noise, high-current driver verified in laboratory environment. Further work required to translate to EFDA, characterize against STE-QUEST performance requirements and subsequent progression to relevant environment.\\
\hline

DFB Laser Driver&
3-4&
5&
Flight-targeted laser driver verified in laboratory environment. Further work required characterize against STE-QUEST performance requirements and subsequent progression to relevant environment. \\
\hline

Optical Shutter Driver&
3-4&
5&
Further definition of flight-targeted electrical interfaces required, no anticipated blockers to TRL raising.\\
\hline

Magnetic Coil Driver&
3-4&
5&
Flight-targeted magnetic coil drive verified in laboratory environment. Further work required to characterize against STE-QUEST performance requirements and subsequent progression to relevant environment.\\
\hline

Heater/Valve Driver&
3-4&
5&
Further definition of flight-targeted electrical interfaces required, no anticipated blockers to TRL raising. \\
\hline

NEG Getter Pump Drive&
3-4&
5&
Further definition of flight-targeted electrical interfaces required, no anticipated blockers to TRL raising. \\
\hline

DMU, Timing \& Synchronization Control&
3-4&
5&
DMU hardware based on established FPGA/processor architectures, no anticipated blockers to TRL raising.  Timing and synchronization proof of concept design work completed.  Further work required to translate to STE-QUEST payload architecture and requirements. \\
\hline

CCD Camera System&
5&
5&
Strong flight heritage (TRL 9) for CCD electronics, some re-design likely required to meet form factor and specific CCD requirements for STE-QUEST (Heritage from SDO and GOES).  Environment remains applicable therefore reduced to TRL 5. \\
\hline
\caption{\it Summary of the technology readiness level for the components of the electronics system and development plan to reach TRL 5/6 in 2026 for all the components.} 
\label{tab:TRL_electronics}
\end{longtable}
\clearpage

\section*{Appendix III: Schedule \label{appendix:shedule}}
\begin{figure}[ht]
\centering
\vspace{-0.5 cm}
\includegraphics[width=0.75\textwidth]{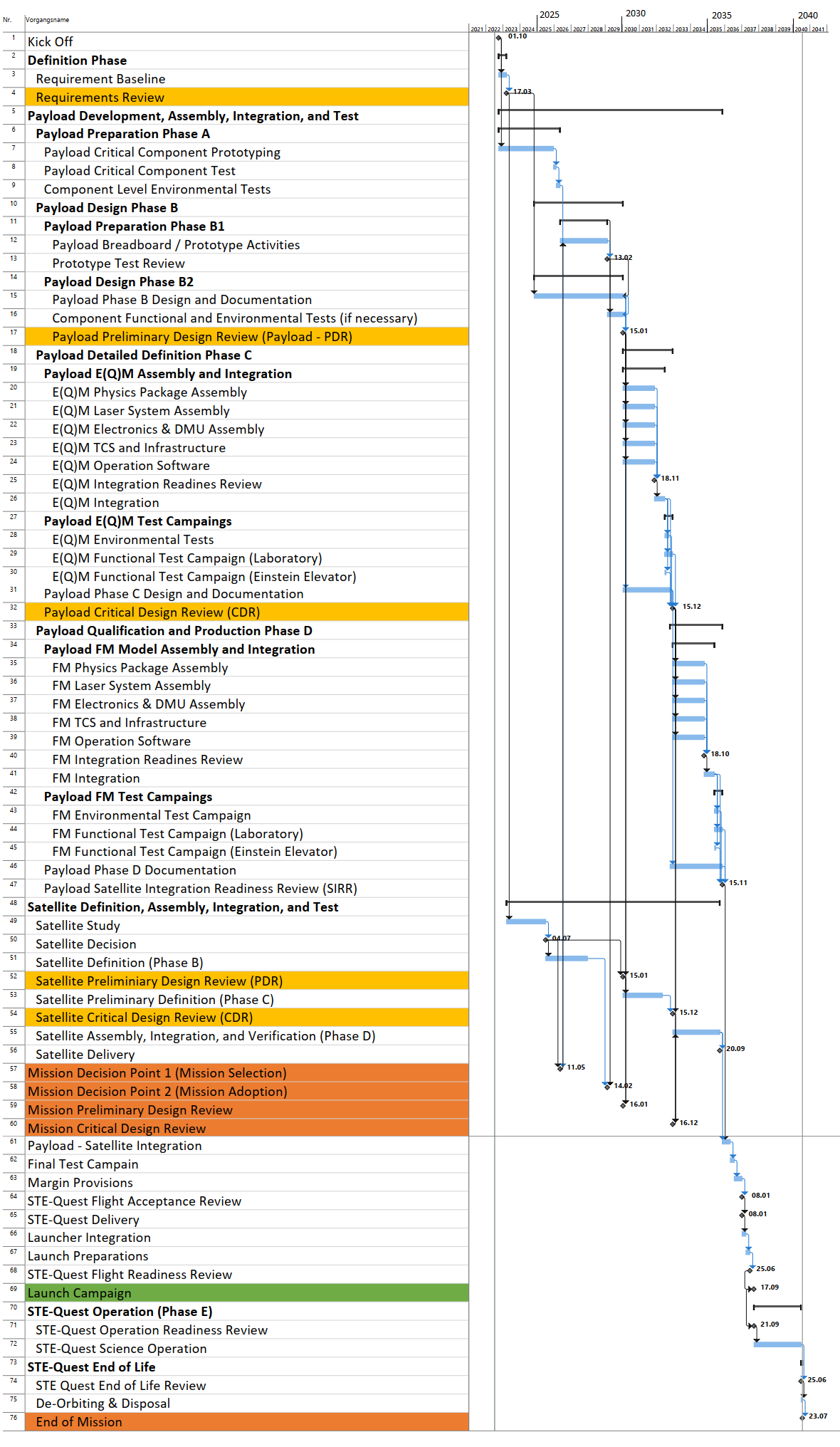}
\vspace{-0.4cm}
\caption{Example of a schedule for STE-QUEST based on the M7 proposal timeline.}
\label{fig:schedule}
\vspace{-110pt}
\end{figure}

\clearpage
\section*{Appendix IV: List of supporters \label{appendix:supporters}}
\foreach \x in {1,2,3,4,5,6,7,8,9,10,11,12,13,14}{  
  \includegraphics[page=\x,width=\textwidth]{STE-QUEST-Supporters-short.pdf}
  \newpage  
}

\end{document}